\documentclass[aps,twocolumn,pre,floatfix]{revtex4-2}

\usepackage{xcolor,hyperref}
\hypersetup{
   colorlinks,
   linkcolor={blue!50!black},
   citecolor={blue!50!black},
   urlcolor={blue!80!black}
}

\usepackage{epsfig,amsmath}
\usepackage{bm}
\usepackage{soul,ulem}
\usepackage{hyperref}
\usepackage{footmisc}
\usepackage{wrapfig}
\DeclareMathAlphabet{\mathitbf}{OML}{cmm}{b}{it}
\newcommand{\pa}{\partial}

\renewcommand{\=}{\!=\!}
\DeclareMathAlphabet\mathbfcal{OMS}{cmsy}{b}{n}
\begin{document}

\title{Yielding and memory in a driven mean-field model of glasses}
\author{Makoto Suda$^1$}
\author{Edan Lerner$^2$}
\author{Eran Bouchbinder$^3$}
\affiliation{$^1$Department of Earth Science, Tohoku University, Sendai, Miyagi 980-8578, Japan\\
$^2$Institute for Theoretical Physics, University of Amsterdam, Science Park 904, 1098 XH Amsterdam, the Netherlands\\
$^3$Chemical and Biological Physics Department, Weizmann Institute of Science, Rehovot 7610001, Israel}

\begin{abstract}
Glassy systems reveal a wide variety of generic behaviors, which lack a unified theoretical description. Here, we study a mean-field model, recently shown to reproduce the universal non-phononic vibrational spectra of glasses, under oscillatory driving forces. The driven mean-field model, featuring a disordered Hamiltonian structure, naturally predicts the salient dynamical phenomena in cyclically deformed glasses. Specifically, it features an oscillatory yielding transition, characterized by an absorbing-to-diffusive transition in the system's microscopic trajectories and large-scale hysteresis. The model also reveals dynamic slowing-down from both sides of the transition, as well as mechanical and thermal annealing effects that mirror their glass counterparts. Finally, we demonstrate a non-equilibrium ensemble equivalence between the driven post-yielding dynamics at fixed quenched disorder and quenched disorder averages of the non-driven system, along with memory formation.
\end{abstract}

\maketitle

\section{B\lowercase{ackground and motivation}}
\vspace{-0.4cm}

Glasses are formed by rapidly cooling (quenching) equilibrium liquids such that crystallization is avoided~\cite{Goldstein1969-ao,Debenedetti2001-an,Ediger2012-xg,Cavagna_pedestrians_2009}. Glasses are intrinsically non-equilibrium and disordered materials, revealing generic properties and dynamical behaviors that differ from their ordered crystalline counterparts. These include universal properties of the low-frequency vibrational spectra of glasses, which affect low temperature thermodynamic and transport properties~\cite{Zeller1971-oj,Phillips1981-ue,Pohl2002-bm}, and strongly driven thermo-mechanical phenomena such as thermal and mechanical annealing, yielding transitions from quiescent to plastically-flowing states, and memory formation~\cite{Utz2000-ry,Viasnoff2002-yk,Lacks2004-bf,Berthier2025-hv,Keim2019-ez,Paulsen2025-av}. All of these are inherently linked to the multiplicity of energy minima of glassy systems, manifested in their complex energy landscapes~\cite{Goldstein1969-ao,Debenedetti2001-an,Ediger2012-xg,Cavagna_pedestrians_2009,Gupta2019-mb}. Despite their basic importance, along with a broad range of engineering applications~\cite{Ojovan2011-pc,Simpson2011-tb,Jones2013-ri},
we currently lack a unified and predictive theory that includes
both sets of features described above.

Recent progress established that the vibrational density of states (VDoS) ${\cal D}(\omega)$ of glasses, experimentally probed through various scattering techniques~\cite{buchenau1999neutron,Hudson2006-sk}, features a non-phononic contribution ${\cal D}_{\rm G}(\omega)$ for small vibrational frequencies $\omega$, i.e., of soft modes~\cite{soft_potential_model_1991,chalker2003,Gurevich2003,Gurevich2007,Lerner2016-ra,Mizuno2017-sc,Lerner2021-en,Moriel2024-lp}. That is, in addition to extended low-frequency phonons, which exist in both glassy and crystalline solids, and follow Debye's VDoS, glasses also feature localized, non-phononic modes that follow a universal quartic VDoS ${\cal D}_{\rm G}(\omega)\!\sim\!\omega^4$ for $\omega\!\to\!0$~\cite{Lerner2016-ra,Mizuno2017-sc,Lerner2021-en,Moriel2024-lp}. Moreover, ${\cal D}_{\rm G}(\omega)$ features a generic peak~\cite{kalampounias2006low,yannopoulos2006analysis,Moriel2024-wp,Moriel2024-lp} at slightly higher frequency $\omega_{\rm p}$ (typically in the THz regime), intrinsically related to the well-known ``boson peak''~\cite{sokolov_1986,ramos1997quantitative,wischnewski1998neutron,surovtsev2003density,parisi_boson_peak_2003,wyart_epl_2005,monaco2006density,baldi2023vibrational,Schirmacher_2013_boson_peak,eric_boson_peak_emt,Gonzalez-Jimenez2023-un}. ${\cal D}_{\rm G}(\omega)$ reflects statistical properties of minima of the underlying complex energy landscape, a manifestation of self-organization during glass formation. Furthermore, dissipative deformation in driven glasses --- i.e., transitions between different minima of the landscape --- corresponds to irreversible (plastic) rearrangements of the localized, non-phononic modes~\cite{Malandro_Lacks,lemaitre2004,Tanguy2010-at,Manning2011-tu,falk_soft_modes_pnas_2014,micromechanics2016,Richard2020-vr}.

In parallel, extensive work has demonstrated that glasses driven by oscillatory forces undergo a plastic yielding transition with increasing driving amplitude~\cite{Fiocco2013-ak,Regev2013-ff,Regev2015-vq,Kawasaki2016-nb,Leishangthem2017-sw,Jin2018-cy,Parmar2019-wj,Schinasi-Lemberg2020-ry,Das2020-cw,Yeh2020-ve,Bhaumik2021-mr,Goswami2025-qv,Liu2022-hq,Parley2022-id,Fielding_prl_oscillatory_shear_2024,Knowlton2014-cl,Hima-Nagamanasa2014-nt,Ghosh2022-fc,van-der-Vaart2013-oa,Denisov2015-ve,Ghosh2017-ed,Aime2022-ac}. The observed yielding transition revealed various generic properties, including an absorbing-to-diffusive transition in the system's microscopic trajectories and large-scale elasto-plastic hysteresis~\cite{Fiocco2013-ak,Regev2013-ff,Regev2015-vq,Kawasaki2016-nb,Leishangthem2017-sw,Jin2018-cy,Parmar2019-wj,Schinasi-Lemberg2020-ry,Das2020-cw,Yeh2020-ve,Bhaumik2021-mr,Goswami2025-qv,Liu2022-hq,Parley2022-id,Fielding_prl_oscillatory_shear_2024,Knowlton2014-cl,Hima-Nagamanasa2014-nt,Ghosh2022-fc}, yet again related to dynamics in the underlying complex energy landscape. In addition, it has been shown that cyclically deformed glasses can be trained to store mechanical memories, which can be subsequently extracted through well-defined reading protocols~\cite{Keim2013-gx,Arceri2021-dt,Mukherji2019-az,Galloway2022-vb,Shohat2023-em,Chen2025-ac,Fiocco2014-jq,Adhikari2018-dl,Mungan2025-wi}.

Here, we show that a Hamiltonian mean-field model of glasses, which remarkably reproduced the observed properties of the non-phononic VDoS ${\cal D}_{\rm G}(\omega)$~\cite{Rainone2021-am,Bouchbinder2021-dh,Moriel2024-wp}, naturally predicts the salient properties of the yielding transition and mechanical memory formation extensively observed in particle-based glasses under driving, with no free parameters relative to the non-driven regime. The model relates statistical-structural properties of quiescent glasses to their elasto-plastic driven dynamics, and opens the way for gaining additional insight into such relations. Taken together, the Hamiltonian mean-field model studied here constitutes a theoretical framework that unifies a broad range of static and dynamic glass properties, and offers a platform for studying additional glass phenomena that currently lack a complete theoretical understanding.\vspace{-0.4cm}

\section{M\lowercase{odel formulation}}
\vspace{-0.2cm}

The mean-field model, originally introduced in~\cite{Rainone2021-am,Bouchbinder2021-dh} and inspired by~\cite{Kuhn_Horstmann_prl_1997,Gurevich2003} (denoted by the acronym `KHGPS' based on the authors' surnames), is defined by the following Hamiltonian for $N$ anharmonic oscillators
\begin{equation}
H\!=\!\frac{1}{2}\sum_i \kappa_i\,x_i^2 + \frac{A}{4!}\sum_i x_i^4 +\sum_{i < j}J_{ij}\,x_i x_j - {f(t)\sum_i x_i}\ ,
\label{eq_sr:model_f}
\end{equation}
where $i,j\=1\text{--}N$. The oscillators, each characterized by a continuous scalar displacement $x_i$, initially represent localized, mesoscopic regions inside an equilibrium, high-temperature liquid, featuring random stiffnesses $\kappa_i$ that are uniformly distributed over $[0,\kappa_0]$ and a deterministic quartic anharmonicity of amplitude $A$. $\kappa_i$ represents instantaneous (short time) and local elasticity in the liquid state, which can be negative as well, hence we take it to attain values down to zero. Upon quenching the liquid, a self-organization process which is mimicked in the framework of the model by the minimization of the Hamiltonian $H$ into one of its local minima, long-range elasticity builds up.

The latter is represented by bilinear interactions between the oscillators, with disordered coupling coefficients $J_{ij}$ that are i.i.d.~Gaussian random variables, with zero mean and variance $J^2/N$ for $i\!\neq\!j$. The mean-field (infinite-dimensional) nature of the model is encapsulated in the disordered interaction coefficients $J_{ij}$, which feature no spatial dependence and couple each oscillator to all other oscillators. It is important to note that as $\kappa_i\!>\!0$ and $A\!>\!0$, all nontrivial properties of the model's glassy landscape emerge from the self-reorganization process induced by the disordered interactions~\cite{Rainone2021-am,Bouchbinder2021-dh}.

The force $f(t)$, which has not been considered previously, is a time-dependent external field that is coupled to the mean oscillators position, $\langle x\rangle\!\equiv\!N^{-1}\!\sum_i x_i$. In previous analyses of the model~\cite{Rainone2021-am,Bouchbinder2021-dh,Moriel2024-wp,Folena2022-tn,Maimbourg2024-lb}, which focused on the statistical properties of the eigenvalues at minima of $H$ and on the localization of the corresponding eigenmodes, $f(t)\=h$ has been used, with $h$ being constant and small. These analyses gave rise to the non-phononic VDoS ${\cal D}_{\rm G}(\omega)$ discussed above. Here, we consider the driven regime, where $f(t)$ is taken to be an oscillatory force of amplitude $f_0$. We study the model in the athermal quasi-static (AQS) limit, where the temperature and driving rate are vanishingly small. Consequently, we do not specify the period of $f(t)$, but rather consider the cyclic sequence $0\!\to\!f_0\!\to\!0\!\to\!-f_0\!\to\!\cdots$, applied through small increments $\Delta{f}$, each followed by a minimization of $H$ (see~{\color{blue}{\em Appendix}}). Therefore, we will quantify the evolution of the system in terms of the number of driving cycles $n$. We measure energy in units of $\kappa_0^2/A$, force in units of $\kappa_0^{3/2}\!/A^{1/2}$ and length in units of $\kappa_0^{1/2}\!/A^{1/2}$. Consequently, the driven model features only two parameters (to be used hereafter in their dimensionless form), the strength of quenched disorder $J$ and the amplitude of the periodic driving $f_0$.
\begin{figure}[h]
    \centering
    \includegraphics[width=1.0\columnwidth]{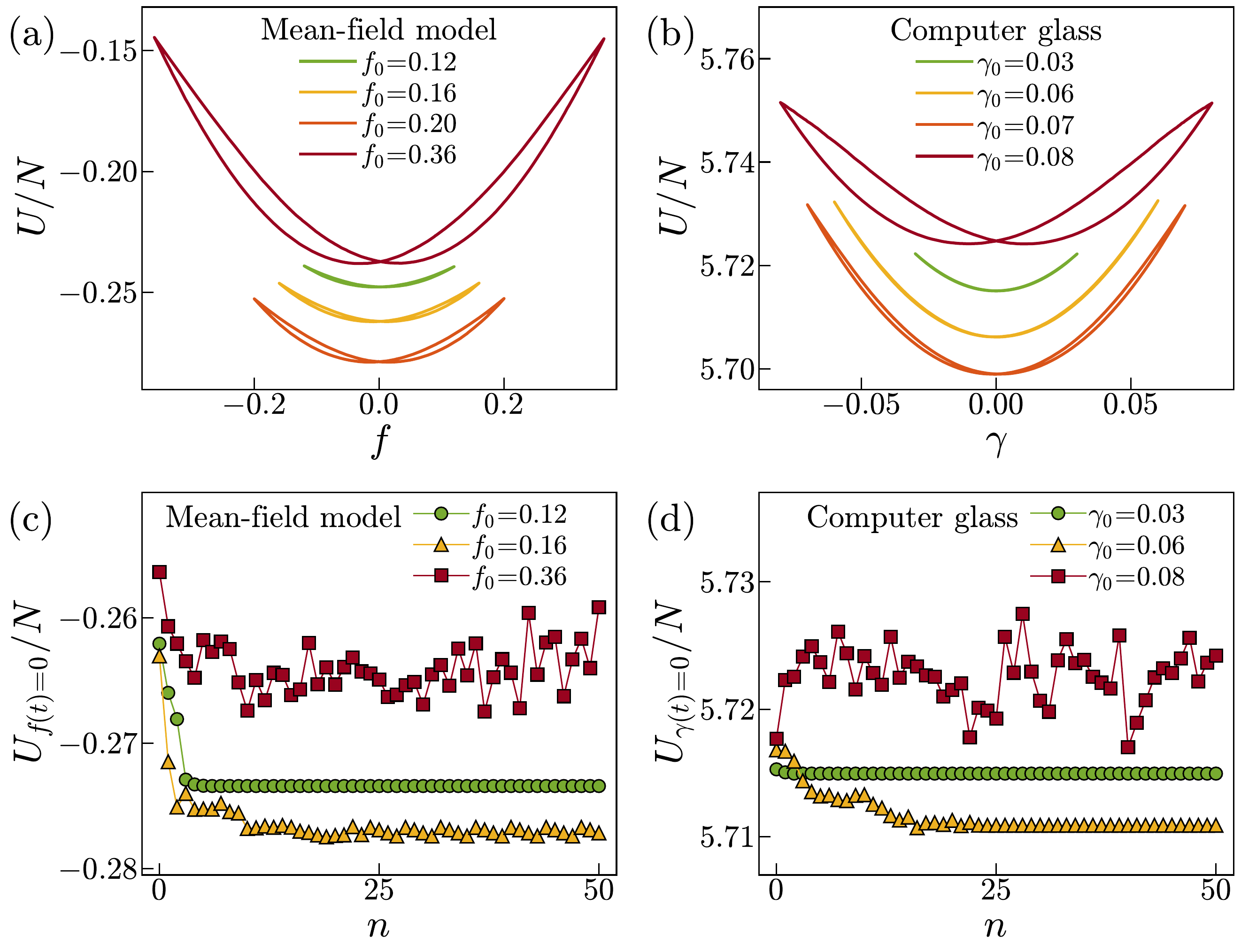}
    \caption{(a) The energy per oscillator $U(t)/N$ vs.~the periodic driving $f(t)$ in steady-state at various amplitudes $f_0$ (see legend), see text for discussion. Here, we used $N\!=\!8192$ and $J\!=\!0.5$, and the results are averaged over more than $M\!=\!20$ realizations of quenched disorder and more than $10$ steady-state cycles. The curves are displaced vertically for visual clarity, but maintain their order (see~{\color{blue}{\em Appendix}}). (b) The same as (a), but for a computer glass driven by an oscillatory shear $\gamma(t)$ (see~{\color{blue}{\em Appendix}} for details). (c) Representative stroboscopic trajectories of the dynamics of the mean-field model at various amplitudes $f_0$ (see legend). Shown is the energy at the end of each cycle, $U_{f(t)=0}$, vs.~the number of cycles $n$, see text for discussion. (d) The corresponding results for the computer glass. Note that at the intermediate $f_0$ value, the model's trajectory reveals multiperiodicity, cf.~panel (c), not observed in the corresponding trajectory of the computer glass, cf.~panel (d).}
    \label{fig:fig1}
\end{figure}

\section{T\lowercase{he yielding transition}}
\vspace{-0.2cm}

A large body of recent work, both computational and experimental~\cite{Fiocco2013-ak,Regev2013-ff,Regev2015-vq,Kawasaki2016-nb,Leishangthem2017-sw,Jin2018-cy,Parmar2019-wj,Schinasi-Lemberg2020-ry,Das2020-cw,Yeh2020-ve,Bhaumik2021-mr,Goswami2025-qv,Liu2022-hq,Parley2022-id,Fielding_prl_oscillatory_shear_2024,van-der-Vaart2013-oa,Knowlton2014-cl,Hima-Nagamanasa2014-nt,Denisov2015-ve,Ghosh2017-ed,Ghosh2022-fc,Aime2022-ac}, demonstrated that glasses under oscillatory shear deformation $\gamma(t)$ undergo a well-defined yielding transition as a function of the amplitude $\gamma_0$ of the applied deformation. The analog of $\gamma_0$ in Eq.~\eqref{eq_sr:model_f} is the amplitude $f_0$ of the driving force $f(t)$. Consequently, our first goal is to test whether the mean-field model features a similar yielding transition as a function of $f_0$. We fix the only other parameter in the model to $J\=0.5$, a choice that will be further discussed below, and drive the system to a steady oscillatory state (i.e., over a large enough number of cycles $n$) at various $f_0$ values.

In Fig.~\ref{fig:fig1}a, we plot the average potential energy $U(t)\!\equiv\!H(t)+f(t)\!\sum_i\!x_i$ per oscillator as a function of $f(t)$ in a steady-state cycle for 4 values of $f_0$ (see legend). It is observed that the $U(f)$ relation is elastic, i.e., revealing no hysteresis, at small $f_0$, but develops an elasto-plastic hysteretic response with increasing $f_0$, indicating an underlying yielding transition. For comparison, we present in Fig.~\ref{fig:fig1}b the corresponding quantities obtained in a particle-based computer glass under AQS oscillatory shear deformation (see figure caption and~{\color{blue}{\em Appendix}}), revealing striking qualitative similarities. Note also that $U(f\=0)$, and likewise $U(\gamma\=0)$, vary non-monotonically with the periodic driving amplitude, which is also a signature of an underlying yielding, as will be further discussed below.

In glassy systems undergoing a yielding transition, the emergence of large-scale hysteresis, as observed in Figs.~\ref{fig:fig1}a-b, is accompanied by an absorbing-to-diffusive transition in the system’s microscopic trajectories~\cite{Fiocco2013-ak,Regev2013-ff,Regev2015-vq,Kawasaki2016-nb,Leishangthem2017-sw,Jin2018-cy,Parmar2019-wj,Schinasi-Lemberg2020-ry,Das2020-cw,Yeh2020-ve,Bhaumik2021-mr,Goswami2025-qv,Liu2022-hq,Parley2022-id,Fielding_prl_oscillatory_shear_2024,Knowlton2014-cl,Hima-Nagamanasa2014-nt,Ghosh2022-fc}. That is, below the transition the system periodically visits exactly the same microscopic state after an integer numbers of cycles, i.e., an absorbing state accompanied by no or minute dissipation, while above the transition the behavior is diffusive, characterized by non-periodic trajectories and significant dissipation. In Fig.~\ref{fig:fig1}c, we plot the (stroboscopic) energy at the end of each deformation cycle as a function of $n$, for 3 of the values of $f_0$ used in Fig.~\ref{fig:fig1}a (see legend). It is observed that the microscopic trajectories in the model undergo an absorbing-to-diffusive transition, indicating that an yielding transition occurs at a driving amplitude slightly larger than $f_0\=0.16$. In Fig.~\ref{fig:fig1}d, we present the corresponding results for the computer glass, yet again revealing striking similarities.
\begin{figure}[h]
    \centering
    \includegraphics[width=1\columnwidth]{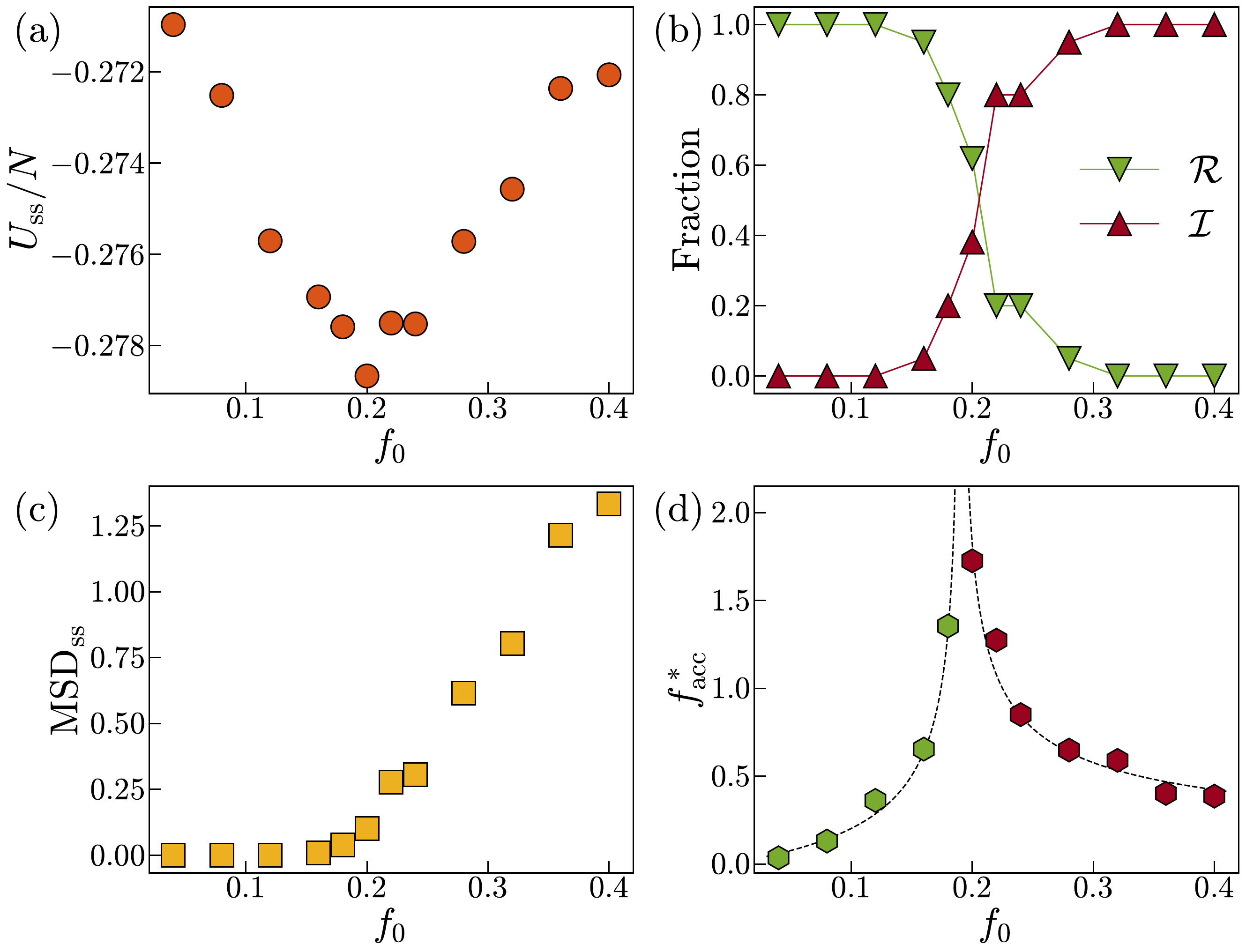}
    \caption{(a) The steady-state stroboscopic energy $U_{\rm ss}(f_0)$, evaluated at the end of each cycle, for the calculations reported on in Fig.~\ref{fig:fig1}a. (b) The fraction of $\mathcal{R}$ and $\mathcal{I}$ steady-states vs.~$f_0$ ($\mathcal{R}$ denotes reversible/periodic steady-states, while $\mathcal{I}$ denotes irreversible/non-periodic ones). (c) The steady-state cycle-to-cycle mean square displacement $\text{MSD}_{\rm ss}$, defined in Eq.~\eqref{eq_sr:MSD_definition}, vs.~$f_0$. (d) The characteristic accumulated driving force $f_{\rm acc}^*$ needed to reach a steady-state as a function of $f_0$, obtained by fitting a stretched exponential function $U_{\rm ss}\!+\!\Delta U\exp\!{[-(f_{\rm acc}/f_{\rm acc}^*)^{1/2}]}$ (where $f_{\rm acc}\!\equiv\!4f_0n$) to storoboscopic trajectories (see~{\color{blue}{\em Appendix}} for details). The dashed lines are guides to the eye.}
    \label{fig:fig2}
\end{figure}

In Fig.~\ref{fig:fig2}a, we plot the average steady-state stroboscopic energy $U_{\rm ss}$ in the model (at $f(t)\=0$, per oscillator) as a function of $f_0$. As implied by Figs.~\ref{fig:fig1}a-b, $U_{\rm ss}(f_0)$ initially decreases with increasing $f_0$, indicating mechanical annealing, a process in which the system visits deeper energy minima, as extensively observed~\cite{Fiocco2013-ak,Leishangthem2017-sw,Parmar2019-wj,Schinasi-Lemberg2020-ry,Das2020-cw,Yeh2020-ve,Bhaumik2021-mr,Goswami2025-qv,Liu2022-hq,Parley2022-id}. Mechanical annealing is maximal at $f_0\!\simeq\!0.2$, close to the above-mentioned estimate of the yielding amplitude, above which $U_{\rm ss}(f_0)$ increases. In Fig.~\ref{fig:fig2}b, we plot the average fraction of reversible/periodic (marked by $\mathcal{R}$) and irreversible/non-periodic (marked by $\mathcal{I}$) steady-states as a function of $f_0$. A crossover between the two types of steady-states occurs close to the position of the minimum of $U_{\rm ss}(f_0)$, which we hereafter denote as the yielding amplitude $f_{\rm y}(J\=0.5)\!\simeq\!0.2$. We verified that this yielding amplitude essentially remains unchanged when $N$ is increased.

Next, we consider fluctuations quantified through the cycle-to-cycle (stroboscopic) mean square displacement of the individual oscillators
\begin{equation}
    \text{MSD}(n)\equiv N^{-1}\sum_{i}{[x_i(n)-x_i(n-1)]^2} \ .
    \label{eq_sr:MSD_definition}
\end{equation}
In Fig.~\ref{fig:fig2}c, we plot the steady-state value of $\text{MSD}(n)$, denoted as $\text{MSD}_{\rm ss}$ (the approach to steady-state is discussed next), against $f_0$. $\text{MSD}_{\rm ss}(f_0)$ significantly grows from zero near the yielding amplitude $f_{\rm y}(J\=0.5)\!\simeq\!0.2$, an observation that mirrors the corresponding behavior of the particle diffusion coefficient in particle-based computer glasses~\cite{Fiocco2013-ak,Kawasaki2016-nb,Parmar2019-wj,Das2020-cw,Schinasi-Lemberg2020-ry}. The latter also reveal a clear signature of the yielding transition in transient (non-steady-state) dynamics, where the number of cycles required to reach steady-state grows significantly as the yielding transition is approached from both sides. To test this in the mean-field model, we define $f_{\rm acc}\!\equiv\!4f_0n$ and extract the characteristic accumulated driving force $f_{\rm acc}^*$ needed to reach steady-state (see~{\color{blue}{\em Appendix}}). $f_{\rm acc}^*(f_0)$ is presented in Fig.~\ref{fig:fig2}d, revealing dynamic slowing-down as $f_{\rm acc}^*\!\to\!f_{\rm y}^{\pm}$, as in computer glasses.

Overall, Figs.~\ref{fig:fig1}-\ref{fig:fig2} establish that the same mean-field model that properly predicts the non-phononic VDoS ${\cal D}_{\rm G}(\omega)$ of glasses in the non-driven regime, undergoes an oscillatory yielding transition that closely resembles that of glassy systems in the driven regime, involving no additional free parameters. Next, we will use the model to shed additional light on the yielding transition. Specifically, we will explore the relations between the statistical-mechanical properties of configurations visited by a single system during the post-yielding driven dynamics and their non-driven counterparts, corresponding to ensemble averaging over quenched disorder. We will also study the effect of the strength of quenched disorder on the yielding transition and its relations to thermal annealing. In~{\color{blue}{\em Appendix}}, we demonstrate mechanical memory formation, in which memories can be stored in the model and subsequently extracted.

\section{S\lowercase{tatistical-mechanical properties of configurations explored during post-yielding dynamics}}
\vspace{-0.2cm}

Some earlier theoretical works suggested that the yielding transition is intrinsically related to the generation of configurational entropy~\cite{bouchbinder2009nonequilibrium,falk2011deformation,Jaiswal2016-mz,Parisi2017-rv}. That is, it has been suggested that upon yielding a large number of glassy configurations in the energy landscape become dynamically accessible. We can use the unifying nature of the mean-field model under discussion to push this line of thinking further and ask: do the statistical-mechanical properties of the minima explored by a driven system of finite size $N$ in the post-yielding regime identify with the corresponding known properties of ensemble averages over realizations of the quenched disorder in the non-driven system?
\begin{figure}[h]
    \centering
    \includegraphics[width=1\columnwidth]{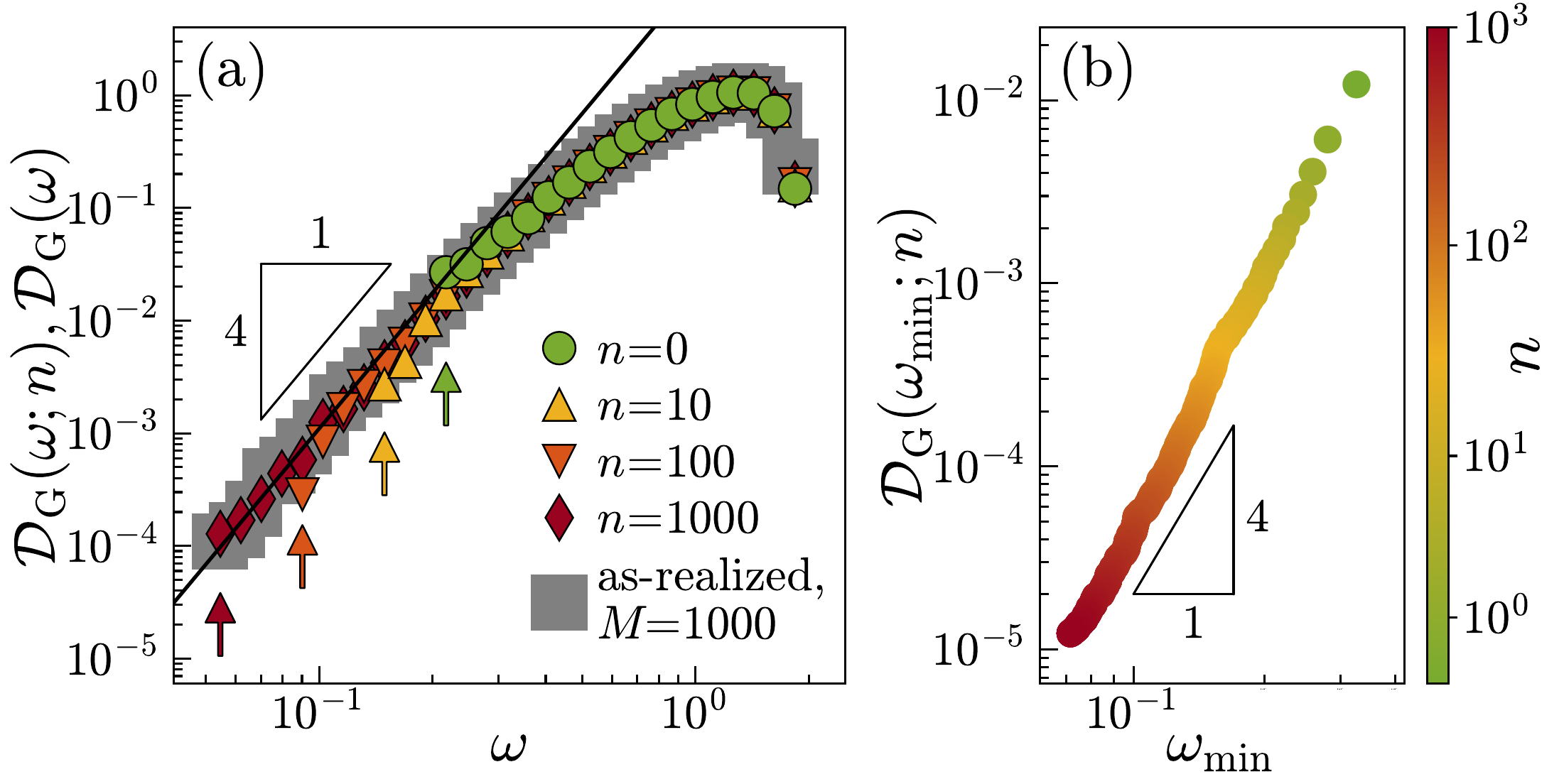}
    \caption{(a) The non-phononic VDoS ${\cal D}_{\rm G}(\omega)\!\sim\!\omega^4$ (see line and power-law triangle) obtained by averaging over $M\!=\!1000$ realizations (with $N\!=\!8192$) of the quenched disorder with $J\!=\!0.5$ (large gray squares) and by periodically driving a single realization with $f_0\!=\!0.28\!>\!f_{\rm y}(J\!=\!0.5)$ over sufficiently large number of cycles $n$ (see legend). It is observed that ${\cal D}_{\rm G}(\omega;n)$ entirely lacks the $\omega^4$ tail initially ($n\!=\!0$) and that the latter progressively builds up with increasing $n$, see text for additional discussion. The arrows mark the edge of the vibrational spectrum per $n$. (b) The number density of the lowest 100 eigenfrequencies, denoted as $\mathcal{D}_{\rm G}(\omega_{\rm min};n)$, with increasing $n$ (see color bar on the right, using the same color code of panel (a)). The progressive build up of the $\omega^4$ tail is highlighted by the power-law triangle.}
    \label{fig:fig3}
\end{figure}

To make this question and the procedure to address it as concrete and quantitative as possible, we consider the non-phononic VDoS ${\cal D}_{\rm G}(\omega)$ and in particular its universal quartic tail ${\cal D}_{\rm G}(\omega)\!\sim\!\omega^4$ of soft modes. For a moderate value of $N$, say $N\=8192$ (as used in Figs.~\ref{fig:fig1}-\ref{fig:fig2}), the vibrational spectrum of an energy minimum of the non-driven system of a single realization of the quenched disorder does not reveal the universal quartic tail ${\cal D}_{\rm G}(\omega)\!\sim\!\omega^4$. This is demonstrated in Fig.~\ref{fig:fig3}a, where ${\cal D}_{\rm G}(\omega,n\=0)$ is plotted (here, $n\=0$ corresponds to a non-driven, single realization of the quenched disorder with $J\=0.5$). The reason for this is that the probability to find a low-frequency eigenmode in the $\omega^4$ tail is very small. The statistically convergent power-law tail ${\cal D}_{\rm G}(\omega)\!\sim\!\omega^4$, down to an $N$-dependent minimal vibrational frequency, is obtained only by averaging over a large enough ensemble of independent realizations of the quenched disorder. This is demonstrated in Fig.~\ref{fig:fig3}a by superposing ${\cal D}_{\rm G}(\omega)$ obtained through $M\=1000$ realizations of the quenched disorder, clearly revealing the universal quartic $\omega^4$ tail.

The question then, rephrasing the above-posed one, is whether driving a single realization with $f_0\!>\!f_{\rm y}$ for a sufficiently large number of cycles $n$ would result in ${\cal D}_{\rm G}(\omega,n)$ converging to ${\cal D}_{\rm G}(\omega)$ at fixed $N$. That is, we ask whether a single system in the ``chaotic'' post-yielding regime is dynamically self-averaging, at least as far as the low-frequency tail of ${\cal D}_{\rm G}(\omega)$ is concerned. In Fig.~\ref{fig:fig3}a, we superpose results for ${\cal D}_{\rm G}(\omega,n)$ for increasing values of driving cycles $n$ (see legend), demonstrating progressive convergence towards ${\cal D}_{\rm G}(\omega)\!\sim\!\omega^4$. The upward pointing arrows indicate the edges of the resolved $\sim\!\omega^4$ tail, clearly demonstrating its systematic shift towards lower frequencies along the power-law tail with increasing $n$. To reiterate the point, we plot in Fig.~\ref{fig:fig3}b the number density of the lowest $100$ eigenfrequencies, denoted by ${\cal D}_{\rm G}(\omega_{\rm min};n)$, which further highlights the progressive build up of the quartic tail with increasing $n$. These results demonstrate the equivalence of the two protocols in the present context, where a larger/smaller $M$ in the non-driven system implies a larger/smaller $n$ in the single driven system at a fixed $N$.

\section{V\lowercase{ariable quenched disorder and thermal annealing}}
\vspace{-0.2cm}


The analysis so far focused on a fixed strength of quenched disorder, characterized by $J\=0.5$. Our next goal is to understand the dependence of the yielding transition under oscillatory driving forces on $J$. In glasses of fixed composition and interparticle interactions, the state of disorder can be varied through thermal annealing~\cite{shi2005strain,shi2007evaluation,vasoya2016notch,misaki_yielding_pnas_2018,Eran_mechanical_glass_transition,david_fracture_mrs_2021,Yeh2020-ve,Bhaumik2021-mr,castellanos2022history}, resulting in self-generated positional --- and consequently also interactional/mechanical --- disorder. We discuss thermal annealing later, but first take advantage of the fact that in the dimensionless formulation of the model, quenched disorder is varied by a single parameter $J$.
\begin{figure*}
    \centering
    \includegraphics[width=1\linewidth]{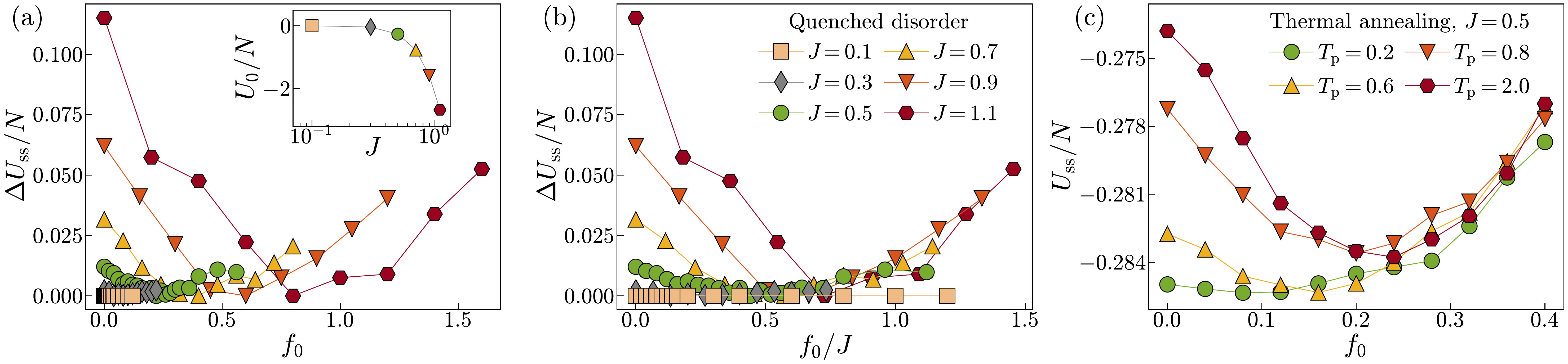}
    \caption{(a) $\Delta{U}_{\rm ss}(f_0)$, similarly to Fig.~\ref{fig:fig2}a but for various levels of quenched disorder quantified by $J$ (see legend in panel (b)). Here, $\Delta{U}_{\rm ss}(f_0)$ is measured relative to the minimum of $U_{\rm ss}(f_0)$ per $J$, with $N\!=\!4096$ and $M\!=\!16$. (inset) The energy $U_0$ of the non-driven system vs.~$J$. (b) The same as panel (a), but vs.~$f_0/J$, see text for discussion. (c) $\Delta{U}_{\rm ss}(f_0)$ for thermally annealed states quantified by the ``parent temperature'' $T_{\rm p}$, see text for discussion and~{\color{blue}{\em Appendix}} for technical details.}
    \label{fig:fig4}
\end{figure*}

In~{\color{blue}{\em Appendix}} (specifically, see Sect.~\ref{sec:J_QLMs} therein), we show that the number of soft modes up to $\omega_{\rm p}$ (defined above), quantified by $\int_0^{\omega_{\rm p}}\!{\cal D}_{\rm G}(\omega)\,d\omega$, is small at small $J$ and increases with $J$ until it levels off above $J\!\simeq\!0.2$. Since the number of soft modes is related to accessible minima in the energy landscape and hence to plastic rearrangements, we expect small values of $J$ to give rise to little mechanical annealibility. In this regime, we expect a ``brittle-like'' behavior with no plastic yielding, which may be analogous to the so-called ``brittle yielding'' observed in particulate systems in finite dimensions~\cite{Yeh2020-ve,Bhaumik2021-mr}. On the other hand, at larger $J$ values, we expect to observe an increasing degree of mechanical annealibility and a ``ductile-like'' yielding transition (already established above for $J\=0.5$). See additional discussion in~{\color{blue}{\em Appendix}}.

In testing these expectations, one should bear in mind that while under thermal annealing the typical energy scale remains the same (inherited from interparticle interactions), this is not the case upon varying $J$ in the model. To stress this point, we plot in the inset of Fig.~\ref{fig:fig4}a the variation of the average energy per oscillator $U_0/N$ of the non-driven model with $J$. It varies from $U_0/N\=-2.65\!\times\!10^{-4}$ to $U_0/N\=-2.68$, i.e., by $4$ orders of magnitude, when $J$ is varied by an order of magnitude. Consequently, in the oscillatory driven case to be considered, we focus on the energy difference $\Delta U_{\rm ss}$ relative to the minimal steady-state energy as a function of $f_0$. The results for $\Delta{U}_{\rm ss}(f_0)$ are presented in Fig.~\ref{fig:fig4}a, for $J\=0.1\!-\!1.1$ (see legend in Fig.~\ref{fig:fig4}b).

An immediate observation in Fig.~\ref{fig:fig4}a is that the yielding amplitude $f_{\rm y}(J)$ is an increasing function of $J$. While we do not discuss here theoretical considerations that may predict this dependence (and in fact indicate a stronger than linear dependence), we replot the results as $\Delta{U}_{\rm ss}(f_0/J)$ in Fig.~\ref{fig:fig4}b. The latter supports the expectations stated above that the degree of mechanical annealibility increases with $J$ and for sufficiently small $J$ a ``brittle-like'' behavior with no yielding emerges. These findings are qualitatively similar to the corresponding phenomenology observed as a function of the degree of thermal annealing~\cite{Leishangthem2017-sw,Schinasi-Lemberg2020-ry,Yeh2020-ve,Bhaumik2021-mr,Goswami2025-qv}.

It has been shown that under a variable degree of thermal annealing, a glass is sensitive to the initial state of disorder for $f_0\!<\!f_{\rm y}$, but is independent of it for $f_0\!\ge\!f_{\rm y}$~\cite{Leishangthem2017-sw,Schinasi-Lemberg2020-ry,Yeh2020-ve,Bhaumik2021-mr,Goswami2025-qv}. Indeed, the different $\Delta{U}_{\rm ss}(f_0/J)$ curves in Fig.~\ref{fig:fig4}b nearly collapse sufficiently above yielding, yet the variability with $J$ does not allow to directly map the effect of quenched disorder to that of thermal annealing. Consequently, and while we do not extensively discuss here thermal dynamics of the model, we thermalized it and generated zero temperature initial states at different degrees of thermal annealing. The latter are quantified by the ``parent temperature'' $T_{\rm p}$ from which zero temperature states are quenched (see~{\color{blue}{\em Appendix}}). In Fig.~\ref{fig:fig4}c, we present $U_{\rm ss}(f_0)$ for 4 values of $T_{\rm p}$, obtained by subjecting the thermally annealed states to oscillatory mechanical driving. The pre-yielding dependence on thermal annealing and its disappearance in the post-yielding regime are observed. Note, though, that for the range of thermal annealing considered here (see~{\color{blue}{\em Appendix}}), a ``brittle-like'' behavior is not observed.

\section{D\lowercase{iscussion}}
\vspace{-0.2cm}


We studied a Hamiltonian mean-field model of glasses, with continuous degrees of freedom, under oscillatory driving forces. In the non-driven regime, the model has been recently shown to correctly predict the low-frequency properties of the non-phononic vibrational density of states ${\cal D}_{\rm G}(\omega)$ of glasses, including its universal quartic tail, the boson peak and the localization of vibrational modes~\cite{Rainone2021-am,Bouchbinder2021-dh,Moriel2024-wp}. We showed that in the periodically driven regime, the model undergoes a yielding transition as a function of the amplitude of the oscillatory force, revealing a wide range of dynamical behaviors that closely mirror observations in particle-based glasses. We also demonstrated mechanical memory formation (see~{\color{blue}{\em Appendix}}), similarly to glasses.

Our analysis treats the model as a predictive theory in the sense that no additional free parameters are introduced relative to the non-driven version, where physically-relevant non-phononic excitations are defined. The latter, missing in discrete Hamiltonian models (e.g.,~\cite{Fiocco2015-qo,Keim2021-pq,Lindeman2025-zd}) and in non-Hamiltonian ones (e.g.,~\cite{Liu2022-hq,Parley2022-id,Fielding_prl_oscillatory_shear_2024,Mungan2019-sd,Mungan2021-lh,Sarkar2025-in}) emerge from the physically-motivated Hamiltonian structure of the model and its continuous degrees of freedom. Overall, the results indicate that the model naturally unifies a broad range glassy phenomena, linked to the underlying complex energy landscape, and offers a generic framework to study the physics of glasses.

\acknowledgements


E.B.~acknowledges support from the Israel Science Foundation (ISF Grant No.~403/24), the Ben May Center for Chemical Theory and Computation and the Harold Perlman Family. M.S.~acknowledges support from the International Joint Graduate Program in Earth and Environmental Sciences. Part of the computations were performed using supercomputing resources at the Cyberscience Center, Tohoku University.
\clearpage

\appendix
\setcounter{figure}{0}
\renewcommand{\thefigure}{A\arabic{figure}}

\section*{A\lowercase{ppendix}}
\vspace{-0.4cm}

The goal of this Appendix is to provide technical details regarding the results presented in the main text, and to offer additional supporting results, most notably about memory formation in Sect.~\ref{sec:memory} and about the relation between the strength of disordered interaction $J$ and the number of soft non-phononic modes, and a ``brittle-like'' behavior, in Sect.~\ref{sec:J_QLMs}.

\vspace{-0.3cm}
\section{N\lowercase{umerical implementation of the model and additional details}}
\vspace{-0.3cm}
\label{sec:KHGPS_preparation}

The mean-field model we studied is formulated through the Hamiltonian $H$ defined in Eq.~\eqref{eq_sr:model_f}, where the parameters and nondimensionalization procedure are also discussed. In most of the results reported in the main text and in this Appendix (specifically, in Figs.~\ref{fig:fig1}-\ref{fig:fig3},~\ref{fig:fig4}a-b and~\ref{fig:fig5} therein), samples were prepared by performing energy minimization in the absence of an applied force, $f(t)\!=\!0$, starting from random oscillator configurations where each $x_i$ is drawn from a uniform distribution over $[-\epsilon/2,\epsilon/2]$ with $\epsilon\!=\!10^{-2}$, independently of $J$. We verified that $\langle x^2 \rangle\!\gg\!\epsilon^2$ after minimization (for all $J$ values employed), demonstrating that the value of $\epsilon$ used does not affect the results. The system size was $N\!=\!8192$ for the results presented in Figs.~\ref{fig:fig1}-\ref{fig:fig3}, \ref{fig:fig5} and $N\!=\!4096$ for those presented in Fig.~\ref{fig:fig4}. We prepared $M\!\geq\!20$ independent realizations of quenched disorder for Figs.~\ref{fig:fig1}-\ref{fig:fig2}, \ref{fig:fig5}, and $M\!=\!16$ realizations for Figs.~\ref{fig:fig4}a-b. We prepared $M\!=\!1$ realizations for the driven minima-sampling presented in Fig.~\ref{fig:fig3}a-b, and $M\!=\!1000$ realizations for the ``as-realized'' data presented in Fig.~\ref{fig:fig3}a.

\vspace{-0.3cm}
\subsection{T\lowercase{he oscillatory driving force}}
\vspace{-0.3cm}
We have carried out oscillatory driving simulations in the athermal quasi-static (AQS) limit, starting from initial zero-force configurations (whose preparation protocol is detailed in Sec.~\ref{sec:KHGPS_preparation}). We increased the applied force $f(t)$ by small increments $\Delta{f}$ and performed energy minimization using the conjugate-gradient method after each increment. The force increment $\Delta{f}$ was selected depending on the value of $J$ to keep the ratio $\Delta{f}/J$ at most $2\!\times\!10^{-3}$. This choice ensures the proper resolution of the oscillators' displacement $\Delta{x}$ in response to a force increment $\Delta{f}$ such that one can accurately detect plastic rearrangements, i.e., transitions between different minima in the complex energy landscape. Moreover, this choice ensures the numerical convergence of our results in the range of parameters under consideration. The values of $\Delta{f}$ per $J$ are provided in Table~\ref{tab:force_increments}. Note that the particular values were chosen such that $f_0/\Delta{f}$ is an integer.

The driving was applied in the sequence $0\!\to\!f_0\!\to\!0\!\to\!-f_0\!\to\!\cdots$ and persisted until the potential energy $U\!\equiv\!H\!+\!f\sum_ix_i$ and mean-squared-displacement $\text{MSD}(n)\!\equiv\!N^{-1}\sum_{i}[x_i(n)\!-\!x_i(n-1)]^2$ reached steady-states, where $U(n)$ and $\text{MSD}(n)$ were calculated at the end of each cycle as functions of the number of driving cycles $n$. Unless otherwise noted, all reported data points result from averaging over $M$ independent realizations, and for steady-state macroscopic quantities, data points are also averaged over $\geq\!10$ steady-state cycles.
\setlength{\tabcolsep}{4pt}
\begin{table}[t]
    \vspace{-0.4cm}
    \caption{Numerical force increment $\Delta{f}(J)$}
    \centering
    \begin{tabular}{ccc}
        \hline
        & \begin{tabular}{@{}c@{}}Interaction strength\\ $J$ \end{tabular} & \begin{tabular}{@{}c@{}}Force increment\\ $\Delta f$ \end{tabular} \\
        \hline
        & 0.1 & $5\!\times\!10^{-5}$ \\
        & 0.3 & $2\!\times\!10^{-4}$ \\
        & 0.5 & $1\!\times\!10^{-3}$ \\
        & 0.7 & $1\!\times\!10^{-3}$ \\
        & 0.9 & $1.5\!\times\!10^{-3}$ \\
        & 1.1 & $2\!\times\!10^{-3}$ \\
        \hline
    \end{tabular}
    \label{tab:force_increments}
\end{table}

\vspace{-0.6cm}
\subsection{Screening multiperiodic steady-state trajectories}
\vspace{-0.3cm}

\begin{figure}[b]
    \centering
    \includegraphics[width=1\columnwidth]{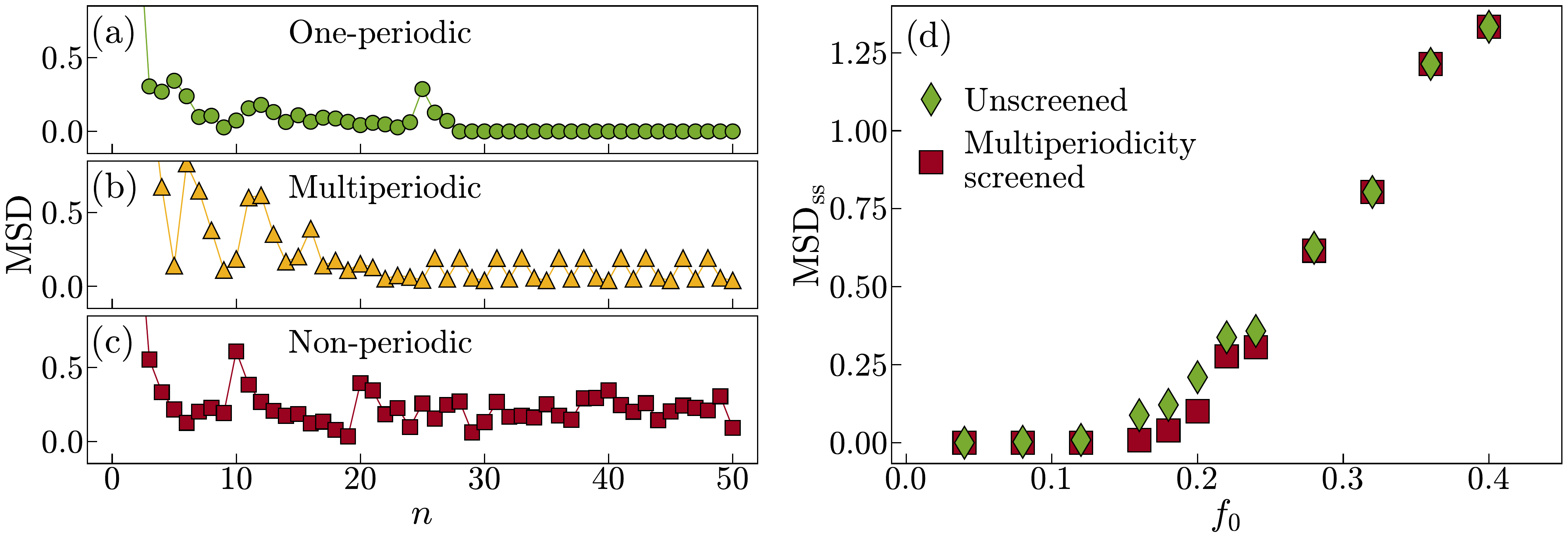}
    \caption{(a) One-periodic, (b) multiperiodic, and (c) non-periodic steady-states observed in $J\!=\!0.5$ (and $N\!=\!8192$) systems driven at $f_0\!=\!0.18$, presented in terms of $\text{MSD}(n)$. (d) The steady-state mean-squared-displacement ${\rm MSD}_{\rm ss}$ as a function of the driving amplitude $f_0$, plotted here for the unscreened (green diamonds) and multiperiodicity screened (brown-red squares) values. See text for definitions and discussion.}
    \label{fig:figS1}
\end{figure}

At driving amplitudes $f_0$ close to the yielding amplitude $f_{\rm y}$, the mean-field model can settle into one-periodic, multiperiodic or non-periodic steady-states. Figures~\ref{fig:figS1}a-c display these representative stroboscopic trajectories in terms of $\text{MSD}(n)$. One- and multiperiodic steady-states form closed loop trajectories, but in the latter the periodicity is a multiple integer of the driving period which results in a finite value of ${\rm MSD}_{\rm ss}$~\cite{Schreck2013-vy,Regev2013-ff,Regev2015-vq,Keim2021-pq,Szulc2022-gx}. Consequently, averaging over an ensemble containing many such trajectories could slightly obscure the emergence of non-periodic steady-states in our analysis. This is not the case in particulate glasses where the particle diffusion is measured, to which multiperiodic trajectories do not contribute. Consequently, in the ${\rm MSD}_{\rm ss}$ results presented in Fig.~\ref{fig:fig2}b, we screened multiperiodic trajectories. For completeness, we present in Fig.~\ref{fig:figS1}d both the ``multiperiodicity screened'' data of Fig.~\ref{fig:fig2}b and the corresponding unscreened data. It is observed, as expected, that differences exist only near $f_{\rm y}$.

In addition, we note that the question of which glass (a coarse-grained model, a molecular computer glass or a laboratory glass) features multiperiodic steady-states near yielding --- and/or under what conditions --- is an interesting open question for future investigation.

\vspace{-0.6cm}
\subsection{Extraction of the characteristic accumulated driving force}
\vspace{-0.3cm}

\begin{figure}[h]
    \centering
    \includegraphics[width=1\columnwidth]{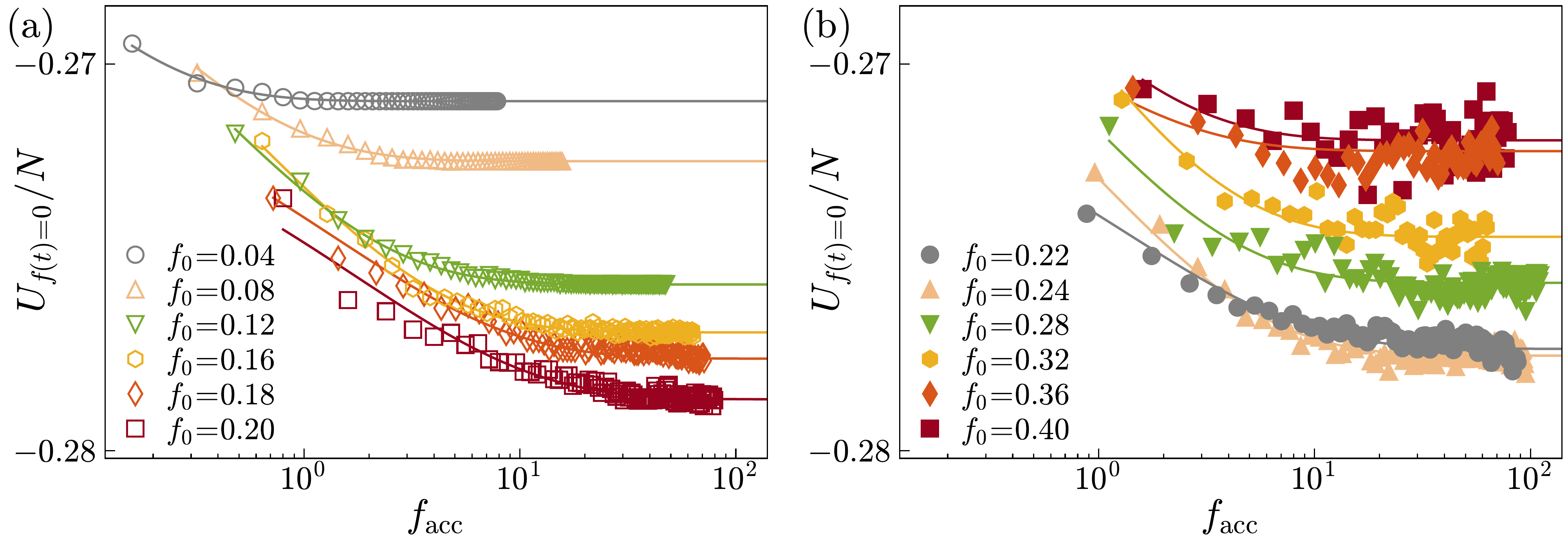}
    \vspace{-0.4cm}
    \caption{(a) The energy per oscillator $U_{f(t)=0}/N$ at the end of each cycle vs.~$f_{\rm acc}\!=\!4f_0n$ (where $n$ is the number of cycles) for $f_0\!\leq\!0.2$ (markers) and fits to a stretched exponential function (solid lines, see text), plotted for $J\!=\!0.5$ and $N\!=\!8192$ systems. (b) Same as (a), but for $f_0\!>\!0.2$.}
    \label{fig:figS2}
\end{figure}

In Fig.~\ref{fig:fig2}d, we report on the characteristic accumulated driving force $f_{\rm acc}^*$ needed to reach a steady-state as a function of $f_0$. Here, we provide detail of the extraction procedure, which was performed as follows: we first plot the stroboscopic trajectories of the potential energy per oscillator $U(n)/N$, and then fit a stretched exponential function of the form $U_{\rm ss}\!+\!\Delta U\exp{[-(f_{\rm acc}/f_{\rm acc}^*)^{1/2}]}$, where $f_{\rm acc}\!\equiv\!4f_0n$. Figures~\ref{fig:figS2}a-b demonstrate this procedure, where the symbols are the data points and the solid lines are the fitted curves. We fixed $U_{\rm ss}$ and $\Delta{U}$, and extracted $f_{\rm acc}^*$ as the sole fitting parameter, which is reported in Fig.~\ref{fig:fig2}d.

\vspace{-0.8cm}
\subsection{Undisplaced full-cycle trajectories}
\vspace{-0.3cm}

\begin{figure}[b]
    \centering
    \includegraphics[width=1\columnwidth]{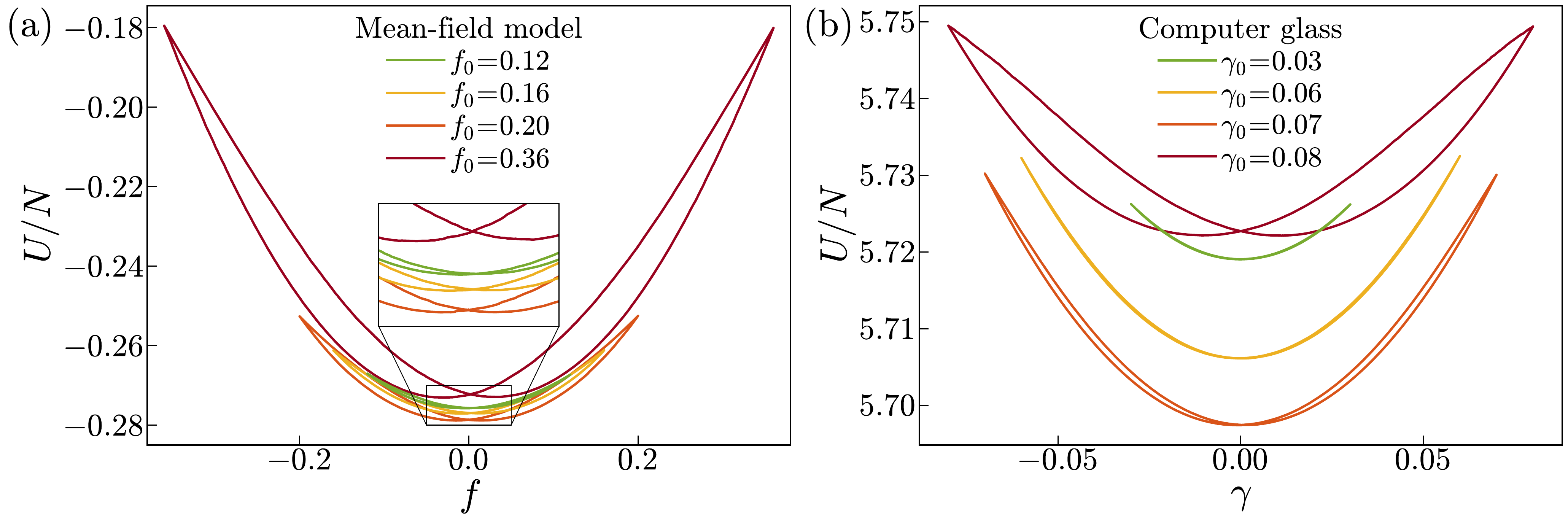}
    \vspace{-0.4cm}
    \caption{The same data as in Figs.~\ref{fig:fig1}a-b, but without the vertical displacement used therein for visual clarity. In panel (a), we also present a zoom in on the small force portion of each curve.}
    \label{fig:figS3}
\end{figure}

In Fig.~\ref{fig:fig1}a-b, we present the energy of the system vs.~the periodic driving in steady-state at various driving amplitudes, for both the mean-field model (panel (a)) and computer glasses (panel (b), details about the latter are provided below in Sect.~\ref{sec:computer_glasses}). As mentioned therein, the curves are displaced vertically for visual clarity, but maintain their order. For completeness, we present in Figs.~\ref{fig:figS3}a-b the undisplaced curves.

\vspace{-0.3cm}
\section{M\lowercase{emory formation}}
\vspace{-0.3cm}
\label{sec:memory}

Glasses are known to reveal rather intriguing memory effects related to their underlying complex landscapes~(see, e.g.,~\cite{Keim2019-ez,Paulsen2025-av} for recent reviews). Our goal here is to demonstrate that the mean-field model defined in Eq.~\eqref{eq_sr:model_f} reveals mechanical memory effects, further strengthening the model's ``glassiness''. Specifically, we set $J\=0.5$ (for which $f_{\rm y}\!\simeq\!0.2$) and follow a procedure known as \textit{parallel read}~\cite{Adhikari2018-dl}. In this protocol, the system is first trained at a specific amplitude $f_\text{tr}$ over $n_\text{tr}$ training cycles. Then, identical copies of the trained system are subjected to a single ``read cycle'' at an amplitude $f_\text{read}$. Memory is quantified through the MSD between configurations before and after a read cycle, i.e., through $\text{MSD}_\text{read}\!\equiv\!N^{-1}\!\sum_i[x_i(f_\text{read})-x_i(f_\text{tr},n_\text{tr})]^2$.

\begin{figure}
    \centering
    \includegraphics[width=1\columnwidth]{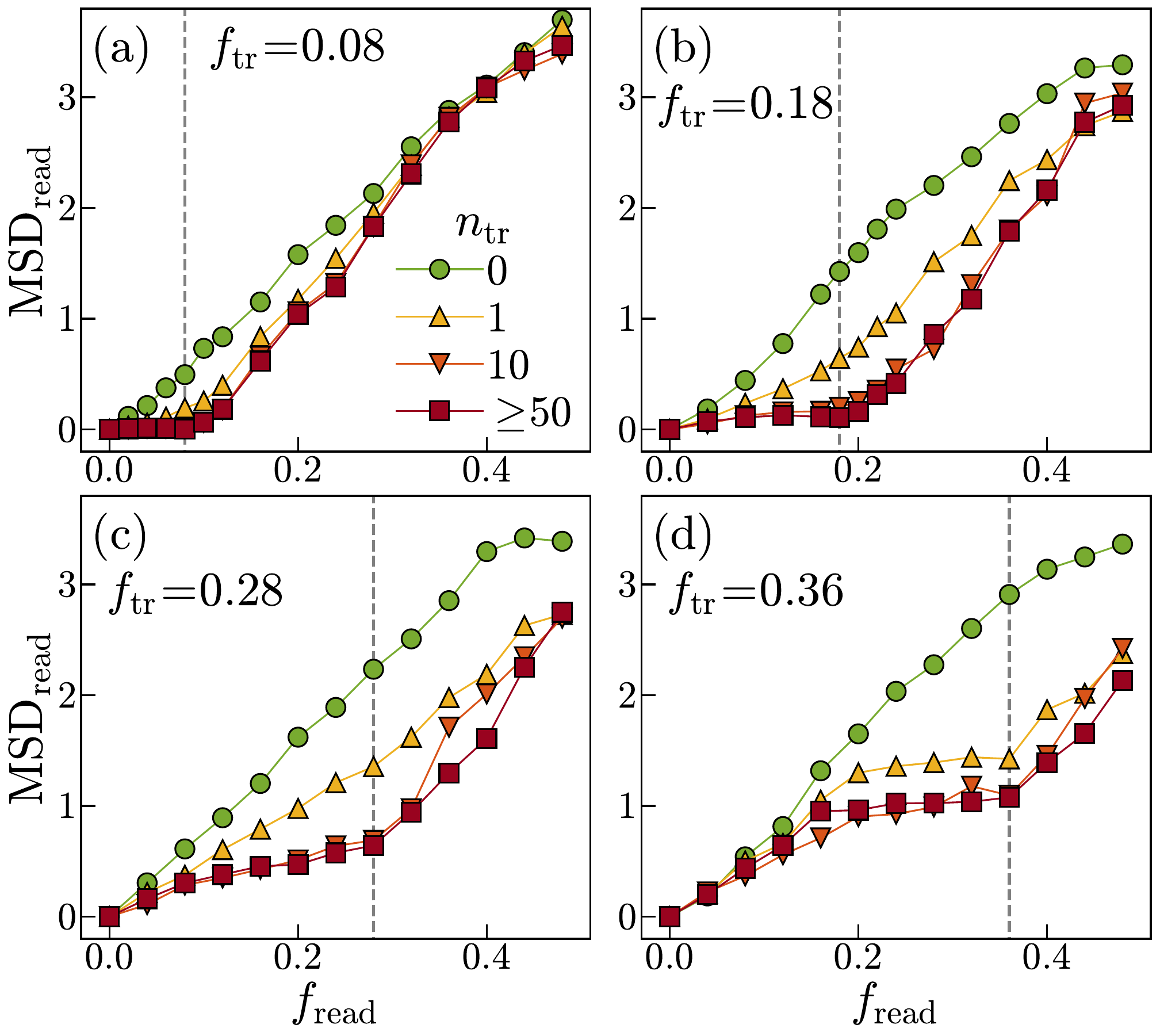}
    \vspace{-0.4cm}
    \caption{The parallel read protocol, demonstrating memory formation in the mean-field model. $\text{MSD}_\text{read}(f_\text{read})$ is plotted for various training cycles $n_\text{tr}$ (see legend in panel (a)) and 4 different training amplitudes (marked by the vertical dashed line), (a) $f_\text{tr}\!=\!0.08$, (b) $f_\text{tr}\!=\!0.18$, (c) $f_\text{tr}\!=\!0.28$ and (d) $f_\text{tr}\!=\!0.36$. See text for definitions and discussion.}
    \label{fig:fig5}
\end{figure}

For $f_\text{tr}\!<\!f_{\rm y}$ and a large enough $n_\text{tr}$, we expect a memory of the amplitude $f_\text{tr}$ to be stored in the system and robustly extracted from it. This is indeed demonstrated in Fig.~\ref{fig:fig5}a, where we plot $\text{MSD}_\text{read}(f_\text{read})$ for $f_\text{tr}\=0.08\!<\!f_{\rm y}\!\simeq\!0.2$, manifested in a strong suppression of $\text{MSD}_\text{read}(f_\text{read})$ for $f_\text{read}\!\leq\! f_\text{tr}$ and a sharp transition at $f_\text{read}\!=\!f_\text{tr}$. A similar pre-yielding behavior is observed for $f_\text{tr}\=0.18\!<\!f_{\rm y}\!\simeq\!0.2$ in Fig.~\ref{fig:fig5}b.

For $f_\text{tr}\!>\!f_{\rm y}$ and a large enough $n_\text{tr}$, we expect a single reading cycle in the range $f_\text{read}\!<\!f_{\rm y}$ to lead to a non-negligible $\text{MSD}_\text{read}$, significantly larger than the corresponding $\text{MSD}_\text{read}$ in Figs.~\ref{fig:fig5}a-b, in view of the accessibility of a large number on different configurations in the post-yielding regime. Yet, it is a priori unclear whether a signature of the training amplitude $f_\text{tr}\!>\!f_{\rm y}$ is entirely washed out as $f_\text{read}$ is further increased or not. In fact, some systems revealed the latter behavior~\cite{Keim2013-gx,Mukherji2019-az,Arceri2021-dt,Galloway2022-vb,Chen2025-ac,Mungan2025-wi} and others the former~\cite{Adhikari2018-dl}.

In Figs.~\ref{fig:fig5}c-d, we present results for the parallel read protocol with $f_\text{tr}\=0.28,\,0.36\!>\!f_{\rm y}\!\simeq\!0.2$, respectively. First, it is observed --- as anticipated --- that $\text{MSD}_\text{read}$ is indeed significantly larger than its counterpart in Figs.~\ref{fig:fig5}a-b for $f_\text{read}\!<\!f_{\rm y}$. Moreover, $\text{MSD}_\text{read}$ appears to reach an $f_\text{tr}$-dependent plateau for $f_\text{read}\!\simeq\!f_{\rm y}$. However, a deviation from the plateau is observed at $f_\text{read}\!\simeq\!f_\text{tr}$ (marked by the vertical dashed line), indicating that some memory of the training amplitude remains encoded in the system also in the post-yielding regime. These results strengthen the findings discussed in the main text, demonstrating that the model naturally captures a broad range of glassy behaviors.

\vspace{-0.4cm}
\section{C\lowercase{omputer glasses}}
\label{sec:computer_glasses}
\vspace{-0.3cm}

In Fig.~\ref{fig:fig1} (right column panels), we present results on computer glasses subjected to oscillatory shear, highlighting the similarities to the model observations (presented on the left column panels). Here, for completeness, we provide technical details about these computer glasses, which are extensively used in the literature (see, for example,~\cite{Lerner2018-me}). We employed a generic glass-forming model in 3D, which consists of a 50:50 binary mixture of $N=16,000$ ``large'' and ``small'' particles of equal mass $m$ enclosed in a cubic box of linear size $L$. The particles interact via a radially symmetric purely repulsive inverse power-law pairwise potentials, given by
\begin{equation}
    \varphi(r_{ij}) =
    \left\{\begin{alignedat}{2}
        \varepsilon\left[\left(\frac{\lambda_{ij}}{r_{ij}}\right)^n\right. &+ \left.\sum_{l=0}^{q}c_{2l}{\left(\frac{r_{ij}}{\lambda_{ij}}\right)^{2l}}\right]\ &, &\ \frac{r_{ij}}{\lambda_{ij}} \leq x_{\rm c} \\ & 0\ &, & \ \frac{r_{ij}}{\lambda_{ij}} > x_{\rm c}
    \end{alignedat}\right.\,,
\label{eq:3dIPL}
\end{equation}
where $r_{ij}$ is the distance between the $i$th and $j$th particles, $\varepsilon$ is an energy scale, and $x_{\rm c}$ is the dimensionless distance for which $\varphi$ vanishes continuously up to $q$ derivatives. Distances are measured in terms of the interaction lengthscale $\lambda$ between two ``small'' particles, and the rest are chosen to be $\lambda_{ij}\!=\!1.18\lambda$ for one ``small'' and one ``large'' particle, and $\lambda_{ij}\!=\!1.4\lambda$ for two ``large'' particles. The coefficients $c_{2l}$ are chosen such that $\varphi$ vanishes continuously up to $q$ derivatives, viz.
\begin{equation}
    c_{2l} = \frac{(-1)^{l+1}}{(2q-2l)!!(2l)!!}\frac{(n+2q)!!}{(n-2)!!(n+2l))}x_c^{-(n+2l)}\ .
\end{equation}
We choose the parameters $x_c\!=\!1.48,n\!=\!10$, and $q\!=\!3$. The density was set to be $N/V\!=\!0.82\lambda^{-3}$, where $V\!\equiv\!L^3$ is the volume. Temperatures are expressed in terms of $\varepsilon/k_{\rm B}$, with $k_{\rm B}$ being the Boltzmann constant, and time in terms of $\sqrt{m\lambda^{2}}/\varepsilon$. This system undergoes a computer glass transition at $T_{\rm g}\!\approx\!0.5$~\cite{Lerner2018-me}.

For Fig.~\ref{fig:fig1}, we prepared $M\!\geq\!50$ samples by performing a continuous quench down to zero temperature at a cooling rate $\dot{T}\!=\!3\!\times\!10^{-3}$, starting from equilibrium configurations at $T\!=\!1.0$. Oscillatory shear AQS deformation is applied as follows; starting from a shear-free configuration, each driving step imposes a uniform shear of $\Delta \gamma \!=\!2\!\times\!10^{-4}$, followed by energy minimization via the conjugate gradient method. All simulations were performed under Lees-Edwards periodic boundary conditions. The driving was applied through the sequence of $0\!\to\!\gamma_0\!\to\!0\!\to\!-\gamma_0\!\to\!\cdots$, and was continued until the potential energy $U\!\equiv\!\sum_{i<j}{\varphi(r_{ij})}$, measured at the end of each cycle, reached a steady-state.

\vspace{-0.4cm}
\section{T\lowercase{hermal annealing}}
\vspace{-0.3cm}
\label{sec:thermal}

Most of the analysis in this work is devoted to applying mechanical driving to mean-field model states (minima of the Hamiltonian $H$) obtained from realizations of the quenched disorder. Yet, in Fig.~\ref{fig:fig4}c, we present results obtained by mechanically driving thermally annealed mean-field model states. Here, we briefly explained how the latter have been obtained.

We considered the following overdamped Langevin dynamics
\begin{equation}
\gamma\frac{dx_i}{dt}\!=\!-\frac{\pa H}{\pa x_i}+\sqrt{2k_{\rm B}T\gamma}\,\zeta_i(t)\,,
\label{eq_SI:Overdampled_Langevin}
\end{equation}
where $H$ is defined in Eq.~\eqref{eq_sr:model_f}, $k_{\rm B}$ is the Boltzmann constant as above, $\gamma$ is a damping coefficient, $\zeta_i(t)$ is a stationary Gaussian process with $\langle\zeta_i(t)\rangle\!=\!0$ and $\langle\zeta_i(t)\zeta_j(t')\rangle\!=\!\delta_{ij}\delta(t\!-\!t')$, where $\delta_{ij}$ is Kronecker's delta, and Roman indices label different oscillators. The external field $f(t)$ is set to zero during the thermal calculations. Temperatures are measured in terms of $\kappa_0^2/k_{\rm B}A$, and time in terms of $\gamma/\kappa_0$. We numerically integrate Eq.~\eqref{eq_SI:Overdampled_Langevin} by considering a small time increment $\Delta{t}\!=\!10^{-2}$. This step size is chosen based on several preliminary $J\!=\!0.5$ simulations at $T\!=\!3.0$ in which integrations were performed at systematically varied $\Delta{t}$, ranging from $10^{-1}$ to $10^{-4}$, until the $N\!=\!4096$ samples reached statistically stationary states. The changes in the resulting steady-states are negligible below $\Delta{t}\!=\!3\!\times\!10^{-2}$, and hence we choose $\Delta t\!=\!10^{-2}$.

Thermally annealed initial zero-force configurations were prepared through the following protocol: systems are initially driven to statistically stationary states at $T\!=\!2.0$, characterized by the time-translational invariance of macroscopic observables. We then subject the samples to an instantaneous quench to the ``parent temperature'' $T_{\rm p}\!<\!2.0$, and continue the simulation for a duration of $10^{5}$ time units (i.e., $10^{7}$ time steps) for $T_{\rm p}\!<\!1.0$, and for a duration of $10^3$ for $T_{\rm p}\!\geq\!1.0$. Finally, we quench all samples to zero temperature by performing energy minimization using the conjugate gradient method. To obtain the results presented in Fig.~\ref{fig:fig4}c, $M\!=\!16$ independent quenched disorder realizations with $N\!=\!4096$ were prepared for $0.2\!\leq\!T_{\rm p}\!\leq\!2.0$. From those, we subjected samples annealed at $T_{\rm p}\!=\!0.2$, $T_{\rm p}\!=\!0.6$, $T_{\rm p}\!=\!0.8$, and $T_{\rm p}\!=\!2.0$ to oscillatory driving, the results of which are reported on in Fig.~\ref{fig:fig4}c.

\begin{figure}[t]
    \centering
    \includegraphics[width=1\columnwidth]{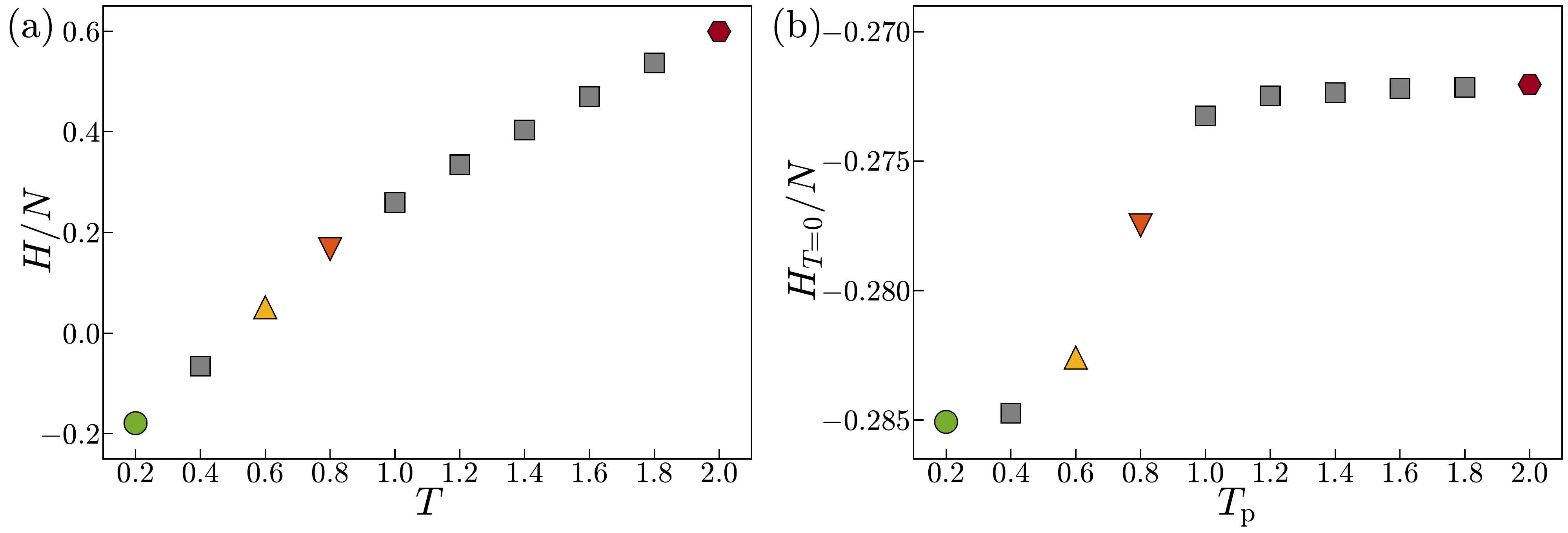}
    \vspace{-0.4cm}
    \caption{(a) The energy per oscillator $H/N$ computed at different temperatures $T$ just before the final instantaneous quench to zero temperature. (b) The energy after a quench to zero temperature, $H_{T=0}/N$, plotted against the parent temperature $T_{\rm p}$ (the temperature $T$ appearing on panel (a) from which the system was quenched to $T\!=\!0$). Colored symbols indicate the samples subjected to oscillatory driving, using the same symbols and colors as in Fig.~\ref{fig:fig4}c.}
    \vspace{-0.2cm}
    \label{fig:figS4}
\end{figure}

In Fig.~\ref{fig:figS4}a, we present $H/N$ computed at different temperatures $T$ just before the final instantaneous quench to zero temperature. The energy after the quench, denoted as $H_{T=0}/N$, is plotted in Fig.~\ref{fig:figS4}b against $T_{\rm p}$, where the latter corresponds to $T$ of panel (a) to which instantaneous $T\=0$ quenches are applied. It is important to note that we employed a simple, well-defined thermalization procedure, which does not guarantee reaching equilibrium at the full range of temperatures considered for the simulation times used. This issue is interesting on its own and should be studied separately.

\vspace{-0.4cm}
\section{T\lowercase{he relation between \uppercase{$J$} and the number of soft non-phononic modes, and a ``brittle-like'' behavior}}
\vspace{-0.3cm}
\label{sec:J_QLMs}

\begin{figure*}[htp]
    \centering
    \includegraphics[width=1.0\linewidth]{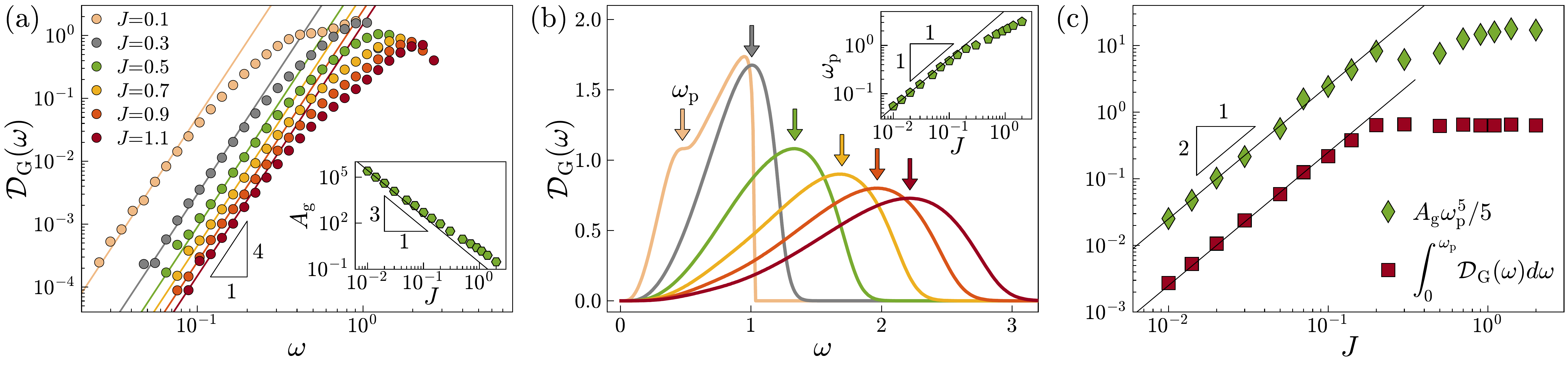}
    \vspace{-0.4cm}
    \caption{(a) The non-phononic VDoS $\mathcal{D}_{\rm G}(\omega)$ for various $J$ values (see legend), plotted on a double-logarithmic scale. The superposed straight lines are fits to $A_{\rm g}\omega^4$ in the low-frequency regime. (inset) The extracted quartic prefactor $A_{\rm g}$ as a function of $J$ (the $A_{\rm g}(J)\!\sim\!J^{-3}$ behavior in the small $J$ regime is highlighted by the superposed line and the power-law triangle). (b) The same as panel (a), but on a linear scale. The arrows mark the first peak frequency $\omega_{\rm p}$ (following the same color code of panel (a)). (inset) $\omega_{\rm p}$ vs.~$J$ (the $\omega_{\rm p}(J)\!\sim\!J$ behavior in the small $J$ regime is highlighted by the superposed line and the power-law triangle). (c) The number of soft non-phononic modes up to the frequency $\omega_{\rm p}$, $\int^{\omega_{\rm p}}_{0}\!\mathcal{D}_{\rm G}(\omega)\,d\omega$, plotted against $J$ (brown-red squares). $\int^{\omega_{\rm p}}_{0}\!A_{\rm g}\omega^4\,d\omega\!=\!A_{\rm g}\omega_{\rm p}^5/5$ vs.~$J$ is plotted as well (green diamonds). The $\sim\!J^2$ behavior in the small $J$ regime is highlighted by the lines superposed on both datasets and the power-law triangle. The non-phononic VDoSs in the non-driven regime analyzed and presented here have been obtained by averaging over $M\!=\!2000$ realizations of the quenched disorder in a system size of $N\!=\!8192$. See text for the discussion of the results.}
    \vspace{-0.2cm}
    \label{fig:figS6}
\end{figure*}

As stressed in the main text, in the dimensionless formulation of the model, quenched disorder is varied by a single parameter $J$. Yet, various physical observables can be affected differently by varying $J$. Of particular interest is the effect of $J$ on the underlying complex energy landscape of the model, which dictates many of its emergent statistical and dynamical properties. Specifically, one is interested in computing the multiplicity of energy landscape minima and the barriers that separate them, which is in general a daunting task. Soft non-phononic modes, as quantified by the vibrational density of states (VDoS) ${\cal D}_{\rm G}(\omega)$ discussed in the main text, may serve as a proxy for these landscape properties. The reason is that they quantify ``soft directions'' in the energy landscape that are dynamically accessible to the system once some driving forces, e.g., mechanical and/or thermal, are applied. Here, we explore this possibility.

As mentioned in the introduction, and as extensively discussed in~\cite{Moriel2024-wp}, ${\cal D}_{\rm G}(\omega)$ features a generic peak at a frequency $\omega_{\rm p}$, which is intrinsically related to the well-known “boson peak” in glasses. In Fig.~\ref{fig:figS6}a, we present the VDoS ${\cal D}_{\rm G}(\omega)$ for the model over the entire range of $J$ values considered in the main text, i.e., $J\=0.1\!-\!1.1$ (see legend), on a double-logarithmic scale. The universal quartic tail, ${\cal D}_{\rm G}(\omega)\!\sim\!\omega^4$, is evident (see the lines going through the small $\omega$ data and the 4:1 power-law triangle). The same data are presented in Fig.~\ref{fig:figS6}b on a linear scale, clearly revealing a peak at $\omega_{\rm p}$, marked by an arrow on each curve (using the same color code as that of Fig.~\ref{fig:figS6}a). The dependence of $\omega_{\rm p}$ on $J$ is presented in the inset over a broader range of $J$; it is important to note that for small $J$, e.g., $J\=0.1$ in the main panel, $\omega_{\rm p}$ corresponds to the first local maximum of ${\cal D}_{\rm G}(\omega)$, where the predominantly linear ${\cal D}_{\rm G}(\omega)\!\sim\!\omega$ for $\omega\!>\!\omega_{\rm p}$ corresponds to initial liquid-like modes that did not undergo interaction-induced ``mixing/reconstruction'', as discussed in~\cite{Moriel2024-wp}.

A natural quantity of interest is the number of soft modes up to $\omega_{\rm p}$, as quantified by
\begin{equation}
    \int_0^{\omega_{\rm p}}{\cal D}_{\rm G}(\omega)\,d\omega \ ,
    \label{eq:D_G_integral}
\end{equation}
which may be viewed as a quantifier of disorder~\cite{Lerner2021-en}, and its variation with $J$. The quantity defined in Eq.~\eqref{eq:D_G_integral} is plotted in Fig.~\ref{fig:figS6}c (brown-red squares). It is observed that $\int_0^{\omega_{\rm p}}{\cal D}_{\rm G}(\omega)\,d\omega$ is very small for small $J$, increases as $\sim\!J^2$ up to $J\!\simeq\!0.2$ and then levels off at larger $J$ values.

These observations can be understood in the following way; first, consider the universal quartic tail ${\cal D}_{\rm G}(\omega)\=A_{\rm g}\,\omega^4$, where the non-universal prefactor $A_{\rm g}$~\cite{Lerner2021-en} is extracted in Fig.~\ref{fig:figS6}a. It is presented in the inset therein and observed to follow $A_{\rm g}(J)\!\sim\!J^{-3}$ at small $J$. Combining the latter with the inset of Fig.~\ref{fig:figS6}b, which indicates that $\omega_{\rm p}(J)\!\sim\!J$ at small $J$, implies $\int_0^{\omega_{\rm p}}A_{\rm g}\,\omega^4\,d\omega\=A_{\rm g}\omega_{\rm p}^5/5\!\sim\!J^2$, exactly as observed for $\int_0^{\omega_{\rm p}}{\cal D}_{\rm G}(\omega)\,d\omega$ at small $J$. $A_{\rm g}\omega_{\rm p}^5/5$ is superposed on Fig.~\ref{fig:figS6}c (green diamonds), revealing the $J^2$ power-law at small $J$ (albeit up shifted, as expected). Moreover, it also reveals a quasi-plateau above $J\!\simeq\!0.2$, similarly to $\int_0^{\omega_{\rm p}}{\cal D}_{\rm G}(\omega)\,d\omega$.

The plateau in $\int_0^{\omega_{\rm p}}{\cal D}_{\rm G}(\omega)\,d\omega$ above $J\!\simeq\!0.2$ indicates that the fraction of soft modes in the frequency range $0\!\le\!\omega\!\le\!\omega_{\rm p}(J)$ becomes nearly $J$ independent for $J\!\gtrsim\!0.2$. The crossover between the $J^2$ power-law and the plateau at $J\!\simeq\!0.2$ corresponds to the strength of disordered interactions $J$ at which all initial liquid-like modes undergo interaction-induced ``mixing/reconstruction''~\cite{Lerner2021-en,Moriel2024-wp}. The small number of soft modes in the $J\!\lesssim\!0.2$ regime appears to indicate a reduced number of accessible energy landscape minima, possibly also echoed in the Replica Method solution of the non-driven model~\cite{Bouchbinder2021-dh} in which the mean-field ``effective potential'' features double-wells with a low probability at small $J$ values.

It is precisely this regime, i.e., $J\!\lesssim\!0.2$, where the driven model (cf.~Fig.~\ref{fig:fig4}a-b) reveals little mechanical annealibility and no yielding. We interpret this regime in the driven model to correspond to a ``brittle-like'' behavior. It is important to stress that the ``brittle-like'' behavior in the model is not accompanied by yielding, unlike the corresponding observations in computer glasses. The reason for this is clear from the structure of the mean-field model in which the Hamiltonian features only the leading order $x^4$ nonlinearity/anharmonicity; that is, in the absence of accessible energy landscape minima for plastic rearrangements, the energy will simply be controlled by the $x^4$ contributions.

Finally, in the plateau $J\!\gtrsim\!0.2$ regime in Fig.~\ref{fig:figS6}c, other physical quantity may feature $J$ dependence. For example, it is evident from Fig.~\ref{fig:figS6}b that modes undergo progressive stiffening with increasing $J$, though their total number up to $\omega_{\rm p}(J)$ is essentially constant. All of these important issues should be carefully addressed in a follow-up investigation.



\begin{thebibliography}{99}%
\makeatletter
\providecommand \@ifxundefined [1]{%
 \@ifx{#1\undefined}
}%
\providecommand \@ifnum [1]{%
 \ifnum #1\expandafter \@firstoftwo
 \else \expandafter \@secondoftwo
 \fi
}%
\providecommand \@ifx [1]{%
 \ifx #1\expandafter \@firstoftwo
 \else \expandafter \@secondoftwo
 \fi
}%
\providecommand \natexlab [1]{#1}%
\providecommand \enquote  [1]{``#1''}%
\providecommand \bibnamefont  [1]{#1}%
\providecommand \bibfnamefont [1]{#1}%
\providecommand \citenamefont [1]{#1}%
\providecommand \href@noop [0]{\@secondoftwo}%
\providecommand \href [0]{\begingroup \@sanitize@url \@href}%
\providecommand \@href[1]{\@@startlink{#1}\@@href}%
\providecommand \@@href[1]{\endgroup#1\@@endlink}%
\providecommand \@sanitize@url [0]{\catcode `\\12\catcode `\$12\catcode
  `\&12\catcode `\#12\catcode `\^12\catcode `\_12\catcode `\%12\relax}%
\providecommand \@@startlink[1]{}%
\providecommand \@@endlink[0]{}%
\providecommand \url  [0]{\begingroup\@sanitize@url \@url }%
\providecommand \@url [1]{\endgroup\@href {#1}{\urlprefix }}%
\providecommand \urlprefix  [0]{URL }%
\providecommand \Eprint [0]{\href }%
\providecommand \doibase [0]{https://doi.org/}%
\providecommand \selectlanguage [0]{\@gobble}%
\providecommand \bibinfo  [0]{\@secondoftwo}%
\providecommand \bibfield  [0]{\@secondoftwo}%
\providecommand \translation [1]{[#1]}%
\providecommand \BibitemOpen [0]{}%
\providecommand \bibitemStop [0]{}%
\providecommand \bibitemNoStop [0]{.\EOS\space}%
\providecommand \EOS [0]{\spacefactor3000\relax}%
\providecommand \BibitemShut  [1]{\csname bibitem#1\endcsname}%
\let\auto@bib@innerbib\@empty
\bibitem [{\citenamefont {Goldstein}(1969)}]{Goldstein1969-ao}%
  \BibitemOpen
  \bibfield  {author} {\bibinfo {author} {\bibfnamefont {M.}~\bibnamefont
  {Goldstein}},\ }\bibfield  {title} {\bibinfo {title} {Viscous liquids and the
  glass transition: A potential energy barrier picture},\ }\href
  {https://doi.org/https://doi.org/10.1063/1.1672587} {\bibfield  {journal}
  {\bibinfo  {journal} {J. Chem. Phys.}\ }\textbf {\bibinfo {volume} {51}},\
  \bibinfo {pages} {3728} (\bibinfo {year} {1969})}\BibitemShut {NoStop}%
\bibitem [{\citenamefont {Debenedetti}\ and\ \citenamefont
  {Stillinger}(2001)}]{Debenedetti2001-an}%
  \BibitemOpen
  \bibfield  {author} {\bibinfo {author} {\bibfnamefont {P.~G.}\ \bibnamefont
  {Debenedetti}}\ and\ \bibinfo {author} {\bibfnamefont {F.~H.}\ \bibnamefont
  {Stillinger}},\ }\bibfield  {title} {\bibinfo {title} {Supercooled liquids
  and the glass transition},\ }\href
  {https://doi.org/https://doi.org/10.1038/35065704} {\bibfield  {journal}
  {\bibinfo  {journal} {Nature}\ }\textbf {\bibinfo {volume} {410}},\ \bibinfo
  {pages} {259} (\bibinfo {year} {2001})}\BibitemShut {NoStop}%
\bibitem [{\citenamefont {Ediger}\ and\ \citenamefont
  {Harrowell}(2012)}]{Ediger2012-xg}%
  \BibitemOpen
  \bibfield  {author} {\bibinfo {author} {\bibfnamefont {M.~D.}\ \bibnamefont
  {Ediger}}\ and\ \bibinfo {author} {\bibfnamefont {P.}~\bibnamefont
  {Harrowell}},\ }\bibfield  {title} {\bibinfo {title} {Perspective:
  Supercooled liquids and glasses},\ }\href
  {https://doi.org/https://doi.org/10.1063/1.4747326} {\bibfield  {journal}
  {\bibinfo  {journal} {J. Chem. Phys.}\ }\textbf {\bibinfo {volume} {137}},\
  \bibinfo {pages} {080901} (\bibinfo {year} {2012})}\BibitemShut {NoStop}%
\bibitem [{\citenamefont {Cavagna}(2009)}]{Cavagna_pedestrians_2009}%
  \BibitemOpen
  \bibfield  {author} {\bibinfo {author} {\bibfnamefont {A.}~\bibnamefont
  {Cavagna}},\ }\bibfield  {title} {\bibinfo {title} {Supercooled liquids for
  pedestrians},\ }\href
  {https://doi.org/https://doi.org/10.1016/j.physrep.2009.03.003} {\bibfield
  {journal} {\bibinfo  {journal} {Phys. Rep.}\ }\textbf {\bibinfo {volume}
  {476}},\ \bibinfo {pages} {51} (\bibinfo {year} {2009})}\BibitemShut
  {NoStop}%
\bibitem [{\citenamefont {Zeller}\ and\ \citenamefont
  {Pohl}(1971)}]{Zeller1971-oj}%
  \BibitemOpen
  \bibfield  {author} {\bibinfo {author} {\bibfnamefont {R.~C.}\ \bibnamefont
  {Zeller}}\ and\ \bibinfo {author} {\bibfnamefont {R.~O.}\ \bibnamefont
  {Pohl}},\ }\bibfield  {title} {\bibinfo {title} {Thermal conductivity and
  specific heat of noncrystalline solids},\ }\href
  {https://doi.org/https://doi.org/10.1103/PhysRevB.4.2029} {\bibfield
  {journal} {\bibinfo  {journal} {Phys. Rev. B}\ }\textbf {\bibinfo {volume}
  {4}},\ \bibinfo {pages} {2029} (\bibinfo {year} {1971})}\BibitemShut
  {NoStop}%
\bibitem [{\citenamefont {Phillips}(1981)}]{Phillips1981-ue}%
  \BibitemOpen
  \bibfield  {author} {\bibinfo {author} {\bibfnamefont {W.~A.}\ \bibnamefont
  {Phillips}},\ }\href
  {https://doi.org/https://doi.org/10.1007/978-3-642-81534-8} {\emph {\bibinfo
  {title} {Amorphous solids: Low-temperature properties}}},\ \bibinfo {edition}
  {1981st}\ ed.,\ edited by\ \bibinfo {editor} {\bibfnamefont {W.~A.}\
  \bibnamefont {Phillips}}\ (\bibinfo  {publisher} {Springer},\ \bibinfo
  {address} {Berlin, Germany},\ \bibinfo {year} {1981})\BibitemShut {NoStop}%
\bibitem [{\citenamefont {Pohl}\ \emph {et~al.}(2002)\citenamefont {Pohl},
  \citenamefont {Liu},\ and\ \citenamefont {Thompson}}]{Pohl2002-bm}%
  \BibitemOpen
  \bibfield  {author} {\bibinfo {author} {\bibfnamefont {R.~O.}\ \bibnamefont
  {Pohl}}, \bibinfo {author} {\bibfnamefont {X.}~\bibnamefont {Liu}},\ and\
  \bibinfo {author} {\bibfnamefont {E.}~\bibnamefont {Thompson}},\ }\bibfield
  {title} {\bibinfo {title} {Low-temperature thermal conductivity and acoustic
  attenuation in amorphous solids},\ }\href
  {https://doi.org/https://doi.org/10.1103/RevModPhys.74.991} {\bibfield
  {journal} {\bibinfo  {journal} {Rev. Mod. Phys.}\ }\textbf {\bibinfo {volume}
  {74}},\ \bibinfo {pages} {991} (\bibinfo {year} {2002})}\BibitemShut
  {NoStop}%
\bibitem [{\citenamefont {Utz}\ \emph {et~al.}(2000)\citenamefont {Utz},
  \citenamefont {Debenedetti},\ and\ \citenamefont {Stillinger}}]{Utz2000-ry}%
  \BibitemOpen
  \bibfield  {author} {\bibinfo {author} {\bibfnamefont {M.}~\bibnamefont
  {Utz}}, \bibinfo {author} {\bibfnamefont {P.~G.}\ \bibnamefont
  {Debenedetti}},\ and\ \bibinfo {author} {\bibfnamefont {F.~H.}\ \bibnamefont
  {Stillinger}},\ }\bibfield  {title} {\bibinfo {title} {Atomistic simulation
  of aging and rejuvenation in glasses},\ }\href
  {https://doi.org/https://doi.org/10.1103/PhysRevLett.84.1471} {\bibfield
  {journal} {\bibinfo  {journal} {Phys. Rev. Lett.}\ }\textbf {\bibinfo
  {volume} {84}},\ \bibinfo {pages} {1471} (\bibinfo {year}
  {2000})}\BibitemShut {NoStop}%
\bibitem [{\citenamefont {Viasnoff}\ and\ \citenamefont
  {Lequeux}(2002)}]{Viasnoff2002-yk}%
  \BibitemOpen
  \bibfield  {author} {\bibinfo {author} {\bibfnamefont {V.}~\bibnamefont
  {Viasnoff}}\ and\ \bibinfo {author} {\bibfnamefont {F.}~\bibnamefont
  {Lequeux}},\ }\bibfield  {title} {\bibinfo {title} {Rejuvenation and
  overaging in a colloidal glass under shear},\ }\href
  {https://doi.org/https://doi.org/10.1103/PhysRevLett.89.065701} {\bibfield
  {journal} {\bibinfo  {journal} {Phys. Rev. Lett.}\ }\textbf {\bibinfo
  {volume} {89}},\ \bibinfo {pages} {065701} (\bibinfo {year}
  {2002})}\BibitemShut {NoStop}%
\bibitem [{\citenamefont {Lacks}\ and\ \citenamefont
  {Osborne}(2004)}]{Lacks2004-bf}%
  \BibitemOpen
  \bibfield  {author} {\bibinfo {author} {\bibfnamefont {D.~J.}\ \bibnamefont
  {Lacks}}\ and\ \bibinfo {author} {\bibfnamefont {M.~J.}\ \bibnamefont
  {Osborne}},\ }\bibfield  {title} {\bibinfo {title} {Energy landscape picture
  of overaging and rejuvenation in a sheared glass},\ }\href
  {https://doi.org/https://doi.org/10.1103/PhysRevLett.93.255501} {\bibfield
  {journal} {\bibinfo  {journal} {Phys. Rev. Lett.}\ }\textbf {\bibinfo
  {volume} {93}},\ \bibinfo {pages} {255501} (\bibinfo {year}
  {2004})}\BibitemShut {NoStop}%
\bibitem [{\citenamefont {Berthier}\ \emph {et~al.}(2025)\citenamefont
  {Berthier}, \citenamefont {Biroli}, \citenamefont {Manning},\ and\
  \citenamefont {Zamponi}}]{Berthier2025-hv}%
  \BibitemOpen
  \bibfield  {author} {\bibinfo {author} {\bibfnamefont {L.}~\bibnamefont
  {Berthier}}, \bibinfo {author} {\bibfnamefont {G.}~\bibnamefont {Biroli}},
  \bibinfo {author} {\bibfnamefont {L.}~\bibnamefont {Manning}},\ and\ \bibinfo
  {author} {\bibfnamefont {F.}~\bibnamefont {Zamponi}},\ }\bibfield  {title}
  {\bibinfo {title} {Yielding and plasticity in amorphous solids},\ }\href
  {https://doi.org/https://doi.org/10.1038/s42254-025-00833-5} {\bibfield
  {journal} {\bibinfo  {journal} {Nat. Rev. Phys.}}
  \textbf{\bibinfo {volume}{7}},\ \bibinfo {pages} {313} (\bibinfo {year} {2025})}\BibitemShut {NoStop}%
\bibitem [{\citenamefont {Keim}\ \emph {et~al.}(2019)\citenamefont {Keim},
  \citenamefont {Paulsen}, \citenamefont {Zeravcic}, \citenamefont {Sastry},\
  and\ \citenamefont {Nagel}}]{Keim2019-ez}%
  \BibitemOpen
  \bibfield  {author} {\bibinfo {author} {\bibfnamefont {N.~C.}\ \bibnamefont
  {Keim}}, \bibinfo {author} {\bibfnamefont {J.~D.}\ \bibnamefont {Paulsen}},
  \bibinfo {author} {\bibfnamefont {Z.}~\bibnamefont {Zeravcic}}, \bibinfo
  {author} {\bibfnamefont {S.}~\bibnamefont {Sastry}},\ and\ \bibinfo {author}
  {\bibfnamefont {S.~R.}\ \bibnamefont {Nagel}},\ }\bibfield  {title} {\bibinfo
  {title} {Memory formation in matter},\ }\href
  {https://doi.org/https://doi.org/10.1103/RevModPhys.91.035002} {\bibfield
  {journal} {\bibinfo  {journal} {Rev. Mod. Phys.}\ }\textbf {\bibinfo {volume}
  {91}},\ \bibinfo {pages} {035002} (\bibinfo {year} {2019})}\BibitemShut
  {NoStop}%
\bibitem [{\citenamefont {Paulsen}\ and\ \citenamefont
  {Keim}(2025)}]{Paulsen2025-av}%
  \BibitemOpen
  \bibfield  {author} {\bibinfo {author} {\bibfnamefont {J.~D.}\ \bibnamefont
  {Paulsen}}\ and\ \bibinfo {author} {\bibfnamefont {N.~C.}\ \bibnamefont
  {Keim}},\ }\bibfield  {title} {\bibinfo {title} {Mechanical memories in
  solids, from disorder to design},\ }\href
  {https://doi.org/https://doi.org/10.1146/annurev-conmatphys-032822-035544}
  {\bibfield  {journal} {\bibinfo  {journal} {Annu. Rev. Condens. Matter
  Phys.}\ }\textbf {\bibinfo {volume} {16}},\ \bibinfo {pages} {61} (\bibinfo
  {year} {2025})}\BibitemShut {NoStop}%
\bibitem [{\citenamefont {Gupta}\ and\ \citenamefont
  {Kob}(2019)}]{Gupta2019-mb}%
  \BibitemOpen
  \bibfield  {author} {\bibinfo {author} {\bibfnamefont {P.~K.}\ \bibnamefont
  {Gupta}}\ and\ \bibinfo {author} {\bibfnamefont {W.}~\bibnamefont {Kob}},\
  }\bibfield  {title} {\bibinfo {title} {Basis glass states: New insights from
  the potential energy landscape},\ }\href
  {https://doi.org/https://doi.org/10.1016/j.nocx.2019.100031} {\bibfield
  {journal} {\bibinfo  {journal} {J. Non-Cryst. Solids: X}\ }\textbf {\bibinfo
  {volume} {3}},\ \bibinfo {pages} {100031} (\bibinfo {year}
  {2019})}\BibitemShut {NoStop}%
\bibitem [{\citenamefont {Ojovan}\ and\ \citenamefont
  {Lee}(2011)}]{Ojovan2011-pc}%
  \BibitemOpen
  \bibfield  {author} {\bibinfo {author} {\bibfnamefont {M.~I.}\ \bibnamefont
  {Ojovan}}\ and\ \bibinfo {author} {\bibfnamefont {W.~E.}\ \bibnamefont
  {Lee}},\ }\bibfield  {title} {\bibinfo {title} {Glassy wasteforms for nuclear
  waste immobilization},\ }\href
  {https://doi.org/https://doi.org/10.1007/s11661-010-0525-7} {\bibfield
  {journal} {\bibinfo  {journal} {Metall. Mater. Trans. A}\ }\textbf {\bibinfo
  {volume} {42}},\ \bibinfo {pages} {837} (\bibinfo {year} {2011})}\BibitemShut
  {NoStop}%
\bibitem [{\citenamefont {Simpson}\ \emph {et~al.}(2011)\citenamefont
  {Simpson}, \citenamefont {Fons}, \citenamefont {Kolobov}, \citenamefont
  {Fukaya}, \citenamefont {Krbal}, \citenamefont {Yagi},\ and\ \citenamefont
  {Tominaga}}]{Simpson2011-tb}%
  \BibitemOpen
  \bibfield  {author} {\bibinfo {author} {\bibfnamefont {R.~E.}\ \bibnamefont
  {Simpson}}, \bibinfo {author} {\bibfnamefont {P.}~\bibnamefont {Fons}},
  \bibinfo {author} {\bibfnamefont {A.~V.}\ \bibnamefont {Kolobov}}, \bibinfo
  {author} {\bibfnamefont {T.}~\bibnamefont {Fukaya}}, \bibinfo {author}
  {\bibfnamefont {M.}~\bibnamefont {Krbal}}, \bibinfo {author} {\bibfnamefont
  {T.}~\bibnamefont {Yagi}},\ and\ \bibinfo {author} {\bibfnamefont
  {J.}~\bibnamefont {Tominaga}},\ }\bibfield  {title} {\bibinfo {title}
  {Interfacial phase-change memory},\ }\href
  {https://doi.org/https://doi.org/10.1038/nnano.2011.96} {\bibfield  {journal}
  {\bibinfo  {journal} {Nat. Nanotechnol.}\ }\textbf {\bibinfo {volume} {6}},\
  \bibinfo {pages} {501} (\bibinfo {year} {2011})}\BibitemShut {NoStop}%
\bibitem [{\citenamefont {Jones}(2013)}]{Jones2013-ri}%
  \BibitemOpen
  \bibfield  {author} {\bibinfo {author} {\bibfnamefont {J.~R.}\ \bibnamefont
  {Jones}},\ }\bibfield  {title} {\bibinfo {title} {Review of bioactive glass:
  from hench to hybrids},\ }\href
  {https://doi.org/https://doi.org/10.1016/j.actbio.2012.08.023} {\bibfield
  {journal} {\bibinfo  {journal} {Acta Biomater.}\ }\textbf {\bibinfo {volume}
  {9}},\ \bibinfo {pages} {4457} (\bibinfo {year} {2013})}\BibitemShut
  {NoStop}%
\bibitem [{\citenamefont {Buchenau}(1999)}]{buchenau1999neutron}%
  \BibitemOpen
  \bibfield  {author} {\bibinfo {author} {\bibfnamefont {U.}~\bibnamefont
  {Buchenau}},\ }\bibfield  {title} {\bibinfo {title} {Neutron and x-ray
  scattering from glasses},\ }\href {https://doi.org/10.1063/1.1301447}
  {\bibfield  {journal} {\bibinfo  {journal} {AIP Conf. Proc.}\ }\textbf
  {\bibinfo {volume} {489}},\ \bibinfo {pages} {3–23} (\bibinfo {year}
  {1999})}\BibitemShut {NoStop}%
\bibitem [{\citenamefont {Hudson}(2006)}]{Hudson2006-sk}%
  \BibitemOpen
  \bibfield  {author} {\bibinfo {author} {\bibfnamefont {B.~S.}\ \bibnamefont
  {Hudson}},\ }\bibfield  {title} {\bibinfo {title} {Vibrational spectroscopy
  using inelastic neutron scattering: Overview and outlook},\ }\href
  {https://doi.org/https://doi.org/10.1016/j.vibspec.2006.04.014} {\bibfield
  {journal} {\bibinfo  {journal} {Vib. Spectrosc.}\ }\textbf {\bibinfo {volume}
  {42}},\ \bibinfo {pages} {25} (\bibinfo {year} {2006})}\BibitemShut {NoStop}%
\bibitem [{\citenamefont {Buchenau}\ \emph {et~al.}(1991)\citenamefont
  {Buchenau}, \citenamefont {Galperin}, \citenamefont {Gurevich},\ and\
  \citenamefont {Schober}}]{soft_potential_model_1991}%
  \BibitemOpen
  \bibfield  {author} {\bibinfo {author} {\bibfnamefont {U.}~\bibnamefont
  {Buchenau}}, \bibinfo {author} {\bibfnamefont {Y.~M.}\ \bibnamefont
  {Galperin}}, \bibinfo {author} {\bibfnamefont {V.~L.}\ \bibnamefont
  {Gurevich}},\ and\ \bibinfo {author} {\bibfnamefont {H.~R.}\ \bibnamefont
  {Schober}},\ }\bibfield  {title} {\bibinfo {title} {Anharmonic potentials and
  vibrational localization in glasses},\ }\href
  {https://doi.org/10.1103/PhysRevB.43.5039} {\bibfield  {journal} {\bibinfo
  {journal} {Phys. Rev. B}\ }\textbf {\bibinfo {volume} {43}},\ \bibinfo
  {pages} {5039} (\bibinfo {year} {1991})}\BibitemShut {NoStop}%
\bibitem [{\citenamefont {Gurarie}\ and\ \citenamefont
  {Chalker}(2003)}]{chalker2003}%
  \BibitemOpen
  \bibfield  {author} {\bibinfo {author} {\bibfnamefont {V.}~\bibnamefont
  {Gurarie}}\ and\ \bibinfo {author} {\bibfnamefont {J.~T.}\ \bibnamefont
  {Chalker}},\ }\bibfield  {title} {\bibinfo {title} {Bosonic excitations in
  random media},\ }\href {https://doi.org/10.1103/PhysRevB.68.134207}
  {\bibfield  {journal} {\bibinfo  {journal} {Phys. Rev. B}\ }\textbf {\bibinfo
  {volume} {68}},\ \bibinfo {pages} {134207} (\bibinfo {year}
  {2003})}\BibitemShut {NoStop}%
\bibitem [{\citenamefont {Gurevich}\ \emph {et~al.}(2003)\citenamefont
  {Gurevich}, \citenamefont {Parshin},\ and\ \citenamefont
  {Schober}}]{Gurevich2003}%
  \BibitemOpen
  \bibfield  {author} {\bibinfo {author} {\bibfnamefont {V.~L.}\ \bibnamefont
  {Gurevich}}, \bibinfo {author} {\bibfnamefont {D.~A.}\ \bibnamefont
  {Parshin}},\ and\ \bibinfo {author} {\bibfnamefont {H.~R.}\ \bibnamefont
  {Schober}},\ }\bibfield  {title} {\bibinfo {title} {Anharmonicity,
  vibrational instability, and the boson peak in glasses},\ }\href
  {https://doi.org/10.1103/PhysRevB.67.094203} {\bibfield  {journal} {\bibinfo
  {journal} {Phys. Rev. B}\ }\textbf {\bibinfo {volume} {67}},\ \bibinfo
  {pages} {094203} (\bibinfo {year} {2003})}\BibitemShut {NoStop}%
\bibitem [{\citenamefont {Parshin}\ \emph {et~al.}(2007)\citenamefont
  {Parshin}, \citenamefont {Schober},\ and\ \citenamefont
  {Gurevich}}]{Gurevich2007}%
  \BibitemOpen
  \bibfield  {author} {\bibinfo {author} {\bibfnamefont {D.~A.}\ \bibnamefont
  {Parshin}}, \bibinfo {author} {\bibfnamefont {H.~R.}\ \bibnamefont
  {Schober}},\ and\ \bibinfo {author} {\bibfnamefont {V.~L.}\ \bibnamefont
  {Gurevich}},\ }\bibfield  {title} {\bibinfo {title} {Vibrational instability,
  two-level systems, and the boson peak in glasses},\ }\href
  {https://doi.org/https://doi.org/10.1103/PhysRevB.76.064206} {\bibfield
  {journal} {\bibinfo  {journal} {Phys. Rev. B}\ }\textbf {\bibinfo {volume}
  {76}},\ \bibinfo {pages} {064206} (\bibinfo {year} {2007})}\BibitemShut
  {NoStop}%
\bibitem [{\citenamefont {Lerner}\ \emph {et~al.}(2016)\citenamefont {Lerner},
  \citenamefont {Düring},\ and\ \citenamefont {Bouchbinder}}]{Lerner2016-ra}%
  \BibitemOpen
  \bibfield  {author} {\bibinfo {author} {\bibfnamefont {E.}~\bibnamefont
  {Lerner}}, \bibinfo {author} {\bibfnamefont {G.}~\bibnamefont {Düring}},\
  and\ \bibinfo {author} {\bibfnamefont {E.}~\bibnamefont {Bouchbinder}},\
  }\bibfield  {title} {\bibinfo {title} {Statistics and properties of
  low-frequency vibrational modes in structural glasses},\ }\href
  {https://doi.org/https://doi.org/10.1103/PhysRevLett.117.035501} {\bibfield
  {journal} {\bibinfo  {journal} {Phys. Rev. Lett.}\ }\textbf {\bibinfo
  {volume} {117}},\ \bibinfo {pages} {035501} (\bibinfo {year}
  {2016})}\BibitemShut {NoStop}%
\bibitem [{\citenamefont {Mizuno}\ \emph {et~al.}(2017)\citenamefont {Mizuno},
  \citenamefont {Shiba},\ and\ \citenamefont {Ikeda}}]{Mizuno2017-sc}%
  \BibitemOpen
  \bibfield  {author} {\bibinfo {author} {\bibfnamefont {H.}~\bibnamefont
  {Mizuno}}, \bibinfo {author} {\bibfnamefont {H.}~\bibnamefont {Shiba}},\ and\
  \bibinfo {author} {\bibfnamefont {A.}~\bibnamefont {Ikeda}},\ }\bibfield
  {title} {\bibinfo {title} {Continuum limit of the vibrational properties of
  amorphous solids},\ }\href
  {https://doi.org/https://doi.org/10.1073/pnas.1709015114} {\bibfield
  {journal} {\bibinfo  {journal} {Proc. Natl. Acad. Sci. U. S. A.}\ }\textbf
  {\bibinfo {volume} {114}},\ \bibinfo {pages} {E9767} (\bibinfo {year}
  {2017})}\BibitemShut {NoStop}%
\bibitem [{\citenamefont {Lerner}\ and\ \citenamefont
  {Bouchbinder}(2021)}]{Lerner2021-en}%
  \BibitemOpen
  \bibfield  {author} {\bibinfo {author} {\bibfnamefont {E.}~\bibnamefont
  {Lerner}}\ and\ \bibinfo {author} {\bibfnamefont {E.}~\bibnamefont
  {Bouchbinder}},\ }\bibfield  {title} {\bibinfo {title} {Low-energy
  quasilocalized excitations in structural glasses},\ }\href
  {https://doi.org/https://doi.org/10.1063/5.0069477} {\bibfield  {journal}
  {\bibinfo  {journal} {J. Chem. Phys.}\ }\textbf {\bibinfo {volume} {155}},\
  \bibinfo {pages} {200901} (\bibinfo {year} {2021})}\BibitemShut {NoStop}%
\bibitem [{\citenamefont {Moriel}\ \emph
  {et~al.}(2024{\natexlab{a}})\citenamefont {Moriel}, \citenamefont {Lerner},\
  and\ \citenamefont {Bouchbinder}}]{Moriel2024-lp}%
  \BibitemOpen
  \bibfield  {author} {\bibinfo {author} {\bibfnamefont {A.}~\bibnamefont
  {Moriel}}, \bibinfo {author} {\bibfnamefont {E.}~\bibnamefont {Lerner}},\
  and\ \bibinfo {author} {\bibfnamefont {E.}~\bibnamefont {Bouchbinder}},\
  }\bibfield  {title} {\bibinfo {title} {Experimental evidence for the
  $\omega^4$ tail of the nonphononic spectra of glasses},\ }\href
  {https://doi.org/https://doi.org/10.1063/5.0246261} {\bibfield  {journal}
  {\bibinfo  {journal} {J. Appl. Phys.}\ }\textbf {\bibinfo {volume} {136}},\
  \bibinfo {pages} {225105} (\bibinfo {year} {2024}{\natexlab{a}})}\BibitemShut
  {NoStop}%
\bibitem [{\citenamefont {Kalampounias}\ \emph {et~al.}(2006)\citenamefont
  {Kalampounias}, \citenamefont {Yannopoulos},\ and\ \citenamefont
  {Papatheodorou}}]{kalampounias2006low}%
  \BibitemOpen
  \bibfield  {author} {\bibinfo {author} {\bibfnamefont {A.}~\bibnamefont
  {Kalampounias}}, \bibinfo {author} {\bibfnamefont {S.}~\bibnamefont
  {Yannopoulos}},\ and\ \bibinfo {author} {\bibfnamefont {G.}~\bibnamefont
  {Papatheodorou}},\ }\bibfield  {title} {\bibinfo {title} {A low-frequency
  raman study of glassy, supercooled and molten silica and the preservation of
  the boson peak in the equilibrium liquid state},\ }\href
  {https://doi.org/10.1016/j.jnoncrysol.2006.02.163} {\bibfield  {journal}
  {\bibinfo  {journal} {J. Non-Cryst. Solids}\ }\textbf {\bibinfo {volume}
  {352}},\ \bibinfo {pages} {4619} (\bibinfo {year} {2006})}\BibitemShut
  {NoStop}%
\bibitem [{\citenamefont {Yannopoulos}\ \emph {et~al.}(2006)\citenamefont
  {Yannopoulos}, \citenamefont {Andrikopoulos},\ and\ \citenamefont
  {Ruocco}}]{yannopoulos2006analysis}%
  \BibitemOpen
  \bibfield  {author} {\bibinfo {author} {\bibfnamefont {S.}~\bibnamefont
  {Yannopoulos}}, \bibinfo {author} {\bibfnamefont {K.}~\bibnamefont
  {Andrikopoulos}},\ and\ \bibinfo {author} {\bibfnamefont {G.}~\bibnamefont
  {Ruocco}},\ }\bibfield  {title} {\bibinfo {title} {On the analysis of the
  vibrational boson peak and low-energy excitations in glasses},\ }\href
  {https://doi.org/10.1016/j.jnoncrysol.2006.02.164} {\bibfield  {journal}
  {\bibinfo  {journal} {J. Non-Cryst. Solids}\ }\textbf {\bibinfo {volume}
  {352}},\ \bibinfo {pages} {4541} (\bibinfo {year} {2006})}\BibitemShut
  {NoStop}%
\bibitem [{\citenamefont {Moriel}\ \emph
  {et~al.}(2024{\natexlab{b}})\citenamefont {Moriel}, \citenamefont {Lerner},\
  and\ \citenamefont {Bouchbinder}}]{Moriel2024-wp}%
  \BibitemOpen
  \bibfield  {author} {\bibinfo {author} {\bibfnamefont {A.}~\bibnamefont
  {Moriel}}, \bibinfo {author} {\bibfnamefont {E.}~\bibnamefont {Lerner}},\
  and\ \bibinfo {author} {\bibfnamefont {E.}~\bibnamefont {Bouchbinder}},\
  }\bibfield  {title} {\bibinfo {title} {Boson peak in the vibrational spectra
  of glasses},\ }\href
  {https://doi.org/https://doi.org/10.1103/PhysRevResearch.6.023053} {\bibfield
   {journal} {\bibinfo  {journal} {Phys. Rev. Res.}\ }\textbf {\bibinfo
  {volume} {6}},\ \bibinfo {pages} {023053} (\bibinfo {year}
  {2024}{\natexlab{b}})}\BibitemShut {NoStop}%
\bibitem [{\citenamefont {Malinovsky}\ and\ \citenamefont
  {Sokolov}(1986)}]{sokolov_1986}%
  \BibitemOpen
  \bibfield  {author} {\bibinfo {author} {\bibfnamefont {V.}~\bibnamefont
  {Malinovsky}}\ and\ \bibinfo {author} {\bibfnamefont {A.}~\bibnamefont
  {Sokolov}},\ }\bibfield  {title} {\bibinfo {title} {The nature of boson peak
  in raman scattering in glasses},\ }\href
  {https://doi.org/https://doi.org/10.1016/0038-1098(86)90854-9} {\bibfield
  {journal} {\bibinfo  {journal} {Solid State Commun.}\ }\textbf {\bibinfo
  {volume} {57}},\ \bibinfo {pages} {757} (\bibinfo {year} {1986})}\BibitemShut
  {NoStop}%
\bibitem [{\citenamefont {Ramos}\ \emph {et~al.}(1997)\citenamefont {Ramos},
  \citenamefont {Vieira}, \citenamefont {Bermejo}, \citenamefont {Dawidowski},
  \citenamefont {Fischer}, \citenamefont {Schober}, \citenamefont {Gonz\'alez},
  \citenamefont {Loong},\ and\ \citenamefont {Price}}]{ramos1997quantitative}%
  \BibitemOpen
  \bibfield  {author} {\bibinfo {author} {\bibfnamefont {M.~A.}\ \bibnamefont
  {Ramos}}, \bibinfo {author} {\bibfnamefont {S.}~\bibnamefont {Vieira}},
  \bibinfo {author} {\bibfnamefont {F.~J.}\ \bibnamefont {Bermejo}}, \bibinfo
  {author} {\bibfnamefont {J.}~\bibnamefont {Dawidowski}}, \bibinfo {author}
  {\bibfnamefont {H.~E.}\ \bibnamefont {Fischer}}, \bibinfo {author}
  {\bibfnamefont {H.}~\bibnamefont {Schober}}, \bibinfo {author} {\bibfnamefont
  {M.~A.}\ \bibnamefont {Gonz\'alez}}, \bibinfo {author} {\bibfnamefont
  {C.~K.}\ \bibnamefont {Loong}},\ and\ \bibinfo {author} {\bibfnamefont
  {D.~L.}\ \bibnamefont {Price}},\ }\bibfield  {title} {\bibinfo {title}
  {Quantitative assessment of the effects of orientational and positional
  disorder on glassy dynamics},\ }\href
  {https://doi.org/10.1103/PhysRevLett.78.82} {\bibfield  {journal} {\bibinfo
  {journal} {Phys. Rev. Lett.}\ }\textbf {\bibinfo {volume} {78}},\ \bibinfo
  {pages} {82} (\bibinfo {year} {1997})}\BibitemShut {NoStop}%
\bibitem [{\citenamefont {Wischnewski}\ \emph {et~al.}(1998)\citenamefont
  {Wischnewski}, \citenamefont {Buchenau}, \citenamefont {Dianoux},
  \citenamefont {Kamitakahara},\ and\ \citenamefont
  {and}}]{wischnewski1998neutron}%
  \BibitemOpen
  \bibfield  {author} {\bibinfo {author} {\bibfnamefont {A.}~\bibnamefont
  {Wischnewski}}, \bibinfo {author} {\bibfnamefont {U.}~\bibnamefont
  {Buchenau}}, \bibinfo {author} {\bibfnamefont {A.~J.}\ \bibnamefont
  {Dianoux}}, \bibinfo {author} {\bibfnamefont {W.~A.}\ \bibnamefont
  {Kamitakahara}},\ and\ \bibinfo {author} {\bibfnamefont {J.~L.~Z.}\
  \bibnamefont {and}},\ }\bibfield  {title} {\bibinfo {title} {Neutron
  scattering analysis of low-frequency modes in silica},\ }\href
  {https://doi.org/10.1080/13642819808204986} {\bibfield  {journal} {\bibinfo
  {journal} {Philos. Mag. B}\ }\textbf {\bibinfo {volume} {77}},\ \bibinfo
  {pages} {579} (\bibinfo {year} {1998})}\BibitemShut {NoStop}%
\bibitem [{\citenamefont {Surovtsev}\ \emph {et~al.}(2003)\citenamefont
  {Surovtsev}, \citenamefont {Shebanin},\ and\ \citenamefont
  {Ramos}}]{surovtsev2003density}%
  \BibitemOpen
  \bibfield  {author} {\bibinfo {author} {\bibfnamefont {N.}~\bibnamefont
  {Surovtsev}}, \bibinfo {author} {\bibfnamefont {A.}~\bibnamefont
  {Shebanin}},\ and\ \bibinfo {author} {\bibfnamefont {M.}~\bibnamefont
  {Ramos}},\ }\bibfield  {title} {\bibinfo {title} {Density of states and
  light-vibration coupling coefficient in $\text{B}_2\text{O}_3$ glasses with
  different thermal history},\ }\href
  {https://doi.org/10.1103/PhysRevB.67.024203} {\bibfield  {journal} {\bibinfo
  {journal} {Phys. Rev. B}\ }\textbf {\bibinfo {volume} {67}},\ \bibinfo
  {pages} {024203} (\bibinfo {year} {2003})}\BibitemShut {NoStop}%
\bibitem [{\citenamefont {Parisi}(2003)}]{parisi_boson_peak_2003}%
  \BibitemOpen
  \bibfield  {author} {\bibinfo {author} {\bibfnamefont {G.}~\bibnamefont
  {Parisi}},\ }\bibfield  {title} {\bibinfo {title} {On the origin of the boson
  peak},\ }\href {https://doi.org/10.1088/0953-8984/15/11/302} {\bibfield
  {journal} {\bibinfo  {journal} {J. Phys. Condens. Matter}\ }\textbf {\bibinfo
  {volume} {15}},\ \bibinfo {pages} {S765} (\bibinfo {year}
  {2003})}\BibitemShut {NoStop}%
\bibitem [{\citenamefont {Wyart}\ \emph {et~al.}(2005)\citenamefont {Wyart},
  \citenamefont {Nagel},\ and\ \citenamefont {Witten}}]{wyart_epl_2005}%
  \BibitemOpen
  \bibfield  {author} {\bibinfo {author} {\bibfnamefont {M.}~\bibnamefont
  {Wyart}}, \bibinfo {author} {\bibfnamefont {S.~R.}\ \bibnamefont {Nagel}},\
  and\ \bibinfo {author} {\bibfnamefont {T.~A.}\ \bibnamefont {Witten}},\
  }\bibfield  {title} {\bibinfo {title} {Geometric origin of excess
  low-frequency vibrational modes in weakly connected amorphous solids},\
  }\href {https://doi.org/10.1209/epl/i2005-10245-5} {\bibfield  {journal}
  {\bibinfo  {journal} {Europhys. Lett.}\ }\textbf {\bibinfo {volume} {72}},\
  \bibinfo {pages} {486} (\bibinfo {year} {2005})}\BibitemShut {NoStop}%
\bibitem [{\citenamefont {Monaco}\ \emph {et~al.}(2006)\citenamefont {Monaco},
  \citenamefont {Chumakov}, \citenamefont {Yue}, \citenamefont {Monaco},
  \citenamefont {Comez}, \citenamefont {Fioretto}, \citenamefont {Crichton},\
  and\ \citenamefont {R{\"u}ffer}}]{monaco2006density}%
  \BibitemOpen
  \bibfield  {author} {\bibinfo {author} {\bibfnamefont {A.}~\bibnamefont
  {Monaco}}, \bibinfo {author} {\bibfnamefont {A.}~\bibnamefont {Chumakov}},
  \bibinfo {author} {\bibfnamefont {Y.-Z.}\ \bibnamefont {Yue}}, \bibinfo
  {author} {\bibfnamefont {G.}~\bibnamefont {Monaco}}, \bibinfo {author}
  {\bibfnamefont {L.}~\bibnamefont {Comez}}, \bibinfo {author} {\bibfnamefont
  {D.}~\bibnamefont {Fioretto}}, \bibinfo {author} {\bibfnamefont
  {W.}~\bibnamefont {Crichton}},\ and\ \bibinfo {author} {\bibfnamefont
  {R.}~\bibnamefont {R{\"u}ffer}},\ }\bibfield  {title} {\bibinfo {title}
  {Density of vibrational states of a hyperquenched glass},\ }\href
  {https://doi.org/10.1103/PhysRevLett.96.205502} {\bibfield  {journal}
  {\bibinfo  {journal} {Phys. Rev. Lett.}\ }\textbf {\bibinfo {volume} {96}},\
  \bibinfo {pages} {205502} (\bibinfo {year} {2006})}\BibitemShut {NoStop}%
\bibitem [{\citenamefont {Baldi}\ \emph {et~al.}(2023)\citenamefont {Baldi},
  \citenamefont {Fontana},\ and\ \citenamefont
  {Monaco}}]{baldi2023vibrational}%
  \BibitemOpen
  \bibfield  {author} {\bibinfo {author} {\bibfnamefont {G.}~\bibnamefont
  {Baldi}}, \bibinfo {author} {\bibfnamefont {A.}~\bibnamefont {Fontana}},\
  and\ \bibinfo {author} {\bibfnamefont {G.}~\bibnamefont {Monaco}},\
  }\bibfield  {title} {\bibinfo {title} {Vibrational dynamics of
  non-crystalline solids},\ }in\ \href
  {https://doi.org/10.1142/9781800612587_0006} {\emph {\bibinfo {booktitle}
  {Low-Temperature Thermal and Vibrational Properties of Disordered Solids: A
  Half-Century of Universal “Anomalies” of Glasses}}}\ (\bibinfo
  {publisher} {World Scientific},\ \bibinfo {year} {2023})\ pp.\ \bibinfo
  {pages} {177--226}\BibitemShut {NoStop}%
\bibitem [{\citenamefont {Marruzzo}\ \emph {et~al.}(2013)\citenamefont
  {Marruzzo}, \citenamefont {Schirmacher}, \citenamefont {Fratalocchi},\ and\
  \citenamefont {Ruocco}}]{Schirmacher_2013_boson_peak}%
  \BibitemOpen
  \bibfield  {author} {\bibinfo {author} {\bibfnamefont {A.}~\bibnamefont
  {Marruzzo}}, \bibinfo {author} {\bibfnamefont {W.}~\bibnamefont
  {Schirmacher}}, \bibinfo {author} {\bibfnamefont {A.}~\bibnamefont
  {Fratalocchi}},\ and\ \bibinfo {author} {\bibfnamefont {G.}~\bibnamefont
  {Ruocco}},\ }\bibfield  {title} {\bibinfo {title} {Heterogeneous shear
  elasticity of glasses: the origin of the boson peak},\ }\href
  {https://doi.org/https://doi.org/10.1038/srep01407} {\bibfield  {journal}
  {\bibinfo  {journal} {Sci. Rep.}\ }\textbf {\bibinfo {volume} {3}},\ \bibinfo
  {pages} {1407} (\bibinfo {year} {2013})}\BibitemShut {NoStop}%
\bibitem [{\citenamefont {DeGiuli}\ \emph {et~al.}(2014)\citenamefont
  {DeGiuli}, \citenamefont {Laversanne-Finot}, \citenamefont {During},
  \citenamefont {Lerner},\ and\ \citenamefont {Wyart}}]{eric_boson_peak_emt}%
  \BibitemOpen
  \bibfield  {author} {\bibinfo {author} {\bibfnamefont {E.}~\bibnamefont
  {DeGiuli}}, \bibinfo {author} {\bibfnamefont {A.}~\bibnamefont
  {Laversanne-Finot}}, \bibinfo {author} {\bibfnamefont {G.}~\bibnamefont
  {During}}, \bibinfo {author} {\bibfnamefont {E.}~\bibnamefont {Lerner}},\
  and\ \bibinfo {author} {\bibfnamefont {M.}~\bibnamefont {Wyart}},\ }\bibfield
   {title} {\bibinfo {title} {Effects of coordination and pressure on sound
  attenuation{,} boson peak and elasticity in amorphous solids},\ }\href
  {https://doi.org/10.1039/C4SM00561A} {\bibfield  {journal} {\bibinfo
  {journal} {Soft Matter}\ }\textbf {\bibinfo {volume} {10}},\ \bibinfo {pages}
  {5628} (\bibinfo {year} {2014})}\BibitemShut {NoStop}%
\bibitem [{\citenamefont {González-Jiménez}\ \emph
  {et~al.}(2023)\citenamefont {González-Jiménez}, \citenamefont {Barnard},
  \citenamefont {Russell}, \citenamefont {Tukachev}, \citenamefont {Javornik},
  \citenamefont {Hayes}, \citenamefont {Farrell}, \citenamefont {Guinane},
  \citenamefont {Senn}, \citenamefont {Smith}, \citenamefont {Wilding},
  \citenamefont {Mali}, \citenamefont {Nakano}, \citenamefont {Miyazaki},
  \citenamefont {McMillan}, \citenamefont {Sosso},\ and\ \citenamefont
  {Wynne}}]{Gonzalez-Jimenez2023-un}%
  \BibitemOpen
  \bibfield  {author} {\bibinfo {author} {\bibfnamefont {M.}~\bibnamefont
  {González-Jiménez}}, \bibinfo {author} {\bibfnamefont {T.}~\bibnamefont
  {Barnard}}, \bibinfo {author} {\bibfnamefont {B.~A.}\ \bibnamefont
  {Russell}}, \bibinfo {author} {\bibfnamefont {N.~V.}\ \bibnamefont
  {Tukachev}}, \bibinfo {author} {\bibfnamefont {U.}~\bibnamefont {Javornik}},
  \bibinfo {author} {\bibfnamefont {L.-A.}\ \bibnamefont {Hayes}}, \bibinfo
  {author} {\bibfnamefont {A.~J.}\ \bibnamefont {Farrell}}, \bibinfo {author}
  {\bibfnamefont {S.}~\bibnamefont {Guinane}}, \bibinfo {author} {\bibfnamefont
  {H.~M.}\ \bibnamefont {Senn}}, \bibinfo {author} {\bibfnamefont {A.~J.}\
  \bibnamefont {Smith}}, \bibinfo {author} {\bibfnamefont {M.}~\bibnamefont
  {Wilding}}, \bibinfo {author} {\bibfnamefont {G.}~\bibnamefont {Mali}},
  \bibinfo {author} {\bibfnamefont {M.}~\bibnamefont {Nakano}}, \bibinfo
  {author} {\bibfnamefont {Y.}~\bibnamefont {Miyazaki}}, \bibinfo {author}
  {\bibfnamefont {P.}~\bibnamefont {McMillan}}, \bibinfo {author}
  {\bibfnamefont {G.~C.}\ \bibnamefont {Sosso}},\ and\ \bibinfo {author}
  {\bibfnamefont {K.}~\bibnamefont {Wynne}},\ }\bibfield  {title} {\bibinfo
  {title} {Understanding the emergence of the boson peak in molecular
  glasses},\ }\href
  {https://doi.org/https://doi.org/10.1038/s41467-023-35878-6} {\bibfield
  {journal} {\bibinfo  {journal} {Nat. Commun.}\ }\textbf {\bibinfo {volume}
  {14}},\ \bibinfo {pages} {215} (\bibinfo {year} {2023})}\BibitemShut
  {NoStop}%
\bibitem [{\citenamefont {Malandro}\ and\ \citenamefont
  {Lacks}(1999)}]{Malandro_Lacks}%
  \BibitemOpen
  \bibfield  {author} {\bibinfo {author} {\bibfnamefont {D.~L.}\ \bibnamefont
  {Malandro}}\ and\ \bibinfo {author} {\bibfnamefont {D.~J.}\ \bibnamefont
  {Lacks}},\ }\bibfield  {title} {\bibinfo {title} {Relationships of
  shear-induced changes in the potential energy landscape to the mechanical
  properties of ductile glasses},\ }\href
  {https://doi.org/http://dx.doi.org/10.1063/1.478340} {\bibfield  {journal}
  {\bibinfo  {journal} {J. Chem. Phys.}\ }\textbf {\bibinfo {volume} {110}},\
  \bibinfo {pages} {4593} (\bibinfo {year} {1999})}\BibitemShut {NoStop}%
\bibitem [{\citenamefont {Maloney}\ and\ \citenamefont
  {Lema\^{\i}tre}(2004)}]{lemaitre2004}%
  \BibitemOpen
  \bibfield  {author} {\bibinfo {author} {\bibfnamefont {C.}~\bibnamefont
  {Maloney}}\ and\ \bibinfo {author} {\bibfnamefont {A.}~\bibnamefont
  {Lema\^{\i}tre}},\ }\bibfield  {title} {\bibinfo {title} {Universal breakdown
  of elasticity at the onset of material failure},\ }\href
  {https://doi.org/10.1103/PhysRevLett.93.195501} {\bibfield  {journal}
  {\bibinfo  {journal} {Phys. Rev. Lett.}\ }\textbf {\bibinfo {volume} {93}},\
  \bibinfo {pages} {195501} (\bibinfo {year} {2004})}\BibitemShut {NoStop}%
\bibitem [{\citenamefont {Tanguy}\ \emph {et~al.}(2010)\citenamefont {Tanguy},
  \citenamefont {Mantisi},\ and\ \citenamefont {Tsamados}}]{Tanguy2010-at}%
  \BibitemOpen
  \bibfield  {author} {\bibinfo {author} {\bibfnamefont {A.}~\bibnamefont
  {Tanguy}}, \bibinfo {author} {\bibfnamefont {B.}~\bibnamefont {Mantisi}},\
  and\ \bibinfo {author} {\bibfnamefont {M.}~\bibnamefont {Tsamados}},\
  }\bibfield  {title} {\bibinfo {title} {Vibrational modes as a predictor for
  plasticity in a model glass},\ }\href
  {https://doi.org/10.1209/0295-5075/90/16004} {\bibfield  {journal} {\bibinfo
  {journal} {Europhys. Lett.}\ }\textbf {\bibinfo {volume} {90}},\ \bibinfo
  {pages} {16004} (\bibinfo {year} {2010})}\BibitemShut {NoStop}%
\bibitem [{\citenamefont {Manning}\ and\ \citenamefont
  {Liu}(2011)}]{Manning2011-tu}%
  \BibitemOpen
  \bibfield  {author} {\bibinfo {author} {\bibfnamefont {M.~L.}\ \bibnamefont
  {Manning}}\ and\ \bibinfo {author} {\bibfnamefont {A.~J.}\ \bibnamefont
  {Liu}},\ }\bibfield  {title} {\bibinfo {title} {Vibrational modes identify
  soft spots in a sheared disordered packing},\ }\href
  {https://doi.org/https://doi.org/10.1103/PhysRevLett.107.108302} {\bibfield
  {journal} {\bibinfo  {journal} {Phys. Rev. Lett.}\ }\textbf {\bibinfo
  {volume} {107}},\ \bibinfo {pages} {108302} (\bibinfo {year}
  {2011})}\BibitemShut {NoStop}%
\bibitem [{\citenamefont {Ding}\ \emph {et~al.}(2014)\citenamefont {Ding},
  \citenamefont {Patinet}, \citenamefont {Falk}, \citenamefont {Cheng},\ and\
  \citenamefont {Ma}}]{falk_soft_modes_pnas_2014}%
  \BibitemOpen
  \bibfield  {author} {\bibinfo {author} {\bibfnamefont {J.}~\bibnamefont
  {Ding}}, \bibinfo {author} {\bibfnamefont {S.}~\bibnamefont {Patinet}},
  \bibinfo {author} {\bibfnamefont {M.~L.}\ \bibnamefont {Falk}}, \bibinfo
  {author} {\bibfnamefont {Y.}~\bibnamefont {Cheng}},\ and\ \bibinfo {author}
  {\bibfnamefont {E.}~\bibnamefont {Ma}},\ }\bibfield  {title} {\bibinfo
  {title} {Soft spots and their structural signature in a metallic glass},\
  }\href {https://doi.org/10.1073/pnas.1412095111} {\bibfield  {journal}
  {\bibinfo  {journal} {Proc. Natl. Acad. Sci. U. S. A.}\ }\textbf {\bibinfo
  {volume} {111}},\ \bibinfo {pages} {14052} (\bibinfo {year}
  {2014})}\BibitemShut {NoStop}%
\bibitem [{\citenamefont {Lerner}(2016)}]{micromechanics2016}%
  \BibitemOpen
  \bibfield  {author} {\bibinfo {author} {\bibfnamefont {E.}~\bibnamefont
  {Lerner}},\ }\bibfield  {title} {\bibinfo {title} {Micromechanics of
  nonlinear plastic modes},\ }\href
  {https://doi.org/10.1103/PhysRevE.93.053004} {\bibfield  {journal} {\bibinfo
  {journal} {Phys. Rev. E}\ }\textbf {\bibinfo {volume} {93}},\ \bibinfo
  {pages} {053004} (\bibinfo {year} {2016})}\BibitemShut {NoStop}%
\bibitem [{\citenamefont {Richard}\ \emph {et~al.}(2020)\citenamefont
  {Richard}, \citenamefont {Ozawa}, \citenamefont {Patinet}, \citenamefont
  {Stanifer}, \citenamefont {Shang}, \citenamefont {Ridout}, \citenamefont
  {Xu}, \citenamefont {Zhang}, \citenamefont {Morse}, \citenamefont {Barrat},
  \citenamefont {Berthier}, \citenamefont {Falk}, \citenamefont {Guan},
  \citenamefont {Liu}, \citenamefont {Martens}, \citenamefont {Sastry},
  \citenamefont {Vandembroucq}, \citenamefont {Lerner},\ and\ \citenamefont
  {Manning}}]{Richard2020-vr}%
  \BibitemOpen
  \bibfield  {author} {\bibinfo {author} {\bibfnamefont {D.}~\bibnamefont
  {Richard}}, \bibinfo {author} {\bibfnamefont {M.}~\bibnamefont {Ozawa}},
  \bibinfo {author} {\bibfnamefont {S.}~\bibnamefont {Patinet}}, \bibinfo
  {author} {\bibfnamefont {E.}~\bibnamefont {Stanifer}}, \bibinfo {author}
  {\bibfnamefont {B.}~\bibnamefont {Shang}}, \bibinfo {author} {\bibfnamefont
  {S.~A.}\ \bibnamefont {Ridout}}, \bibinfo {author} {\bibfnamefont
  {B.}~\bibnamefont {Xu}}, \bibinfo {author} {\bibfnamefont {G.}~\bibnamefont
  {Zhang}}, \bibinfo {author} {\bibfnamefont {P.~K.}\ \bibnamefont {Morse}},
  \bibinfo {author} {\bibfnamefont {J.-L.}\ \bibnamefont {Barrat}}, \bibinfo
  {author} {\bibfnamefont {L.}~\bibnamefont {Berthier}}, \bibinfo {author}
  {\bibfnamefont {M.~L.}\ \bibnamefont {Falk}}, \bibinfo {author}
  {\bibfnamefont {P.}~\bibnamefont {Guan}}, \bibinfo {author} {\bibfnamefont
  {A.~J.}\ \bibnamefont {Liu}}, \bibinfo {author} {\bibfnamefont
  {K.}~\bibnamefont {Martens}}, \bibinfo {author} {\bibfnamefont
  {S.}~\bibnamefont {Sastry}}, \bibinfo {author} {\bibfnamefont
  {D.}~\bibnamefont {Vandembroucq}}, \bibinfo {author} {\bibfnamefont
  {E.}~\bibnamefont {Lerner}},\ and\ \bibinfo {author} {\bibfnamefont {M.~L.}\
  \bibnamefont {Manning}},\ }\bibfield  {title} {\bibinfo {title} {Predicting
  plasticity in disordered solids from structural indicators},\ }\href
  {https://doi.org/https://doi.org/10.1103/PhysRevMaterials.4.113609}
  {\bibfield  {journal} {\bibinfo  {journal} {Phys. Rev. Mater.}\ }\textbf
  {\bibinfo {volume} {4}},\ \bibinfo {pages} {113609} (\bibinfo {year}
  {2020})}\BibitemShut {NoStop}%
\bibitem [{\citenamefont {Fiocco}\ \emph {et~al.}(2013)\citenamefont {Fiocco},
  \citenamefont {Foffi},\ and\ \citenamefont {Sastry}}]{Fiocco2013-ak}%
  \BibitemOpen
  \bibfield  {author} {\bibinfo {author} {\bibfnamefont {D.}~\bibnamefont
  {Fiocco}}, \bibinfo {author} {\bibfnamefont {G.}~\bibnamefont {Foffi}},\ and\
  \bibinfo {author} {\bibfnamefont {S.}~\bibnamefont {Sastry}},\ }\bibfield
  {title} {\bibinfo {title} {Oscillatory athermal quasistatic deformation of a
  model glass},\ }\href
  {https://doi.org/https://doi.org/10.1103/PhysRevE.88.020301} {\bibfield
  {journal} {\bibinfo  {journal} {Phys. Rev. E}\ }\textbf {\bibinfo {volume}
  {88}},\ \bibinfo {pages} {020301} (\bibinfo {year} {2013})}\BibitemShut
  {NoStop}%
\bibitem [{\citenamefont {Regev}\ \emph {et~al.}(2013)\citenamefont {Regev},
  \citenamefont {Lookman},\ and\ \citenamefont {Reichhardt}}]{Regev2013-ff}%
  \BibitemOpen
  \bibfield  {author} {\bibinfo {author} {\bibfnamefont {I.}~\bibnamefont
  {Regev}}, \bibinfo {author} {\bibfnamefont {T.}~\bibnamefont {Lookman}},\
  and\ \bibinfo {author} {\bibfnamefont {C.}~\bibnamefont {Reichhardt}},\
  }\bibfield  {title} {\bibinfo {title} {Onset of irreversibility and chaos in
  amorphous solids under periodic shear},\ }\href
  {https://doi.org/https://doi.org/10.1103/PhysRevE.88.062401} {\bibfield
  {journal} {\bibinfo  {journal} {Phys. Rev. E}\ }\textbf {\bibinfo {volume}
  {88}},\ \bibinfo {pages} {062401} (\bibinfo {year} {2013})}\BibitemShut
  {NoStop}%
\bibitem [{\citenamefont {Regev}\ \emph {et~al.}(2015)\citenamefont {Regev},
  \citenamefont {Weber}, \citenamefont {Reichhardt}, \citenamefont {Dahmen},\
  and\ \citenamefont {Lookman}}]{Regev2015-vq}%
  \BibitemOpen
  \bibfield  {author} {\bibinfo {author} {\bibfnamefont {I.}~\bibnamefont
  {Regev}}, \bibinfo {author} {\bibfnamefont {J.}~\bibnamefont {Weber}},
  \bibinfo {author} {\bibfnamefont {C.}~\bibnamefont {Reichhardt}}, \bibinfo
  {author} {\bibfnamefont {K.~A.}\ \bibnamefont {Dahmen}},\ and\ \bibinfo
  {author} {\bibfnamefont {T.}~\bibnamefont {Lookman}},\ }\bibfield  {title}
  {\bibinfo {title} {Reversibility and criticality in amorphous solids},\
  }\href {https://doi.org/https://doi.org/10.1038/ncomms9805} {\bibfield
  {journal} {\bibinfo  {journal} {Nat. Commun.}\ }\textbf {\bibinfo {volume}
  {6}},\ \bibinfo {pages} {8805} (\bibinfo {year} {2015})}\BibitemShut
  {NoStop}%
\bibitem [{\citenamefont {Kawasaki}\ and\ \citenamefont
  {Berthier}(2016)}]{Kawasaki2016-nb}%
  \BibitemOpen
  \bibfield  {author} {\bibinfo {author} {\bibfnamefont {T.}~\bibnamefont
  {Kawasaki}}\ and\ \bibinfo {author} {\bibfnamefont {L.}~\bibnamefont
  {Berthier}},\ }\bibfield  {title} {\bibinfo {title} {Macroscopic yielding in
  jammed solids is accompanied by a nonequilibrium first-order transition in
  particle trajectories},\ }\href
  {https://doi.org/https://doi.org/10.1103/PhysRevE.94.022615} {\bibfield
  {journal} {\bibinfo  {journal} {Phys. Rev. E.}\ }\textbf {\bibinfo {volume}
  {94}},\ \bibinfo {pages} {022615} (\bibinfo {year} {2016})}\BibitemShut
  {NoStop}%
\bibitem [{\citenamefont {Leishangthem}\ \emph {et~al.}(2017)\citenamefont
  {Leishangthem}, \citenamefont {Parmar},\ and\ \citenamefont
  {Sastry}}]{Leishangthem2017-sw}%
  \BibitemOpen
  \bibfield  {author} {\bibinfo {author} {\bibfnamefont {P.}~\bibnamefont
  {Leishangthem}}, \bibinfo {author} {\bibfnamefont {A.~D.~S.}\ \bibnamefont
  {Parmar}},\ and\ \bibinfo {author} {\bibfnamefont {S.}~\bibnamefont
  {Sastry}},\ }\bibfield  {title} {\bibinfo {title} {The yielding transition in
  amorphous solids under oscillatory shear deformation},\ }\href
  {https://doi.org/https://doi.org/10.1038/ncomms14653} {\bibfield  {journal}
  {\bibinfo  {journal} {Nat.~Commun.}\ }\textbf {\bibinfo {volume} {8}},\
  \bibinfo {pages} {14653} (\bibinfo {year} {2017})}\BibitemShut {NoStop}%
\bibitem [{\citenamefont {Jin}\ \emph {et~al.}(2018)\citenamefont {Jin},
  \citenamefont {Urbani}, \citenamefont {Zamponi},\ and\ \citenamefont
  {Yoshino}}]{Jin2018-cy}%
  \BibitemOpen
  \bibfield  {author} {\bibinfo {author} {\bibfnamefont {Y.}~\bibnamefont
  {Jin}}, \bibinfo {author} {\bibfnamefont {P.}~\bibnamefont {Urbani}},
  \bibinfo {author} {\bibfnamefont {F.}~\bibnamefont {Zamponi}},\ and\ \bibinfo
  {author} {\bibfnamefont {H.}~\bibnamefont {Yoshino}},\ }\bibfield  {title}
  {\bibinfo {title} {A stability-reversibility map unifies elasticity,
  plasticity, yielding, and jamming in hard sphere glasses},\ }\href
  {https://doi.org/10.1126/sciadv.aat638} {\bibfield  {journal} {\bibinfo
  {journal} {Sci. Adv.}\ }\textbf {\bibinfo {volume} {4}},\ \bibinfo {pages}
  {eaat6387} (\bibinfo {year} {2018})}\BibitemShut {NoStop}%
\bibitem [{\citenamefont {Parmar}\ \emph {et~al.}(2019)\citenamefont {Parmar},
  \citenamefont {Kumar},\ and\ \citenamefont {Sastry}}]{Parmar2019-wj}%
  \BibitemOpen
  \bibfield  {author} {\bibinfo {author} {\bibfnamefont {A.~D.~S.}\
  \bibnamefont {Parmar}}, \bibinfo {author} {\bibfnamefont {S.}~\bibnamefont
  {Kumar}},\ and\ \bibinfo {author} {\bibfnamefont {S.}~\bibnamefont
  {Sastry}},\ }\bibfield  {title} {\bibinfo {title} {Strain localization above
  the yielding point in cyclically deformed glasses},\ }\href
  {https://doi.org/https://doi.org/10.1103/PhysRevX.9.021018} {\bibfield
  {journal} {\bibinfo  {journal} {Phys. Rev. X.}\ }\textbf {\bibinfo {volume}
  {9}},\ \bibinfo {pages} {021018} (\bibinfo {year} {2019})}\BibitemShut
  {NoStop}%
\bibitem [{\citenamefont {Schinasi-Lemberg}\ and\ \citenamefont
  {Regev}(2020)}]{Schinasi-Lemberg2020-ry}%
  \BibitemOpen
  \bibfield  {author} {\bibinfo {author} {\bibfnamefont {E.}~\bibnamefont
  {Schinasi-Lemberg}}\ and\ \bibinfo {author} {\bibfnamefont {I.}~\bibnamefont
  {Regev}},\ }\bibfield  {title} {\bibinfo {title} {Annealing and rejuvenation
  in a two-dimensional model amorphous solid under oscillatory shear},\ }\href
  {https://doi.org/https://doi.org/10.1103/PhysRevE.101.012603} {\bibfield
  {journal} {\bibinfo  {journal} {Phys. Rev. E.}\ }\textbf {\bibinfo {volume}
  {101}},\ \bibinfo {pages} {012603} (\bibinfo {year} {2020})}\BibitemShut
  {NoStop}%
\bibitem [{\citenamefont {Das}\ \emph {et~al.}(2020)\citenamefont {Das},
  \citenamefont {Vinutha},\ and\ \citenamefont {Sastry}}]{Das2020-cw}%
  \BibitemOpen
  \bibfield  {author} {\bibinfo {author} {\bibfnamefont {P.}~\bibnamefont
  {Das}}, \bibinfo {author} {\bibfnamefont {H.~A.}\ \bibnamefont {Vinutha}},\
  and\ \bibinfo {author} {\bibfnamefont {S.}~\bibnamefont {Sastry}},\
  }\bibfield  {title} {\bibinfo {title} {Unified phase diagram of
  reversible-irreversible, jamming, and yielding transitions in cyclically
  sheared soft-sphere packings},\ }\href
  {https://doi.org/https://doi.org/10.1103/PhysRevE.101.012603} {\bibfield
  {journal} {\bibinfo  {journal} {Proc. Natl. Acad. Sci. U. S. A.}\ }\textbf
  {\bibinfo {volume} {117}},\ \bibinfo {pages} {10203} (\bibinfo {year}
  {2020})}\BibitemShut {NoStop}%
\bibitem [{\citenamefont {Yeh}\ \emph {et~al.}(2020)\citenamefont {Yeh},
  \citenamefont {Ozawa}, \citenamefont {Miyazaki}, \citenamefont {Kawasaki},\
  and\ \citenamefont {Berthier}}]{Yeh2020-ve}%
  \BibitemOpen
  \bibfield  {author} {\bibinfo {author} {\bibfnamefont {W.-T.}\ \bibnamefont
  {Yeh}}, \bibinfo {author} {\bibfnamefont {M.}~\bibnamefont {Ozawa}}, \bibinfo
  {author} {\bibfnamefont {K.}~\bibnamefont {Miyazaki}}, \bibinfo {author}
  {\bibfnamefont {T.}~\bibnamefont {Kawasaki}},\ and\ \bibinfo {author}
  {\bibfnamefont {L.}~\bibnamefont {Berthier}},\ }\bibfield  {title} {\bibinfo
  {title} {Glass stability changes the nature of yielding under oscillatory
  shear},\ }\href
  {https://doi.org/https://doi.org/10.1103/PhysRevLett.124.225502} {\bibfield
  {journal} {\bibinfo  {journal} {Phys. Rev. Lett.}\ }\textbf {\bibinfo
  {volume} {124}},\ \bibinfo {pages} {225502} (\bibinfo {year}
  {2020})}\BibitemShut {NoStop}%
\bibitem [{\citenamefont {Bhaumik}\ \emph {et~al.}(2021)\citenamefont
  {Bhaumik}, \citenamefont {Foffi},\ and\ \citenamefont
  {Sastry}}]{Bhaumik2021-mr}%
  \BibitemOpen
  \bibfield  {author} {\bibinfo {author} {\bibfnamefont {H.}~\bibnamefont
  {Bhaumik}}, \bibinfo {author} {\bibfnamefont {G.}~\bibnamefont {Foffi}},\
  and\ \bibinfo {author} {\bibfnamefont {S.}~\bibnamefont {Sastry}},\
  }\bibfield  {title} {\bibinfo {title} {The role of annealing in determining
  the yielding behavior of glasses under cyclic shear deformation},\ }\href
  {https://doi.org/https://doi.org/10.1073/pnas.2100227118} {\bibfield
  {journal} {\bibinfo  {journal} {Proc. Natl. Acad. Sci. U. S. A.}\ }\textbf
  {\bibinfo {volume} {118}},\ \bibinfo {pages} {e2100227118} (\bibinfo {year}
  {2021})}\BibitemShut {NoStop}%
\bibitem [{\citenamefont {Goswami}\ \emph {et~al.}(2025)\citenamefont
  {Goswami}, \citenamefont {Shivashankar},\ and\ \citenamefont
  {Sastry}}]{Goswami2025-qv}%
  \BibitemOpen
  \bibfield  {author} {\bibinfo {author} {\bibfnamefont {Y.}~\bibnamefont
  {Goswami}}, \bibinfo {author} {\bibfnamefont {G.~V.}\ \bibnamefont
  {Shivashankar}},\ and\ \bibinfo {author} {\bibfnamefont {S.}~\bibnamefont
  {Sastry}},\ }\bibfield  {title} {\bibinfo {title} {Yielding behaviour of
  active particles in bulk and in confinement},\ }\href
  {https://doi.org/https://doi.org/10.1038/s41567-025-02843-7} {\bibfield
  {journal} {\bibinfo  {journal} {Nat. Phys.}\ }\textbf {\bibinfo {volume}
  {21}},\ \bibinfo {pages} {817} (\bibinfo {year} {2025})}\BibitemShut
  {NoStop}%
\bibitem [{\citenamefont {Liu}\ \emph {et~al.}(2022)\citenamefont {Liu},
  \citenamefont {Ferrero}, \citenamefont {Jagla}, \citenamefont {Martens},
  \citenamefont {Rosso},\ and\ \citenamefont {Talon}}]{Liu2022-hq}%
  \BibitemOpen
  \bibfield  {author} {\bibinfo {author} {\bibfnamefont {C.}~\bibnamefont
  {Liu}}, \bibinfo {author} {\bibfnamefont {E.~E.}\ \bibnamefont {Ferrero}},
  \bibinfo {author} {\bibfnamefont {E.~A.}\ \bibnamefont {Jagla}}, \bibinfo
  {author} {\bibfnamefont {K.}~\bibnamefont {Martens}}, \bibinfo {author}
  {\bibfnamefont {A.}~\bibnamefont {Rosso}},\ and\ \bibinfo {author}
  {\bibfnamefont {L.}~\bibnamefont {Talon}},\ }\bibfield  {title} {\bibinfo
  {title} {The fate of shear-oscillated amorphous solids},\ }\href
  {https://doi.org/https://doi.org/10.1063/5.0079460} {\bibfield  {journal}
  {\bibinfo  {journal} {J. Chem. Phys.}\ }\textbf {\bibinfo {volume} {156}},\
  \bibinfo {pages} {104902} (\bibinfo {year} {2022})}\BibitemShut {NoStop}%
\bibitem [{\citenamefont {Parley}\ \emph {et~al.}(2022)\citenamefont {Parley},
  \citenamefont {Sastry},\ and\ \citenamefont {Sollich}}]{Parley2022-id}%
  \BibitemOpen
  \bibfield  {author} {\bibinfo {author} {\bibfnamefont {J.~T.}\ \bibnamefont
  {Parley}}, \bibinfo {author} {\bibfnamefont {S.}~\bibnamefont {Sastry}},\
  and\ \bibinfo {author} {\bibfnamefont {P.}~\bibnamefont {Sollich}},\
  }\bibfield  {title} {\bibinfo {title} {Mean-field theory of yielding under
  oscillatory shear},\ }\href
  {https://doi.org/https://doi.org/10.1103/PhysRevLett.128.198001} {\bibfield
  {journal} {\bibinfo  {journal} {Phys. Rev. Lett.}\ }\textbf {\bibinfo
  {volume} {128}},\ \bibinfo {pages} {198001} (\bibinfo {year}
  {2022})}\BibitemShut {NoStop}%
\bibitem [{\citenamefont {Cochran}\ \emph {et~al.}(2024)\citenamefont
  {Cochran}, \citenamefont {Callaghan}, \citenamefont {Caven},\ and\
  \citenamefont {Fielding}}]{Fielding_prl_oscillatory_shear_2024}%
  \BibitemOpen
  \bibfield  {author} {\bibinfo {author} {\bibfnamefont {J.~O.}\ \bibnamefont
  {Cochran}}, \bibinfo {author} {\bibfnamefont {G.~L.}\ \bibnamefont
  {Callaghan}}, \bibinfo {author} {\bibfnamefont {M.~J.~G.}\ \bibnamefont
  {Caven}},\ and\ \bibinfo {author} {\bibfnamefont {S.~M.}\ \bibnamefont
  {Fielding}},\ }\bibfield  {title} {\bibinfo {title} {Slow fatigue and highly
  delayed yielding via shear banding in oscillatory shear},\ }\href
  {https://doi.org/10.1103/PhysRevLett.132.168202} {\bibfield  {journal}
  {\bibinfo  {journal} {Phys. Rev. Lett.}\ }\textbf {\bibinfo {volume} {132}},\
  \bibinfo {pages} {168202} (\bibinfo {year} {2024})}\BibitemShut {NoStop}%
\bibitem [{\citenamefont {Knowlton}\ \emph {et~al.}(2014)\citenamefont
  {Knowlton}, \citenamefont {Pine},\ and\ \citenamefont
  {Cipelletti}}]{Knowlton2014-cl}%
  \BibitemOpen
  \bibfield  {author} {\bibinfo {author} {\bibfnamefont {E.~D.}\ \bibnamefont
  {Knowlton}}, \bibinfo {author} {\bibfnamefont {D.~J.}\ \bibnamefont {Pine}},\
  and\ \bibinfo {author} {\bibfnamefont {L.}~\bibnamefont {Cipelletti}},\
  }\bibfield  {title} {\bibinfo {title} {A microscopic view of the yielding
  transition in concentrated emulsions},\ }\href
  {https://doi.org/https://doi.org/10.1039/C4SM00531G} {\bibfield  {journal}
  {\bibinfo  {journal} {Soft Matter}\ }\textbf {\bibinfo {volume} {10}},\
  \bibinfo {pages} {6931} (\bibinfo {year} {2014})}\BibitemShut {NoStop}%
\bibitem [{\citenamefont {Hima~Nagamanasa}\ \emph {et~al.}(2014)\citenamefont
  {Hima~Nagamanasa}, \citenamefont {Gokhale}, \citenamefont {Sood},\ and\
  \citenamefont {Ganapathy}}]{Hima-Nagamanasa2014-nt}%
  \BibitemOpen
  \bibfield  {author} {\bibinfo {author} {\bibfnamefont {K.}~\bibnamefont
  {Hima~Nagamanasa}}, \bibinfo {author} {\bibfnamefont {S.}~\bibnamefont
  {Gokhale}}, \bibinfo {author} {\bibfnamefont {A.~K.}\ \bibnamefont {Sood}},\
  and\ \bibinfo {author} {\bibfnamefont {R.}~\bibnamefont {Ganapathy}},\
  }\bibfield  {title} {\bibinfo {title} {Experimental signatures of a
  nonequilibrium phase transition governing the yielding of a soft glass},\
  }\href {https://doi.org/https://doi.org/10.1103/PhysRevE.89.062308}
  {\bibfield  {journal} {\bibinfo  {journal} {Phys. Rev. E}\ }\textbf {\bibinfo
  {volume} {89}},\ \bibinfo {pages} {062308} (\bibinfo {year}
  {2014})}\BibitemShut {NoStop}%
\bibitem [{\citenamefont {Ghosh}\ \emph {et~al.}(2022)\citenamefont {Ghosh},
  \citenamefont {Radhakrishnan}, \citenamefont {Chaikin}, \citenamefont
  {Levine},\ and\ \citenamefont {Ghosh}}]{Ghosh2022-fc}%
  \BibitemOpen
  \bibfield  {author} {\bibinfo {author} {\bibfnamefont {A.}~\bibnamefont
  {Ghosh}}, \bibinfo {author} {\bibfnamefont {J.}~\bibnamefont
  {Radhakrishnan}}, \bibinfo {author} {\bibfnamefont {P.~M.}\ \bibnamefont
  {Chaikin}}, \bibinfo {author} {\bibfnamefont {D.}~\bibnamefont {Levine}},\
  and\ \bibinfo {author} {\bibfnamefont {S.}~\bibnamefont {Ghosh}},\ }\bibfield
   {title} {\bibinfo {title} {Coupled dynamical phase transitions in driven
  disk packings},\ }\href
  {https://doi.org/https://doi.org/10.1103/PhysRevLett.129.188002} {\bibfield
  {journal} {\bibinfo  {journal} {Phys. Rev. Lett.}\ }\textbf {\bibinfo
  {volume} {129}},\ \bibinfo {pages} {188002} (\bibinfo {year}
  {2022})}\BibitemShut {NoStop}%
\bibitem [{\citenamefont {van~der Vaart}\ \emph {et~al.}(2013)\citenamefont
  {van~der Vaart}, \citenamefont {Rahmani}, \citenamefont {Zargar},
  \citenamefont {Hu}, \citenamefont {Bonn},\ and\ \citenamefont
  {Schall}}]{van-der-Vaart2013-oa}%
  \BibitemOpen
  \bibfield  {author} {\bibinfo {author} {\bibfnamefont {K.}~\bibnamefont
  {van~der Vaart}}, \bibinfo {author} {\bibfnamefont {Y.}~\bibnamefont
  {Rahmani}}, \bibinfo {author} {\bibfnamefont {R.}~\bibnamefont {Zargar}},
  \bibinfo {author} {\bibfnamefont {Z.}~\bibnamefont {Hu}}, \bibinfo {author}
  {\bibfnamefont {D.}~\bibnamefont {Bonn}},\ and\ \bibinfo {author}
  {\bibfnamefont {P.}~\bibnamefont {Schall}},\ }\bibfield  {title} {\bibinfo
  {title} {Rheology of concentrated soft and hard-sphere suspensions},\ }\href
  {https://doi.org/https://doi.org/10.1122/1.4808054} {\bibfield  {journal}
  {\bibinfo  {journal} {J. Rheol.}\ }\textbf {\bibinfo {volume} {57}},\
  \bibinfo {pages} {1195} (\bibinfo {year} {2013})}\BibitemShut {NoStop}%
\bibitem [{\citenamefont {Denisov}\ \emph {et~al.}(2015)\citenamefont
  {Denisov}, \citenamefont {Dang}, \citenamefont {Struth}, \citenamefont
  {Zaccone}, \citenamefont {Wegdam},\ and\ \citenamefont
  {Schall}}]{Denisov2015-ve}%
  \BibitemOpen
  \bibfield  {author} {\bibinfo {author} {\bibfnamefont {D.~V.}\ \bibnamefont
  {Denisov}}, \bibinfo {author} {\bibfnamefont {M.~T.}\ \bibnamefont {Dang}},
  \bibinfo {author} {\bibfnamefont {B.}~\bibnamefont {Struth}}, \bibinfo
  {author} {\bibfnamefont {A.}~\bibnamefont {Zaccone}}, \bibinfo {author}
  {\bibfnamefont {G.~H.}\ \bibnamefont {Wegdam}},\ and\ \bibinfo {author}
  {\bibfnamefont {P.}~\bibnamefont {Schall}},\ }\bibfield  {title} {\bibinfo
  {title} {Sharp symmetry-change marks the mechanical failure transition of
  glasses},\ }\href {https://doi.org/https://doi.org/10.1038/srep14359}
  {\bibfield  {journal} {\bibinfo  {journal} {Sci. Rep.}\ }\textbf {\bibinfo
  {volume} {5}},\ \bibinfo {pages} {14359} (\bibinfo {year}
  {2015})}\BibitemShut {NoStop}%
\bibitem [{\citenamefont {Ghosh}\ \emph {et~al.}(2017)\citenamefont {Ghosh},
  \citenamefont {Budrikis}, \citenamefont {Chikkadi}, \citenamefont {Sellerio},
  \citenamefont {Zapperi},\ and\ \citenamefont {Schall}}]{Ghosh2017-ed}%
  \BibitemOpen
  \bibfield  {author} {\bibinfo {author} {\bibfnamefont {A.}~\bibnamefont
  {Ghosh}}, \bibinfo {author} {\bibfnamefont {Z.}~\bibnamefont {Budrikis}},
  \bibinfo {author} {\bibfnamefont {V.}~\bibnamefont {Chikkadi}}, \bibinfo
  {author} {\bibfnamefont {A.~L.}\ \bibnamefont {Sellerio}}, \bibinfo {author}
  {\bibfnamefont {S.}~\bibnamefont {Zapperi}},\ and\ \bibinfo {author}
  {\bibfnamefont {P.}~\bibnamefont {Schall}},\ }\bibfield  {title} {\bibinfo
  {title} {Direct observation of percolation in the yielding transition of
  colloidal glasses},\ }\href
  {https://doi.org/https://doi.org/10.1103/PhysRevLett.118.148001} {\bibfield
  {journal} {\bibinfo  {journal} {Phys. Rev. Lett.}\ }\textbf {\bibinfo
  {volume} {118}},\ \bibinfo {pages} {148001} (\bibinfo {year}
  {2017})}\BibitemShut {NoStop}%
\bibitem [{\citenamefont {Aime}\ \emph {et~al.}(2022)\citenamefont {Aime},
  \citenamefont {Truzzolillo}, \citenamefont {Pine}, \citenamefont {Ramos},\
  and\ \citenamefont {Cipelletti}}]{Aime2022-ac}%
  \BibitemOpen
  \bibfield  {author} {\bibinfo {author} {\bibfnamefont {S.}~\bibnamefont
  {Aime}}, \bibinfo {author} {\bibfnamefont {D.}~\bibnamefont {Truzzolillo}},
  \bibinfo {author} {\bibfnamefont {D.}~\bibnamefont {Pine}}, \bibinfo {author}
  {\bibfnamefont {L.}~\bibnamefont {Ramos}},\ and\ \bibinfo {author}
  {\bibfnamefont {L.}~\bibnamefont {Cipelletti}},\ }\bibfield  {title}
  {\bibinfo {title} {A unified state diagram for the yielding transition of
  soft colloids},\ }\href
  {https://doi.org/https://doi.org/10.1038/s41567-023-02153-w} {\bibfield
  {journal} {\bibinfo  {journal} {Nat. Phys.}\ }\textbf {\bibinfo {volume}
  {19}},\ \bibinfo {pages} {1673} (\bibinfo {year} {2022})}\BibitemShut
  {NoStop}%
\bibitem [{\citenamefont {Keim}\ \emph {et~al.}(2013)\citenamefont {Keim},
  \citenamefont {Paulsen},\ and\ \citenamefont {Nagel}}]{Keim2013-gx}%
  \BibitemOpen
  \bibfield  {author} {\bibinfo {author} {\bibfnamefont {N.~C.}\ \bibnamefont
  {Keim}}, \bibinfo {author} {\bibfnamefont {J.~D.}\ \bibnamefont {Paulsen}},\
  and\ \bibinfo {author} {\bibfnamefont {S.~R.}\ \bibnamefont {Nagel}},\
  }\bibfield  {title} {\bibinfo {title} {Multiple transient memories in sheared
  suspensions: robustness, structure, and routes to plasticity},\ }\href
  {https://doi.org/https://doi.org/10.1103/PhysRevE.88.032306} {\bibfield
  {journal} {\bibinfo  {journal} {Phys. Rev. E}\ }\textbf {\bibinfo {volume}
  {88}},\ \bibinfo {pages} {032306} (\bibinfo {year} {2013})}\BibitemShut
  {NoStop}%
\bibitem [{\citenamefont {Arceri}\ \emph {et~al.}(2021)\citenamefont {Arceri},
  \citenamefont {Corwin},\ and\ \citenamefont {Hagh}}]{Arceri2021-dt}%
  \BibitemOpen
  \bibfield  {author} {\bibinfo {author} {\bibfnamefont {F.}~\bibnamefont
  {Arceri}}, \bibinfo {author} {\bibfnamefont {E.~I.}\ \bibnamefont {Corwin}},\
  and\ \bibinfo {author} {\bibfnamefont {V.~F.}\ \bibnamefont {Hagh}},\
  }\bibfield  {title} {\bibinfo {title} {Marginal stability in memory training
  of jammed solids},\ }\href
  {https://doi.org/https://doi.org/10.1103/PhysRevE.104.044907} {\bibfield
  {journal} {\bibinfo  {journal} {Phys. Rev. E.}\ }\textbf {\bibinfo {volume}
  {104}},\ \bibinfo {pages} {044907} (\bibinfo {year} {2021})}\BibitemShut
  {NoStop}%
\bibitem [{\citenamefont {Mukherji}\ \emph {et~al.}(2019)\citenamefont
  {Mukherji}, \citenamefont {Kandula}, \citenamefont {Sood},\ and\
  \citenamefont {Ganapathy}}]{Mukherji2019-az}%
  \BibitemOpen
  \bibfield  {author} {\bibinfo {author} {\bibfnamefont {S.}~\bibnamefont
  {Mukherji}}, \bibinfo {author} {\bibfnamefont {N.}~\bibnamefont {Kandula}},
  \bibinfo {author} {\bibfnamefont {A.~K.}\ \bibnamefont {Sood}},\ and\
  \bibinfo {author} {\bibfnamefont {R.}~\bibnamefont {Ganapathy}},\ }\bibfield
  {title} {\bibinfo {title} {Strength of mechanical memories is maximal at the
  yield point of a soft glass},\ }\href
  {https://doi.org/https://doi.org/10.1103/PhysRevLett.122.158001} {\bibfield
  {journal} {\bibinfo  {journal} {Phys. Rev. Lett.}\ }\textbf {\bibinfo
  {volume} {122}},\ \bibinfo {pages} {158001} (\bibinfo {year}
  {2019})}\BibitemShut {NoStop}%
\bibitem [{\citenamefont {Galloway}\ \emph {et~al.}(2022)\citenamefont
  {Galloway}, \citenamefont {Teich}, \citenamefont {Ma}, \citenamefont
  {Kammer}, \citenamefont {Graham}, \citenamefont {Keim}, \citenamefont
  {Reina}, \citenamefont {Jerolmack}, \citenamefont {Yodh},\ and\ \citenamefont
  {Arratia}}]{Galloway2022-vb}%
  \BibitemOpen
  \bibfield  {author} {\bibinfo {author} {\bibfnamefont {K.~L.}\ \bibnamefont
  {Galloway}}, \bibinfo {author} {\bibfnamefont {E.~G.}\ \bibnamefont {Teich}},
  \bibinfo {author} {\bibfnamefont {X.~G.}\ \bibnamefont {Ma}}, \bibinfo
  {author} {\bibfnamefont {C.}~\bibnamefont {Kammer}}, \bibinfo {author}
  {\bibfnamefont {I.~R.}\ \bibnamefont {Graham}}, \bibinfo {author}
  {\bibfnamefont {N.~C.}\ \bibnamefont {Keim}}, \bibinfo {author}
  {\bibfnamefont {C.}~\bibnamefont {Reina}}, \bibinfo {author} {\bibfnamefont
  {D.~J.}\ \bibnamefont {Jerolmack}}, \bibinfo {author} {\bibfnamefont {A.~G.}\
  \bibnamefont {Yodh}},\ and\ \bibinfo {author} {\bibfnamefont {P.~E.}\
  \bibnamefont {Arratia}},\ }\bibfield  {title} {\bibinfo {title}
  {Relationships between structure, memory and flow in sheared disordered
  materials},\ }\href
  {https://doi.org/https://doi.org/10.1038/s41567-022-01536-9} {\bibfield
  {journal} {\bibinfo  {journal} {Nat. Phys.}\ }\textbf {\bibinfo {volume}
  {18}},\ \bibinfo {pages} {565} (\bibinfo {year} {2022})}\BibitemShut
  {NoStop}%
\bibitem [{\citenamefont {Shohat}\ and\ \citenamefont
  {Lahini}(2023)}]{Shohat2023-em}%
  \BibitemOpen
  \bibfield  {author} {\bibinfo {author} {\bibfnamefont {D.}~\bibnamefont
  {Shohat}}\ and\ \bibinfo {author} {\bibfnamefont {Y.}~\bibnamefont
  {Lahini}},\ }\bibfield  {title} {\bibinfo {title} {Dissipation indicates
  memory formation in driven disordered systems},\ }\href
  {https://doi.org/https://doi.org/10.1103/PhysRevLett.130.048202} {\bibfield
  {journal} {\bibinfo  {journal} {Phys. Rev. Lett.}\ }\textbf {\bibinfo
  {volume} {130}},\ \bibinfo {pages} {048202} (\bibinfo {year}
  {2023})}\BibitemShut {NoStop}%
\bibitem [{\citenamefont {Chen}\ \emph {et~al.}(2025)\citenamefont {Chen},
  \citenamefont {Rogers}, \citenamefont {Narayanan}, \citenamefont {Harden},\
  and\ \citenamefont {Leheny}}]{Chen2025-ac}%
  \BibitemOpen
  \bibfield  {author} {\bibinfo {author} {\bibfnamefont {Y.}~\bibnamefont
  {Chen}}, \bibinfo {author} {\bibfnamefont {S.~A.}\ \bibnamefont {Rogers}},
  \bibinfo {author} {\bibfnamefont {S.}~\bibnamefont {Narayanan}}, \bibinfo
  {author} {\bibfnamefont {J.~L.}\ \bibnamefont {Harden}},\ and\ \bibinfo
  {author} {\bibfnamefont {R.~L.}\ \bibnamefont {Leheny}},\ }\bibfield  {title}
  {\bibinfo {title} {Microstructural and rheological training and memory of
  nanocolloidal soft glasses under cyclic shear},\ }\href
  {https://doi.org/https://doi.org/10.1103/PhysRevMaterials.9.025601}
  {\bibfield  {journal} {\bibinfo  {journal} {Phys. Rev. Mater.}\ }\textbf
  {\bibinfo {volume} {9}},\ \bibinfo {pages} {025601} (\bibinfo {year}
  {2025})}\BibitemShut {NoStop}%
\bibitem [{\citenamefont {Fiocco}\ \emph {et~al.}(2014)\citenamefont {Fiocco},
  \citenamefont {Foffi},\ and\ \citenamefont {Sastry}}]{Fiocco2014-jq}%
  \BibitemOpen
  \bibfield  {author} {\bibinfo {author} {\bibfnamefont {D.}~\bibnamefont
  {Fiocco}}, \bibinfo {author} {\bibfnamefont {G.}~\bibnamefont {Foffi}},\ and\
  \bibinfo {author} {\bibfnamefont {S.}~\bibnamefont {Sastry}},\ }\bibfield
  {title} {\bibinfo {title} {Encoding of memory in sheared amorphous solids},\
  }\href {https://doi.org/https://doi.org/10.1103/PhysRevLett.112.025702}
  {\bibfield  {journal} {\bibinfo  {journal} {Phys. Rev. Lett.}\ }\textbf
  {\bibinfo {volume} {112}},\ \bibinfo {pages} {025702} (\bibinfo {year}
  {2014})}\BibitemShut {NoStop}%
\bibitem [{\citenamefont {Adhikari}\ and\ \citenamefont
  {Sastry}(2018)}]{Adhikari2018-dl}%
  \BibitemOpen
  \bibfield  {author} {\bibinfo {author} {\bibfnamefont {M.}~\bibnamefont
  {Adhikari}}\ and\ \bibinfo {author} {\bibfnamefont {S.}~\bibnamefont
  {Sastry}},\ }\bibfield  {title} {\bibinfo {title} {Memory formation in
  cyclically deformed amorphous solids and sphere assemblies},\ }\href
  {https://doi.org/https://doi.org/10.1140/epje/i2018-11717-5} {\bibfield
  {journal} {\bibinfo  {journal} {Eur. Phys. J. E}\ }\textbf {\bibinfo {volume}
  {41}},\ \bibinfo {pages} {105} (\bibinfo {year} {2018})}\BibitemShut
  {NoStop}%
\bibitem [{\citenamefont {Mungan}\ \emph {et~al.}(2025)\citenamefont {Mungan},
  \citenamefont {Kumar}, \citenamefont {Patinet},\ and\ \citenamefont
  {Vandembroucq}}]{Mungan2025-wi}%
  \BibitemOpen
  \bibfield  {author} {\bibinfo {author} {\bibfnamefont {M.}~\bibnamefont
  {Mungan}}, \bibinfo {author} {\bibfnamefont {D.}~\bibnamefont {Kumar}},
  \bibinfo {author} {\bibfnamefont {S.}~\bibnamefont {Patinet}},\ and\ \bibinfo
  {author} {\bibfnamefont {D.}~\bibnamefont {Vandembroucq}},\ }\bibfield
  {title} {\bibinfo {title} {Self-organization and memory in a disordered solid
  subject to random driving},\ }\href
  {https://doi.org/https://doi.org/10.1103/PhysRevLett.134.178203} {\bibfield
  {journal} {\bibinfo  {journal} {Phys. Rev. Lett.}\ }\textbf {\bibinfo
  {volume} {134}},\ \bibinfo {pages} {178203} (\bibinfo {year}
  {2025})}\BibitemShut {NoStop}%
\bibitem [{\citenamefont {Rainone}\ \emph {et~al.}(2021)\citenamefont
  {Rainone}, \citenamefont {Urbani}, \citenamefont {Zamponi}, \citenamefont
  {Lerner},\ and\ \citenamefont {Bouchbinder}}]{Rainone2021-am}%
  \BibitemOpen
  \bibfield  {author} {\bibinfo {author} {\bibfnamefont {C.}~\bibnamefont
  {Rainone}}, \bibinfo {author} {\bibfnamefont {P.}~\bibnamefont {Urbani}},
  \bibinfo {author} {\bibfnamefont {F.}~\bibnamefont {Zamponi}}, \bibinfo
  {author} {\bibfnamefont {E.}~\bibnamefont {Lerner}},\ and\ \bibinfo {author}
  {\bibfnamefont {E.}~\bibnamefont {Bouchbinder}},\ }\bibfield  {title}
  {\bibinfo {title} {Mean-field model of interacting quasilocalized excitations
  in glasses},\ }\href
  {https://doi.org/https://doi.org/10.21468/SciPostPhysCore.4.2.008} {\bibfield
   {journal} {\bibinfo  {journal} {SciPost Phys. Core}\ }\textbf {\bibinfo
  {volume} {4}},\ \bibinfo {pages} {008} (\bibinfo {year} {2021})}\BibitemShut
  {NoStop}%
\bibitem [{\citenamefont {Bouchbinder}\ \emph {et~al.}(2021)\citenamefont
  {Bouchbinder}, \citenamefont {Lerner}, \citenamefont {Rainone}, \citenamefont
  {Urbani},\ and\ \citenamefont {Zamponi}}]{Bouchbinder2021-dh}%
  \BibitemOpen
  \bibfield  {author} {\bibinfo {author} {\bibfnamefont {E.}~\bibnamefont
  {Bouchbinder}}, \bibinfo {author} {\bibfnamefont {E.}~\bibnamefont {Lerner}},
  \bibinfo {author} {\bibfnamefont {C.}~\bibnamefont {Rainone}}, \bibinfo
  {author} {\bibfnamefont {P.}~\bibnamefont {Urbani}},\ and\ \bibinfo {author}
  {\bibfnamefont {F.}~\bibnamefont {Zamponi}},\ }\bibfield  {title} {\bibinfo
  {title} {Low-frequency vibrational spectrum of mean-field disordered
  systems},\ }\href
  {https://doi.org/https://doi.org/10.1103/PhysRevB.103.174202} {\bibfield
  {journal} {\bibinfo  {journal} {Phys. Rev. B}\ }\textbf {\bibinfo {volume}
  {103}},\ \bibinfo {pages} {174202} (\bibinfo {year} {2021})}\BibitemShut
  {NoStop}%
\bibitem [{\citenamefont {K\"uhn}\ and\ \citenamefont
  {Horstmann}(1997)}]{Kuhn_Horstmann_prl_1997}%
  \BibitemOpen
  \bibfield  {author} {\bibinfo {author} {\bibfnamefont {R.}~\bibnamefont
  {K\"uhn}}\ and\ \bibinfo {author} {\bibfnamefont {U.}~\bibnamefont
  {Horstmann}},\ }\bibfield  {title} {\bibinfo {title} {Random matrix approach
  to glassy physics: Low temperatures and beyond},\ }\href
  {https://doi.org/https://doi.org/10.1103/PhysRevLett.78.4067} {\bibfield
  {journal} {\bibinfo  {journal} {Phys. Rev. Lett.}\ }\textbf {\bibinfo
  {volume} {78}},\ \bibinfo {pages} {4067} (\bibinfo {year}
  {1997})}\BibitemShut {NoStop}%
\bibitem [{\citenamefont {Folena}\ and\ \citenamefont
  {Urbani}(2022)}]{Folena2022-tn}%
  \BibitemOpen
  \bibfield  {author} {\bibinfo {author} {\bibfnamefont {G.}~\bibnamefont
  {Folena}}\ and\ \bibinfo {author} {\bibfnamefont {P.}~\bibnamefont
  {Urbani}},\ }\bibfield  {title} {\bibinfo {title} {Marginal stability of soft
  anharmonic mean field spin glasses},\ }\href
  {https://doi.org/https://doi.org/10.1088/1742-5468/ac6253} {\bibfield
  {journal} {\bibinfo  {journal} {J. Stat. Mech.}\ }\textbf {\bibinfo {volume}
  {2022}},\ \bibinfo {pages} {053301} (\bibinfo {year} {2022})}\BibitemShut
  {NoStop}%
\bibitem [{\citenamefont {Maimbourg}(2024)}]{Maimbourg2024-lb}%
  \BibitemOpen
  \bibfield  {author} {\bibinfo {author} {\bibfnamefont {T.}~\bibnamefont
  {Maimbourg}},\ }\bibfield  {title} {\bibinfo {title} {Two-level systems and
  harmonic excitations in a mean-field anharmonic quantum glass},\ }\href
  {https://doi.org/https://doi.org/10.1103/PhysRevB.110.064203} {\bibfield
  {journal} {\bibinfo  {journal} {Phys. Rev. B.}\ }\textbf {\bibinfo {volume}
  {110}},\ \bibinfo {pages} {064203} (\bibinfo {year} {2024})}\BibitemShut
  {NoStop}%
\bibitem [{\citenamefont {Bouchbinder}\ and\ \citenamefont
  {Langer}(2009)}]{bouchbinder2009nonequilibrium}%
  \BibitemOpen
  \bibfield  {author} {\bibinfo {author} {\bibfnamefont {E.}~\bibnamefont
  {Bouchbinder}}\ and\ \bibinfo {author} {\bibfnamefont {J.~S.}\ \bibnamefont
  {Langer}},\ }\bibfield  {title} {\bibinfo {title} {Nonequilibrium
  thermodynamics of driven amorphous materials. III. shear-transformation-zone
  plasticity},\ }\href {https://doi.org/10.1103/PhysRevE.80.031133} {\bibfield
  {journal} {\bibinfo  {journal} {Phys. Rev. E}\ }\textbf {\bibinfo {volume}
  {80}},\ \bibinfo {pages} {031133} (\bibinfo {year} {2009})}\BibitemShut
  {NoStop}%
\bibitem [{\citenamefont {Falk}\ and\ \citenamefont
  {Langer}(2011)}]{falk2011deformation}%
  \BibitemOpen
  \bibfield  {author} {\bibinfo {author} {\bibfnamefont {M.~L.}\ \bibnamefont
  {Falk}}\ and\ \bibinfo {author} {\bibfnamefont {J.~S.}\ \bibnamefont
  {Langer}},\ }\bibfield  {title} {\bibinfo {title} {Deformation and failure of
  amorphous, solidlike materials},\ }\href
  {https://doi.org/https://doi.org/10.1146/annurev-conmatphys-062910-140452}
  {\bibfield  {journal} {\bibinfo  {journal} {Annu. Rev. Condens. Matter
  Phys.}\ }\textbf {\bibinfo {volume} {2}},\ \bibinfo {pages} {353} (\bibinfo
  {year} {2011})}\BibitemShut {NoStop}%
\bibitem [{\citenamefont {Jaiswal}\ \emph {et~al.}(2016)\citenamefont
  {Jaiswal}, \citenamefont {Procaccia}, \citenamefont {Rainone},\ and\
  \citenamefont {Singh}}]{Jaiswal2016-mz}%
  \BibitemOpen
  \bibfield  {author} {\bibinfo {author} {\bibfnamefont {P.~K.}\ \bibnamefont
  {Jaiswal}}, \bibinfo {author} {\bibfnamefont {I.}~\bibnamefont {Procaccia}},
  \bibinfo {author} {\bibfnamefont {C.}~\bibnamefont {Rainone}},\ and\ \bibinfo
  {author} {\bibfnamefont {M.}~\bibnamefont {Singh}},\ }\bibfield  {title}
  {\bibinfo {title} {Mechanical yield in amorphous solids: A first-order phase
  transition},\ }\href
  {https://doi.org/https://doi.org/10.1103/PhysRevLett.116.085501} {\bibfield
  {journal} {\bibinfo  {journal} {Phys. Rev. Lett.}\ }\textbf {\bibinfo
  {volume} {116}},\ \bibinfo {pages} {085501} (\bibinfo {year}
  {2016})}\BibitemShut {NoStop}%
\bibitem [{\citenamefont {Parisi}\ \emph {et~al.}(2017)\citenamefont {Parisi},
  \citenamefont {Procaccia}, \citenamefont {Rainone},\ and\ \citenamefont
  {Singh}}]{Parisi2017-rv}%
  \BibitemOpen
  \bibfield  {author} {\bibinfo {author} {\bibfnamefont {G.}~\bibnamefont
  {Parisi}}, \bibinfo {author} {\bibfnamefont {I.}~\bibnamefont {Procaccia}},
  \bibinfo {author} {\bibfnamefont {C.}~\bibnamefont {Rainone}},\ and\ \bibinfo
  {author} {\bibfnamefont {M.}~\bibnamefont {Singh}},\ }\bibfield  {title}
  {\bibinfo {title} {Shear bands as manifestation of a criticality in yielding
  amorphous solids},\ }\href
  {https://doi.org/https://doi.org/10.1073/pnas.1700075114} {\bibfield
  {journal} {\bibinfo  {journal} {Proc. Natl. Acad. Sci. U. S. A.}\ }\textbf
  {\bibinfo {volume} {114}},\ \bibinfo {pages} {5577} (\bibinfo {year}
  {2017})}\BibitemShut {NoStop}%
\bibitem [{\citenamefont {Shi}\ and\ \citenamefont
  {Falk}(2005)}]{shi2005strain}%
  \BibitemOpen
  \bibfield  {author} {\bibinfo {author} {\bibfnamefont {Y.}~\bibnamefont
  {Shi}}\ and\ \bibinfo {author} {\bibfnamefont {M.~L.}\ \bibnamefont {Falk}},\
  }\bibfield  {title} {\bibinfo {title} {Strain localization and percolation of
  stable structure in amorphous solids},\ }\href
  {https://doi.org/https://doi.org/10.1103/PhysRevLett.95.095502} {\bibfield
  {journal} {\bibinfo  {journal} {Phys. Rev. Lett.}\ }\textbf {\bibinfo
  {volume} {95}},\ \bibinfo {pages} {095502} (\bibinfo {year}
  {2005})}\BibitemShut {NoStop}%
\bibitem [{\citenamefont {Shi}\ \emph {et~al.}(2007)\citenamefont {Shi},
  \citenamefont {Katz}, \citenamefont {Li},\ and\ \citenamefont
  {Falk}}]{shi2007evaluation}%
  \BibitemOpen
  \bibfield  {author} {\bibinfo {author} {\bibfnamefont {Y.}~\bibnamefont
  {Shi}}, \bibinfo {author} {\bibfnamefont {M.~B.}\ \bibnamefont {Katz}},
  \bibinfo {author} {\bibfnamefont {H.}~\bibnamefont {Li}},\ and\ \bibinfo
  {author} {\bibfnamefont {M.~L.}\ \bibnamefont {Falk}},\ }\bibfield  {title}
  {\bibinfo {title} {Evaluation of the disorder temperature and free-volume
  formalisms via simulations of shear banding in amorphous solids},\ }\href
  {https://doi.org/10.1103/PhysRevLett.98.185505} {\bibfield  {journal}
  {\bibinfo  {journal} {Phys. Rev. Lett.}\ }\textbf {\bibinfo {volume} {98}},\
  \bibinfo {pages} {185505} (\bibinfo {year} {2007})}\BibitemShut {NoStop}%
\bibitem [{\citenamefont {Vasoya}\ \emph {et~al.}(2016)\citenamefont {Vasoya},
  \citenamefont {Rycroft},\ and\ \citenamefont
  {Bouchbinder}}]{vasoya2016notch}%
  \BibitemOpen
  \bibfield  {author} {\bibinfo {author} {\bibfnamefont {M.}~\bibnamefont
  {Vasoya}}, \bibinfo {author} {\bibfnamefont {C.~H.}\ \bibnamefont
  {Rycroft}},\ and\ \bibinfo {author} {\bibfnamefont {E.}~\bibnamefont
  {Bouchbinder}},\ }\bibfield  {title} {\bibinfo {title} {Notch fracture
  toughness of glasses: Dependence on rate, age, and geometry},\ }\href
  {https://doi.org/https://doi.org/10.1103/PhysRevApplied.6.024008} {\bibfield
  {journal} {\bibinfo  {journal} {Phys. Rev. Appl.}\ }\textbf {\bibinfo
  {volume} {6}},\ \bibinfo {pages} {024008} (\bibinfo {year}
  {2016})}\BibitemShut {NoStop}%
\bibitem [{\citenamefont {Ozawa}\ \emph {et~al.}(2018)\citenamefont {Ozawa},
  \citenamefont {Berthier}, \citenamefont {Biroli}, \citenamefont {Rosso},\
  and\ \citenamefont {Tarjus}}]{misaki_yielding_pnas_2018}%
  \BibitemOpen
  \bibfield  {author} {\bibinfo {author} {\bibfnamefont {M.}~\bibnamefont
  {Ozawa}}, \bibinfo {author} {\bibfnamefont {L.}~\bibnamefont {Berthier}},
  \bibinfo {author} {\bibfnamefont {G.}~\bibnamefont {Biroli}}, \bibinfo
  {author} {\bibfnamefont {A.}~\bibnamefont {Rosso}},\ and\ \bibinfo {author}
  {\bibfnamefont {G.}~\bibnamefont {Tarjus}},\ }\bibfield  {title} {\bibinfo
  {title} {Random critical point separates brittle and ductile yielding
  transitions in amorphous materials},\ }\href
  {https://doi.org/10.1073/pnas.1806156115} {\bibfield  {journal} {\bibinfo
  {journal} {Proc. Natl. Acad. Sci. U.S.A.}\ }\textbf {\bibinfo {volume}
  {115}},\ \bibinfo {pages} {6656} (\bibinfo {year} {2018})}\BibitemShut
  {NoStop}%
\bibitem [{\citenamefont {Ketkaew}\ \emph {et~al.}(2018)\citenamefont
  {Ketkaew}, \citenamefont {Chen}, \citenamefont {Wang}, \citenamefont {Datye},
  \citenamefont {Fan}, \citenamefont {Pereira}, \citenamefont {Schwarz},
  \citenamefont {Liu}, \citenamefont {Yamada}, \citenamefont {Dmowski},
  \citenamefont {Shattuck}, \citenamefont {O'Hern}, \citenamefont {Egami},
  \citenamefont {Bouchbinder},\ and\ \citenamefont
  {Schroers}}]{Eran_mechanical_glass_transition}%
  \BibitemOpen
  \bibfield  {author} {\bibinfo {author} {\bibfnamefont {J.}~\bibnamefont
  {Ketkaew}}, \bibinfo {author} {\bibfnamefont {W.}~\bibnamefont {Chen}},
  \bibinfo {author} {\bibfnamefont {H.}~\bibnamefont {Wang}}, \bibinfo {author}
  {\bibfnamefont {A.}~\bibnamefont {Datye}}, \bibinfo {author} {\bibfnamefont
  {M.}~\bibnamefont {Fan}}, \bibinfo {author} {\bibfnamefont {G.}~\bibnamefont
  {Pereira}}, \bibinfo {author} {\bibfnamefont {U.~D.}\ \bibnamefont
  {Schwarz}}, \bibinfo {author} {\bibfnamefont {Z.}~\bibnamefont {Liu}},
  \bibinfo {author} {\bibfnamefont {R.}~\bibnamefont {Yamada}}, \bibinfo
  {author} {\bibfnamefont {W.}~\bibnamefont {Dmowski}}, \bibinfo {author}
  {\bibfnamefont {M.~D.}\ \bibnamefont {Shattuck}}, \bibinfo {author}
  {\bibfnamefont {C.~S.}\ \bibnamefont {O'Hern}}, \bibinfo {author}
  {\bibfnamefont {T.}~\bibnamefont {Egami}}, \bibinfo {author} {\bibfnamefont
  {E.}~\bibnamefont {Bouchbinder}},\ and\ \bibinfo {author} {\bibfnamefont
  {J.}~\bibnamefont {Schroers}},\ }\bibfield  {title} {\bibinfo {title}
  {Mechanical glass transition revealed by the fracture toughness of metallic
  glasses},\ }\href {https://doi.org/10.1038/s41467-018-05682-8} {\bibfield
  {journal} {\bibinfo  {journal} {Nat. Commun.}\ }\textbf {\bibinfo {volume}
  {9}},\ \bibinfo {pages} {3271} (\bibinfo {year} {2018})}\BibitemShut
  {NoStop}%
\bibitem [{\citenamefont {Richard}\ \emph {et~al.}(2021)\citenamefont
  {Richard}, \citenamefont {Lerner},\ and\ \citenamefont
  {Bouchbinder}}]{david_fracture_mrs_2021}%
  \BibitemOpen
  \bibfield  {author} {\bibinfo {author} {\bibfnamefont {D.}~\bibnamefont
  {Richard}}, \bibinfo {author} {\bibfnamefont {E.}~\bibnamefont {Lerner}},\
  and\ \bibinfo {author} {\bibfnamefont {E.}~\bibnamefont {Bouchbinder}},\
  }\bibfield  {title} {\bibinfo {title} {Brittle-to-ductile transitions in
  glasses: Roles of soft defects and loading geometry},\ }\href
  {https://doi.org/10.1557/s43577-021-00171-8} {\bibfield  {journal} {\bibinfo
  {journal} {MRS Bull.}\ }\textbf {\bibinfo {volume} {46}},\ \bibinfo {pages}
  {902} (\bibinfo {year} {2021})}\BibitemShut {NoStop}%
\bibitem [{\citenamefont {Castellanos}\ \emph {et~al.}(2022)\citenamefont
  {Castellanos}, \citenamefont {Roux},\ and\ \citenamefont
  {Patinet}}]{castellanos2022history}%
  \BibitemOpen
  \bibfield  {author} {\bibinfo {author} {\bibfnamefont {D.~F.}\ \bibnamefont
  {Castellanos}}, \bibinfo {author} {\bibfnamefont {S.}~\bibnamefont {Roux}},\
  and\ \bibinfo {author} {\bibfnamefont {S.}~\bibnamefont {Patinet}},\
  }\bibfield  {title} {\bibinfo {title} {History dependent plasticity of glass:
  A mapping between atomistic and elasto-plastic models},\ }\href
  {https://doi.org/https://doi.org/10.1016/j.actamat.2022.118405} {\bibfield
  {journal} {\bibinfo  {journal} {Acta Mater.}\ }\textbf {\bibinfo {volume}
  {241}},\ \bibinfo {pages} {118405} (\bibinfo {year} {2022})}\BibitemShut
  {NoStop}%
\bibitem [{\citenamefont {Fiocco}\ \emph {et~al.}(2015)\citenamefont {Fiocco}, \citenamefont {Foffi},\ and\ \citenamefont {Sastry}}]{Fiocco2015-qo}%
  \BibitemOpen
  \bibfield  {author} {\bibinfo {author} {\bibfnamefont {D.}~\bibnamefont {Fiocco}}, \bibinfo {author} {\bibfnamefont {G.}~\bibnamefont {Foffi}},\ and\ \bibinfo {author} {\bibfnamefont {S.}~\bibnamefont {Sastry}},\ }\bibfield  {title} {\bibinfo {title} {Memory effects in schematic models of glasses subjected to oscillatory deformation},\ }\href {https://doi.org/10.1088/0953-8984/27/19/194130} {\bibfield  {journal} {\bibinfo  {journal} {J. Phys. Condens. Matter}\ }\textbf {\bibinfo {volume} {27}},\ \bibinfo {pages} {194130} (\bibinfo {year} {2015})}\BibitemShut {NoStop}%
\bibitem [{\citenamefont {Keim}\ and\ \citenamefont
  {Paulsen}(2021)}]{Keim2021-pq}%
  \BibitemOpen
  \bibfield  {author} {\bibinfo {author} {\bibfnamefont {N.~C.}\ \bibnamefont
  {Keim}}\ and\ \bibinfo {author} {\bibfnamefont {J.~D.}\ \bibnamefont
  {Paulsen}},\ }\bibfield  {title} {\bibinfo {title} {Multiperiodic orbits from
  interacting soft spots in cyclically sheared amorphous solids},\ }\href
  {https://doi.org/10.1126/sciadv.abg7685} {\bibfield  {journal} {\bibinfo
  {journal} {Sci. Adv.}\ }\textbf {\bibinfo {volume} {7}},\ \bibinfo {pages}
  {eabg7685} (\bibinfo {year} {2021})}\BibitemShut {NoStop}%
\bibitem [{\citenamefont {Lindeman}\ \emph {et~al.}(2025)\citenamefont {Lindeman}, \citenamefont {Jalowiec},\ and\ \citenamefont {Keim}}]{Lindeman2025-zd}%
  \BibitemOpen
  \bibfield  {author} {\bibinfo {author} {\bibfnamefont {C.~W.}\ \bibnamefont {Lindeman}}, \bibinfo {author} {\bibfnamefont {T.~R.}\ \bibnamefont {Jalowiec}},\ and\ \bibinfo {author} {\bibfnamefont {N.~C.}\ \bibnamefont {Keim}},\ }\bibfield  {title} {\bibinfo {title} {Generalizing multiple memories from a single drive: The hysteron latch},\ }\href {https://doi.org/10.1126/sciadv.adr5933} {\bibfield  {journal} {\bibinfo  {journal} {Sci. Adv.}\ }\textbf {\bibinfo {volume} {11}},\ \bibinfo {pages} {eadr5933} (\bibinfo {year} {2025})}\BibitemShut {NoStop}%
\bibitem [{\citenamefont {Mungan}\ and\ \citenamefont {Witten}(2019)}]{Mungan2019-sd}%
  \BibitemOpen
  \bibfield  {author} {\bibinfo {author} {\bibfnamefont {M.}~\bibnamefont {Mungan}}\ and\ \bibinfo {author} {\bibfnamefont {T.~A.}\ \bibnamefont {Witten}},\ }\bibfield  {title} {\bibinfo {title} {Cyclic annealing as an iterated random map},\ }\href {https://doi.org/10.1103/PhysRevE.99.052132} {\bibfield  {journal} {\bibinfo  {journal} {Phys. Rev. E.}\ }\textbf {\bibinfo {volume} {99}},\ \bibinfo {pages} {052132} (\bibinfo {year} {2019})}\BibitemShut {NoStop}%
\bibitem [{\citenamefont {Mungan}\ and\ \citenamefont {Sastry}(2021)}]{Mungan2021-lh}%
  \BibitemOpen
  \bibfield  {author} {\bibinfo {author} {\bibfnamefont {M.}~\bibnamefont {Mungan}}\ and\ \bibinfo {author} {\bibfnamefont {S.}~\bibnamefont {Sastry}},\ }\bibfield  {title} {\bibinfo {title} {Metastability as a mechanism for yielding in amorphous solids under cyclic shear},\ }\href {https://doi.org/10.1103/PhysRevLett.127.248002} {\bibfield  {journal} {\bibinfo  {journal} {Phys. Rev. Lett.}\ }\textbf {\bibinfo {volume} {127}},\ \bibinfo {pages} {248002} (\bibinfo {year} {2021})}\BibitemShut {NoStop}%
\bibitem [{\citenamefont {Sarkar}\ \emph {et~al.}(2025)\citenamefont {Sarkar}, \citenamefont {Nampoothiri}, \citenamefont {Mungan}, \citenamefont {Parley}, \citenamefont {Sollich},\ and\ \citenamefont {Sastry}}]{Sarkar2025-in}%
  \BibitemOpen
  \bibfield  {author} {\bibinfo {author} {\bibfnamefont {D.}~\bibnamefont {Sarkar}}, \bibinfo {author} {\bibfnamefont {J.~N.}\ \bibnamefont {Nampoothiri}}, \bibinfo {author} {\bibfnamefont {M.}~\bibnamefont {Mungan}}, \bibinfo {author} {\bibfnamefont {J.~T.}\ \bibnamefont {Parley}}, \bibinfo {author} {\bibfnamefont {P.}~\bibnamefont {Sollich}},\ and\ \bibinfo {author} {\bibfnamefont {S.}~\bibnamefont {Sastry}},\ }\bibfield  {title} {\bibinfo {title} {Coarse grained descriptions of the dynamics of yielding of amorphous solids under cyclic shear},\ }\href {http://arxiv.org/abs/2505.14912} {\bibfield  {journal} {\bibinfo  {journal} {arXiv [cond-mat.stat-mech]}\ } (\bibinfo {year} {2025})}\BibitemShut {NoStop}%
\bibitem [{\citenamefont {Schreck}\ \emph {et~al.}(2013)\citenamefont
  {Schreck}, \citenamefont {Hoy}, \citenamefont {Shattuck},\ and\ \citenamefont
  {O'Hern}}]{Schreck2013-vy}%
  \BibitemOpen
  \bibfield  {author} {\bibinfo {author} {\bibfnamefont {C.~F.}\ \bibnamefont
  {Schreck}}, \bibinfo {author} {\bibfnamefont {R.~S.}\ \bibnamefont {Hoy}},
  \bibinfo {author} {\bibfnamefont {M.~D.}\ \bibnamefont {Shattuck}},\ and\
  \bibinfo {author} {\bibfnamefont {C.~S.}\ \bibnamefont {O'Hern}},\ }\bibfield
   {title} {\bibinfo {title} {Particle-scale reversibility in athermal
  particulate media below jamming},\ }\href
  {https://doi.org/10.1103/PhysRevE.88.052205} {\bibfield  {journal} {\bibinfo
  {journal} {Phys. Rev. E}\ }\textbf {\bibinfo {volume} {88}},\ \bibinfo
  {pages} {052205} (\bibinfo {year} {2013})}\BibitemShut {NoStop}%
\bibitem [{\citenamefont {Szulc}\ \emph {et~al.}(2022)\citenamefont {Szulc},
  \citenamefont {Mungan},\ and\ \citenamefont {Regev}}]{Szulc2022-gx}%
  \BibitemOpen
  \bibfield  {author} {\bibinfo {author} {\bibfnamefont {A.}~\bibnamefont
  {Szulc}}, \bibinfo {author} {\bibfnamefont {M.}~\bibnamefont {Mungan}},\ and\
  \bibinfo {author} {\bibfnamefont {I.}~\bibnamefont {Regev}},\ }\bibfield
  {title} {\bibinfo {title} {Cooperative effects driving the multi-periodic
  dynamics of cyclically sheared amorphous solids},\ }\href
  {https://doi.org/10.1063/5.0087164} {\bibfield  {journal} {\bibinfo
  {journal} {J. Chem. Phys.}\ }\textbf {\bibinfo {volume} {156}},\ \bibinfo
  {pages} {164506} (\bibinfo {year} {2022})}\BibitemShut {NoStop}%
\bibitem [{\citenamefont {Lerner}\ and\ \citenamefont
  {Bouchbinder}(2018)}]{Lerner2018-me}%
  \BibitemOpen
  \bibfield  {author} {\bibinfo {author} {\bibfnamefont {E.}~\bibnamefont
  {Lerner}}\ and\ \bibinfo {author} {\bibfnamefont {E.}~\bibnamefont
  {Bouchbinder}},\ }\bibfield  {title} {\bibinfo {title} {A characteristic
  energy scale in glasses},\ }\href {https://doi.org/10.1063/1.5024776}
  {\bibfield  {journal} {\bibinfo  {journal} {J. Chem. Phys.}\ }\textbf
  {\bibinfo {volume} {148}},\ \bibinfo {pages} {214502} (\bibinfo {year}
  {2018})}\BibitemShut {NoStop}%
\end{thebibliography}


%

\end{document}